\documentclass{article}
\usepackage{amsfonts,amssymb}
\usepackage{xy}
\usepackage[breaklinks=true]{hyperref}
\usepackage{float}
\usepackage{afterpage}
\usepackage{graphicx}
\usepackage{hyperref}

\input xy
\xyoption{all}

\newcommand{\BOX}{\hbox {$\sqcap$ \kern -1em $\sqcup$}}
\newcommand{\qed}{\hskip 3em \hbox{\BOX} \vskip 2ex}

\newcommand{\maps}{\colon}    
\newcommand{\R}{{\mathbb R}}  
\newcommand{\C}{{\mathbb C}}  
\newcommand{\Z}{{\mathbb Z}}  

\renewcommand{\O}{{\rm O}}    
\newcommand{\U}{{\rm U}}    
\renewcommand{\S}{{\rm S}}    
\newcommand{\SO}{{\rm SO}}    
\newcommand{\SU}{{\rm SU}}    
\newcommand{\GL}{{\rm{GL}}}  
\newcommand{\SL}{{\rm{SL}}}  
\newcommand{\Spin}{{\rm Spin}}    

\renewcommand{\sl}{\mathfrak{sl}} 
\newcommand{\so}{{\mathfrak{so}}}  
\newcommand{\su}{{\mathfrak{su}}}  
\newcommand{\gl}{{\mathfrak{gl}}}  
\renewcommand{\u}{{\mathfrak{u}}}  
\newcommand{\g}{{\mathfrak{g}}}  

\newcommand{\Ex}{\Lambda} 
\newcommand{\Exev}{\Lambda^{\rm ev}} 
\newcommand{\Exodd}{\Lambda^{\rm odd}} 
\newcommand{\End}{{\rm End}} 
\newcommand{\Sym}{{\rm Sym}} 
\newcommand{\Cliff}{{\rm Cliff}}    

\newcommand{\inclusion}{\hookrightarrow}
\newcommand{\iso}{\cong} 

\newcommand{\half}{\frac{1}{2}} 
\newcommand{\third}{\frac{1}{3}} 
\newcommand{\twothirds}{\frac{2}{3}} 
\newcommand{\fourthirds}{\frac{4}{3}} 
\newcommand{\rev}{\rho_{\rm{ev}}}
\newcommand{\rodd}{\rho_{\rm{odd}}}

\newcommand{\GSM}{{G_{\mbox{\rm SM}}}}  

\newcommand{\ubar}{\overline{u}} 
\newcommand{\dbar}{\overline{d}} 
\newcommand{\nubar}{\overline{\nu}} 
\newcommand{\rbar}{{\overline{r}}} 
\newcommand{\gbar}{{\overline{g}}} 
\newcommand{\bbar}{{\overline{b}}} 
\newcommand{\wbar}{{\overline{w}}} 
\newcommand{\cbar}{{\overline{c}}} 
\newcommand{\lep}{\left( \! \begin{array}{c} \nu_L \\ e^-_L \end{array} \! \right)} 
\newcommand{\anglep}{\left\langle \! \begin{array}{c} \nu_L \\ e^-_L \end{array} \! \right\rangle} 
\newcommand{\angantilep}{\left\langle \! \begin{array}{c} e^+_R \\ \nubar_R \end{array} \! \right\rangle} 
\newcommand{\quark}{\left( \! \begin{array}{c} u_L \\ d_L \end{array} \! \right)} 
\newcommand{\angquark}{\left\langle \! \begin{array}{c} u_L \\ d_L \end{array} \! \right\rangle} 
\newcommand{\angantiquark}{\left\langle \! \begin{array}{c} \dbar_R \\ \ubar_R \end{array} \! \right\rangle} 
\newcommand{\quarkwithcolor}{\renewcommand{\arraystretch}{1.2} \left( \! \begin{array}{c} u^r_L, u^g_L, u^b_L \\ d^r_L, d^g_L, d^b_L \end{array} \! \right) \renewcommand{\arraystretch}{1}} 

\newcommand{\define}[1]{{\bf #1}}

\newcommand{\et}{\hspace{-0.08in}{\bf .}\hspace{0.1in}}

\newtheorem{thm}{Theorem}

        \newcommand{\be}{\begin{equation}}
        \newcommand{\ee}{\end{equation}}
        \newcommand{\ba}{\begin{eqnarray}}
        \newcommand{\ea}{\end{eqnarray}}
        \newcommand{\ban}{\begin{eqnarray*}}
        \newcommand{\ean}{\end{eqnarray*}}
        \newcommand{\barr}{\begin{array}}
        \newcommand{\earr}{\end{array}}

\title{The Algebra of Grand Unified Theories} 

\author{John Baez and John Huerta \\
\\
Department of Mathematics \\
University of California \\
Riverside, CA 92521 USA 
}
\date{\today}

\begin{document}

\maketitle

\begin{abstract}
The Standard Model is the best tested and most widely accepted theory of
elementary particles we have today.  It may seem complicated and arbitrary, but
it has hidden patterns that are revealed by the relationship between three
`grand unified theories': theories that unify forces and particles by extending
the Standard Model symmetry group $\U(1) \times \SU(2) \times \SU(3)$ to a
larger group.  These three are Georgi and Glashow's $\SU(5)$ theory,
Georgi's theory based on the group $\Spin(10)$, and the Pati--Salam model based
on the group $\SU(2) \times \SU(2) \times \SU(4)$.  In this expository account
for mathematicians, we explain only the portion of these theories that involves
finite-dimensional group representations.  This allows us to reduce the
prerequisites to a bare minimum while still giving a taste of the profound
puzzles that physicists are struggling to solve.
\end{abstract}

\section{Introduction} \label{sec:introduction}

The Standard Model of particle physics is one of the greatest triumphs
of physics.  This theory is our best attempt to describe all the
particles and all the forces of nature... \textit{except} gravity.  It
does a great job of fitting experiments we can do in the lab.  But
physicists are dissatisfied with it.  There are three main reasons.
First, it leaves out gravity: that force is described by Einstein's
theory of general relativity, which has not yet been reconciled with
the Standard Model.  Second, astronomical observations suggest that
there may be forms of matter not covered by the Standard Model---most
notably, `dark matter'.  And third, the Standard Model is complicated
and seemingly arbitrary.  This goes against the cherished notion that
the laws of nature, when deeply understood, are simple and beautiful.

For the modern theoretical physicist, looking beyond the Standard
Model has been an endeavor both exciting and frustrating.  Most modern
attempts are based on string theory.  There are also other interesting
approaches, such as loop quantum gravity and theories based on
noncommutative geometry.  But back in the mid 1970's, before any of
the currently popular approaches rose to prominence, physicists
pursued a program called `grand unification'.  This sought to unify
the forces and particles of the Standard Model using the mathematics
of Lie groups, Lie algebras, and their representations.  Ideas from
this era remain influential, because grand unification is still one of
the most fascinating attempts to find order and beauty lurking within
the Standard Model.

This paper is a gentle introduction to the group representations that
describe particles in the Standard Model and the most famous grand
unified theories.  To make the material more approachable for
mathematicians, we limit our scope by not discussing
particle interactions or `symmetry breaking'---the way a theory with
a large symmetry group can mimic one with a smaller group at low energies.  
These topics lie at the very heart of particle physics.  But by omitting
them, we can focus on ideas from algebra that many mathematicians will
find familiar, while introducing the unfamiliar ways that physicists
use these ideas.

In fact, the essential simplicity of the representation theory
involved in the Standard Model and grand unified theories is quite
striking.  The usual textbook approach to particle physics proceeds
through quantum field theory and gauge theory.  While these subjects
are very important to modern mathematics, learning them is a
major undertaking.  We have chosen to focus on the algebra of grand
unified theories because many mathematicians have the 
prerequisites to understand it with only a little work.

A full-fledged treatment of particle physics requires
quantum field theory, which uses representations of a noncompact
Lie group called the Poincar\'e group on \emph{infinite-dimensional} 
Hilbert spaces.  This brings in a lot of analytical subtleties, 
which make it hard to formulate theories of 
particle physics in a mathematically rigorous way.  In fact, no one 
has yet succeeded in doing this for the Standard Model.  But by 
neglecting the all-important topic of particle interactions, we 
can restrict attention to \emph{finite-dimensional} Hilbert spaces:
that is, finite-dimensional complex inner product spaces.  
This makes our discussion purely algebraic in flavor.

Every theory we consider has an `internal symmetry group'
or `gauge group'.   This is a compact Lie group, say $G$.  
Particles then live in representations of $G$ on a finite-dimensional 
Hilbert space $V$.  More precisely: $V$ can always be 
decomposed as a direct sum of irreducible representations, or 
\textbf{irreps}---and for our limited purposes, \textsl{particles are 
basis vectors of irreps}.  This provides a way to organize particles, 
which physicists have been exploiting since the 1960s.

The idea of `unification' has a clear meaning in these terms.
Suppose $V$ is a representation, not only of $G$, but also 
some larger group $H$ having $G$ as a subgroup.  Then we expect 
$V$ to decompose into fewer irreps as a representation of $H$ than 
as a representation of $G$, because elements of $H$ can mix
different irreps of $G$.  So: by introducing a larger symmetry 
group, particles can be unified into larger irreps.

`Grand unification' occurs when the compact Lie group $G$ is simple,
and thus not a product of other groups.  A gauge theory based on $G$
requires an invariant inner product on its Lie algebra.  When $G$ is
simple, this form is unique up to a scale factor, which physicists
call a `coupling constant': this measures the strength of the force
corresponding to $G$.  When $G$ is the product of simple factors, 
there is one coupling constant for each factor of $G$. So, by 
using a simple Lie group as gauge group, we minimize the number 
of coupling constants.  

In this paper, we give an account of the algebra behind the Standard
Model and three attempts at unification: Georgi and Glashow's
$\SU(5)$ theory, Georgi's theory based on the group $\Spin(10)$
(physicists call this the $\SO(10)$ theory), and the Pati--Salam
model.  All three date to around 1974.  The first two are known
as grand unified theories, or GUTs, because they are based on
simple Lie groups.  The Pati--Salam model is different: while it is
called a GUT by some authors, and does indeed involve unification,
it is based on the Lie group $\SU(2) \times \SU(2) \times \SU(4)$,
which is merely semisimple.

It is important to note that these theories have their problems.  The
$\SU(5)$ theory predicts that protons will decay more quickly than
they do, and it requires certain trends to hold among the relative
strengths of forces at high energies---trends which the data do not
support.  The $\SO(10)$ theory may still be viable, especially if at
low enough energies it reduces to the Pati--Salam model.  However, the issues
involved are complex.  For details, see the paper by Bertolini \textsl{et al}
and the many references therein \cite{BLM}.

Nonetheless, it is still very much worthwhile for mathematicians to
study the algebra of grand unified theories.  First, even apart from
their physical significance, these theories are intrinsically
beautiful mathematical structures.  Second, they provide a nice way
for mathematicians to get some sense of the jigsaw puzzle that
physicists are struggling to solve.  It is certainly hopeless trying
to understand what physicists are trying to accomplish with string
theory without taking a look at grand unified theories.  Finally,
grand unified theories can be generalized by adding
`supersymmetry'---and the resulting generalizations are considered
serious contenders for describing the real world.  For some recent
overviews of their prospects, see Pati~\cite{pati:decay,pati:probing}
and Peskin \cite{Peskin}.

This is how we shall proceed. In Section~\ref{sec:sm} we start by
describing the Standard Model.  After a brief nod to the electron and
photon, we explain some nuclear physics in Section~\ref{sec:isospin}.
We start with Heisenberg's old attempt to think of the proton
and neutron as two states of a single particle, the `nucleon', 
described by a 2-dimensional representation of  $\SU(2)$.  The 
idea of unification through representation theory traces its origins 
back to this notion.

After this warmup we tour the Standard Model in its current form.
In Section~\ref{sec:fermions} we describe the particles called
`fundamental fermions', which constitute matter.  In
Section~\ref{sec:forces} we describe the particles called `gauge
bosons', which carry forces.  Apart from the elusive Higgs boson, all
particles in the Standard Model are of these two kinds.   In 
Section~\ref{sec:smrep} we give a more mathematical treatment of
these ideas: the gauge bosons are determined by the Standard Model
gauge group 
\[      \GSM = \U(1) \times \SU(2) \times \SU(3) , \]
while the fundamental fermions and their antiparticles are basis
vectors of a highly reducible representation of this group,
which we call $F \oplus F^*$.  Here $F$ describes the fermions, while
$F^*$ describes their antiparticles.
  
Amazingly, using the ideas of gauge theory and quantum field theory, plus 
the `Higgs mechanism' for symmetry breaking, we can recover the dynamical
laws obeyed by these particles from the representation of $\GSM$ on $F
\oplus F^*$.  This information is enough to decode the physics of these
particles and make predictions about what is seen in the gigantic
accelerators that experimental physicists use to probe the natural
world at high energies.  Unfortunately, to explain all this would 
go far beyond the modest goals of this paper.  For a guide
to further study, see Section \ref{subsec:reading}.

Having acquainted the reader with the Standard Model of particle
physics in Section~\ref{sec:sm}, we then go on to talk about grand
unified theories in Section~\ref{sec:guts}. These theories go beyond
the Standard Model by `extending' the gauge group. That is, we pick a
way to include $\GSM$ in some larger group $G$, and choose a 
representation $V$ of $G$ which reduces to the Standard Model
representation $F \oplus F^*$ when we restrict it to $\GSM$.  We
describe how this works for the $\SU(5)$ theory
(Section~\ref{sec:su(5)}), the $\SO(10)$ theory
(Section~\ref{sec:so(10)}), and the Pati--Salam model
(Section~\ref{sec:g(2,2,4)}).
Of course, since we do not discuss the dynamics, a lot will go
unsaid about these GUTs. 

As we proceed, we explain how the $\SU(5)$ theory and the Pati--Salam
model are based on two distinct visions about how to extend the
Standard Model.  However, we will see that the $\SO(10)$ theory is an
extension of both the $\SU(5)$ theory (Section~\ref{sec:so(10)}) and
the Pati--Salam model (Section~\ref{sec:route}). Moreover, these two
routes to the $\SO(10)$ theory are compatible in a precise sense: we
get a commuting square of groups, and a commuting square of
representations, which fit together to form a commuting cube
(Section~\ref{sec:compatibility}).

In Section~\ref{sec:conclusion}, we conclude by discussing what this
means for physics: namely, how the Standard Model reconciles the 
two visions of physics lying behind the
$\SU(5)$ theory and the Pati--Salam model.  In a sense, it is the
intersection of the $\SU(5)$ theory and the Pati--Salam model within
their common unification, $\SO(10)$.

Throughout the course of the paper, we occasionally summarize our 
progress in theorems, most phrased in terms of commutative
diagrams:

\begin{itemize}
\item Section 3.1, Theorem \ref{thm:su(5)}: the $\SU(5)$ theory extends the
Standard Model.
\item Section 3.2, Theorem \ref{thm:spinor}: the $\Spin(10)$ theory extends
the $\SU(5)$ theory.
\item Section 3.3, Theorem \ref{thm:Pati--Salam}: the Pati--Salam model 
extends the Standard Model.
\item Section 3.4, Theorem \ref{thm:Pati--Salam2}: the Pati--Salam model 
is isomorphic to a theory involving $\Spin(4) \times \Spin(6)$.
\item Section 3.4, Theorem \ref{thm:Pati--Salam3}: the $\Spin(4) \times 
\Spin(6)$ theory extends the Standard Model.
\item Section 3.4, Theorem \ref{thm:Spin(10)}: the $\Spin(10)$ theory extends
the $\Spin(4) \times \Spin(6)$ theory.
\item Section 3.5, Theorem \ref{thm:cube}: the Standard Model, the $\SU(5)$
theory, the $\Spin(4) \times \Spin(6)$ theory and the $\Spin(10)$
theory fit together in a commutative cube.
\item Section 4, Theorem \ref{thm:pullbackSO(10)}: The true gauge group of
the Standard Model is the intersection of $\SU(5)$ and $\SO(4) \times
\SO(6)$ in $\SO(10)$.
\item Section 4, Theorem \ref{thm:pullbackSpin(10)}: The true gauge group of
the Standard Model is the intersection of $\SU(5)$ and $(\Spin(4) \times
\Spin(6))/\mathbb{Z}_2$ in $\Spin(10)$.
\end{itemize}

\subsection{Guide to Further Reading} \label{subsec:reading}

We have tried to limit our prerequisites to the bare
minimum.  This includes basic facts about Lie groups, Lie
algebras, and their representations---especially finite-dimensional 
unitary representations of compact Lie groups.  We will not need 
the structure theory for simple Lie groups.  We do, however, 
assume a little familiarity with the classical Lie groups 
$\GL(n)$, $\SL(n)$, $\O(n)$, $\SO(n)$, $\U(n)$, and $\SU(n)$, 
as well as their Lie algebras.

There are countless books on Lie groups, Lie algebras and their 
representations, but the text by Hall \cite{hall} has everything we 
need, and more.  Sternberg's introduction to group theory and physics
\cite{sternberg} includes an excellent account of applications 
to particle physics.  To see the subject more through the eyes
of a physicist, try the books by Lipkin \cite{lipkin} or 
Tinkham \cite{tinkham}.  Georgi's text \cite{georgi:lie} shows 
how the subject looks to one of the inventors of grand unified theories. 

Starting in Section \ref{sec:su(5)} we assume familiarity
with exterior algebras, and in Section \ref{sec:so(10)} we
also use Clifford algebras.  For what we need, 
Chevalley's book \cite{Chevalley} is more than sufficient.

For the interested reader, there are many introductions to particle physics
where one can learn the dynamics after getting a taste of the algebra here.  
It might be good to start by reading Griffiths' introductory
book~\cite{griffiths:intro} together with Sudbery's text specially designed for
mathematicians \cite{sudbery}.  The book by Huang~\cite{huang:qlgf} delves as
deep as one can go into the Standard Model without a heavy dose of quantum
field theory, and the book by Lee~\cite{lee:pp} is full of practical wisdom.
For more information on grand unified theories, see the textbooks by
Ross~\cite{ross:gut} and Mohapatra~\cite{mohapatra:us}.

Particle physics relies heavily on quantum field theory. 
There are many books on this subject, none of which make it easy.
Prerequisites include a good understanding of classical mechanics, 
classical field theory and quantum mechanics.  Many physicists 
consider the books by Brown \cite{Brown} and Ryder \cite{Ryder} to 
be the most approachable.  The text by Peskin and 
Schroeder~\cite{PeskinSchroeder:qft} offers a lot of physical insight, 
and we have also found Zee's book~\cite{zee:nutshell} 
very useful in this respect.  Srednicki's text~\cite{srednicki:qft} is clear 
about many details that other books gloss over---and it costs nothing!  Of 
course, these books are geared toward physicists: mathematicians may
find the lack of rigor frustrating.  Ticciati~\cite{ticciati:qft}
provides a nice introduction for mathematicians, but anyone serious
about this subject should quickly accept the fact that quantum field 
theory has not been made rigorous: this is a project for the century to come.

Particle physics also relies heavily on geometry, especially gauge
theory.  This subject is easier to develop in a rigorous way, so there are
plenty of texts that describe the applications to physics, but which a
mathematician can easily understand.  Naber's books are a great place
to start \cite{naber:foundations, naber:interactions}, and one of us
has also written an elementary introduction \cite{BaezMuniain}.  Isham's
text is elegant and concise \cite{isham}, and many people swear by
Nakahara \cite{nakahara}.  The quantum field theory texts mentioned
above also discuss gauge theory, but in language less familiar to 
mathematicians.

Finally, few things are more enjoyable than the history of nuclear
and particle physics---a romantic tale full of heroic figures
and tightly linked to the dark drama of World War II, the Manhattan Project,
and the ensuing Cold War.  Crease and Mann~\cite{CreaseMann:sc} 
give a very readable introduction.  To dig deeper, try the book 
by Segr{\`e}~\cite{segre:modern}, or the still more detailed 
treatments by Pais~\cite{pais:ib} and Hoddeson \textsl{et al}~\cite{hoddeson}. 

\section{The Standard Model} \label{sec:sm}

Today, most educated people know that the world is made of atoms, and
that atoms, in turn, are made of electrons, protons, and neutrons. The
electrons orbit a dense nucleus made of protons and neutrons, and as
the outermost layer of any atom's structure, they are responsible for
all chemistry. They are held close to the nucleus by electromagnetic
forces: the electrons carry a negative electric charge, and protons
carry a positive charge. Opposite charges attract, and this keeps the
electrons and the nucleus together.

At one point in time, electrons, protons, and neutrons were all
believed to be fundamental and without any constituent parts, just as
atoms themselves were once believed to be, before the discovery of the
electron. Electrons are the only one of these subatomic particles
still considered fundamental, and it is with this venerable particle
that we begin a table of the basic constituents of matter, called
`fundamental fermions'.  We will see more soon.

\vskip 1em
\begin{center}
	\begin{tabular}{lcc}
		\hline
		\multicolumn{3}{|c|}{\bf{Fundamental Fermions (first try)}} \\
		\hline
		Name & Symbol & Charge  \\
		\hline
		Electron & $e^-$ & -1  \\
		\hline
\end{tabular}
\end{center}
\vskip 1em

Since the electron is charged, it participates in electromagnetic
interactions.  From the modern perspective of
quantum field theory, electromagnetic interactions are mediated by the exchange
of virtual photons, particles of light that we never see in the lab, but whose
effects we witness whenever like charges are repelled or opposite charges are
attracted.  We depict this process with a diagram:

\vskip 1em
\begin{center}
	\includegraphics[scale=0.75]{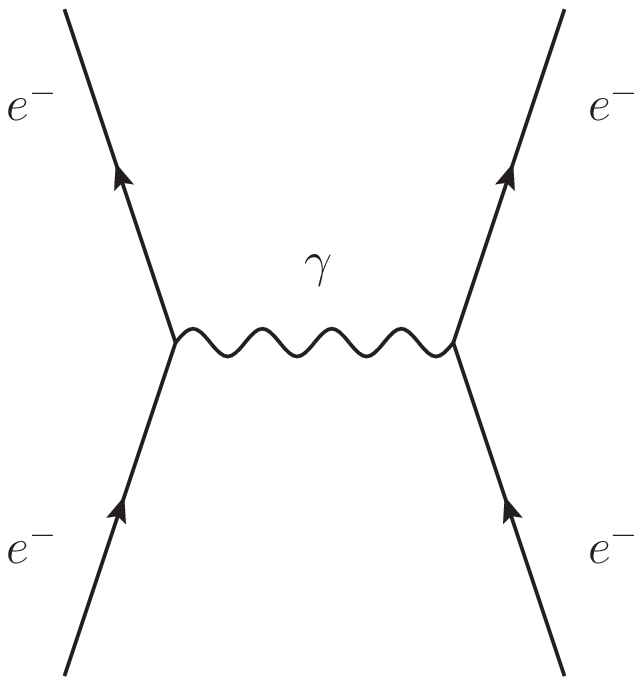}
\end{center}
\vskip 1em

Here, time runs up the page. Two electrons come in,
exchange a photon, and leave, slightly repelled from each other by the process.

The photon is our next example of a fundamental particle, though it is of a
different character than the electron and quarks. As a mediator of forces, the
photon is known as a \textbf{gauge boson} in modern parlance. It is massless, 
and interacts only with charged particles, though it carries no charge itself.
So, we begin our list of gauge bosons as follows:

\vskip 1em
\begin{center}
	\begin{tabular}{llc}
		\hline
		\multicolumn{3}{|c|}{\bf{Gauge Bosons (first try)}} \\
		\hline
		Force & Gauge Boson & Symbol \\
		\hline
		Electromagnetism & Photon & $\gamma$ \\
		\hline
	\end{tabular}
\end{center}
\vskip 1em

\subsection{Isospin and {\rm{SU(2)}}} \label{sec:isospin}

Because like charges repel, it is remarkable that the atomic nucleus
stays together. After all, the protons are all positively charged and
are repelled from each other electrically.  To hold these particles so
closely together, physicists hypothesized a new force, the
\textbf{strong force}, strong enough to overcome the electric
repulsion of the protons.  It must be strongest only at short
distances (about $10^{-15}$ m), and then it must fall off rapidly, for
protons are repelled electrically unless their separation is that
small.  Neutrons must also experience it, because they are bound to
the nucleus as well.

Physicists spent several decades trying to understand the strong force; it was
one of the principal problems in physics in the mid-twentieth century. About
1932, Werner Heisenberg, pioneer in quantum mechanics, discovered one of the
first clues to its nature. He proposed, in \cite{heisenberg:77}, that the
proton and neutron might really be two states of the same particle, now
called the \textbf{nucleon}.  In modern terms, he attempted to unify the 
proton and neutron. 

To understand how, we need to know a little quantum mechanics.
In quantum mechanics, the state of any physical system is given by a 
unit vector in a complex Hilbert space, and it is possible to take 
complex linear combinations of the system in different states.  For 
example, the state for a quantum system, like a particle on a line, 
is a complex-valued function 
\[	\psi \in L^2(\R) , 	\]
or if the particle is confined to a 1-dimensional box, so that its
position lies in the unit interval $[0,1]$, then its state lives 
in the Hilbert space $L^2([0,1])$.

We have special rules for combining quantum systems. If, say, we have two
particles in a box, particle 1 and particle 2, then the state is a
function of both particle 1's position and particle 2's:
\[	\psi \in L^2([0,1] \times [0,1]),	\]
but this is isomorphic to the tensor product of particle 1's Hilbert space with
particle 2's:
\[	L^2( [0,1] \times [0,1] ) \iso L^2([0,1] \otimes L^2([0,1]).	\]
This is how we combine systems in general. If a system consists of one part
with Hilbert space $V$ \emph{and} another part with Hilbert space $W$, 
their tensor product $V \otimes W$ is the Hilbert space of the combined system.
Heuristically, 
\[	\textit{and} = \otimes.	\]
We just discussed the Hilbert space for two particles in a single box. We now
consider the Hilbert space for a single particle in two boxes, by which we mean
a particle that is in one box, say $[0,1]$, \emph{or} in another box, say 
$[2,3]$.  The Hilbert space here is
\[	L^2( [0,1] \cup [2,3]) \iso L^2([0,1]) \oplus L^2([2,3]).	\]
In general, if a system's state can lie in a Hilbert space $V$ \emph{or} in a 
Hilbert space $W$, the total Hilbert space is then
\[	V \oplus W.	\]
Heuristically,
\[	\textit{or} = \oplus.	\]

Back to nucleons. According to Heisenberg's theory, a nucleon is a
proton \textit{or} a neutron.  If we use the simplest nontrivial Hilbert
space for both the proton and neutron, namely $\C$, then the Hilbert
space for the nucleon should be
\[       \C^2 \iso \C \oplus \C.	\]
The proton and neutron then correspond to basis vectors of this Hilbert
space:
\[	p = \left( \begin{array}{c} 1 \\ 0 \end{array} \right) \in \C^2	\]
and
\[	n = \left( \begin{array}{c} 0 \\ 1 \end{array} \right) \in \C^2.	\]
But, we can also have a nucleon in a linear combination of 
these states.  More precisely, the state of the nucleon can be represented 
by any  unit vector in $\C^2$. 

The inner product in $\C^2$ then allows us to compute probabilities, using the
following rule coming from quantum mechanics: the probability that a system in
state $\psi \in H$, a given Hilbert space, will be observed in state $\phi \in
H$ is
\[	\left| \langle \psi , \phi \rangle \right|^2.	\]
Since $p$ and $n$ are orthogonal, there is no chance of seeing a proton as
a neutron or vice versa, but for a nucleon in the state
\[	\alpha p + \beta n \in \C^2,	\]
there is probability $| \alpha |^2$ that
measurement will result in finding a proton, and $|\beta|^2$ that 
measurement will result in finding a neutron.  The condition that our
state be a unit vector ensures that these probabilities add to 1.

In order for this to be interesting, however, there must be
processes that can turn protons and neutrons into different states of
the nucleon. Otherwise, there would be no point in having the full
$\C^2$ space of states.  Conversely, if there are processes which can
change protons into neutrons and back, it turns out we need all of 
$\C^2$ to describe them.

Heisenberg believed in such processes, because of an analogy between
nuclear physics and atomic physics. The analogy turned out to be poor,
based on the faulty notion that the neutron was composed of a proton
and an electron, but the idea of the nucleon with states in $\C^2$
proved to be a breakthrough.

The reason is that in 1936 a paper by Cassen and
Condon~\cite{CassenCondon:nuclearforces} appeared suggesting that the
nucleon's Hilbert space $\C^2$ is acted on by the symmetry group $\SU(2)$. 
They emphasized the analogy between this and the spin of the electron,
which is also described by vectors in $\C^2$, acted on by the double
cover of the 3d rotation group, which is also $\SU(2)$.  In keeping with 
this analogy, the property that distinguishes the proton
and neutron states of a nucleon was dubbed \textbf{isospin}.  The
proton was declared the \textbf{isospin up} state or $I_3 = \half$
state, and the neutron was declared the \textbf{isospin down} or $I_3 =
-\half$ state. Cassen and Condon's paper put isospin on its way to
becoming a useful tool in nuclear physics.

Isospin proved useful because it formalized the following idea, which
emerged from empirical data around the time of Cassen and Condon's
paper.  Namely: the strong force, unlike the electromagnetic force, is
the same whether the particles involved are protons or neutrons.
Protons and neutrons are interchangeable, as long as we neglect the
small difference in their mass, and most importantly, as long as we
neglect electromagnetic effects.  One can phrase this idea in terms of
group representation theory as follows: the strong force is
\emph{invariant} under the action of $\SU(2)$.

Though this idea was later seen to be an oversimplification, 
it foreshadowed modern ideas about unification. The proton, living in the
representation $\C$ of the trivial group, and the neutron, living in a
different representation $\C$ of the trivial group, were unified into the
nucleon, with representation $\C^2$ of $\SU(2)$.  These symmetries hold 
for the strong force, but not for electromagnetism: we say this force
`breaks' $\SU(2)$ symmetry.

But what does it mean, exactly, to say that a force is invariant under the 
action of some group?    It means that when we are studying particles 
interacting via this force, the Hilbert space of each particle
should be equipped with a unitary representation of this group.
Moreover, any physical process caused by this force should be described 
by an `intertwining operator': that is, a linear operator that
respects the action of this group.  A bit more precisely, suppose
$V$ and $W$ are finite-dimensional Hilbert spaces on which some group 
$G$ acts as unitary operators.  Then a linear operator $F \maps V \to W$ is
an \textbf{intertwining operator} if 
\[            F(g \psi) = gF(\psi)   \]
for every $\psi \in V$ and $g \in G$.

Quite generally, symmetries give rise to conserved quantities.  
In quantum mechanics this works as follows.  Suppose that $G$ is
a Lie group with a unitary representation on the finite-dimensional
Hilbert spaces $V$ and $W$.  Then $V$ and $W$ automatically become 
representations of $\g$, the Lie algebra of $G$, and any intertwining operator 
$F \maps V \to W$ respects the action of $\g$.  In other words,
\[              F(T \psi) = T F(\psi)  \]
for every $\psi \in V$ and $T \in \g$.
Next suppose that $\psi \in V$ is an eigenvector of $T$:
\[                  T\psi = i \lambda \psi  \]
for some real number $\lambda$.  Then it is easy to check $F(\psi)$ 
is again an eigenvector of $T$ with the same eigenvalue:
\[                 T F(\psi) = i \lambda F(\psi)  .\]
So, the number $\lambda$ is `conserved' by the operator $F$. 

The element $T \in \g$ will act as a skew-adjoint operator on any
unitary representation of $G$.  Physicists prefer to work with 
self-adjoint operators since these have real eigenvalues.  
In quantum mechanics, self-adjoint operators are called `observables'.  
We can get an observable by dividing $T$ by $i$.  

In Casson and Condon's isospin theory of the strong interaction, the symmetry
group $G$ is $\SU(2)$.  Here isospin, or more precisely $I_3$, arises
as above: it is just the eigenvalue of a certain element of $\su(2)$,
divided by $i$ to get a real number.  Because any physical process 
caused by the strong force is described by an intertwining operator, 
\emph{isospin is conserved}.  So, the total $I_3$ of any system 
remains unchanged after a process that involves only strong interactions.

Nevertheless, for the states in $\C^2$ that mix protons and neutrons to have
any meaning, there must be a mechanism which can convert protons into neutrons
and vice versa. Mathematically, we have a way to do this: the action of
$\SU(2)$.  What does this correspond to, physically?

The answer originates in the work of Hideki Yukawa. In the early 1930s, he
predicted the existence of a particle that mediates the strong force,
much as the photon mediates the electromagnetic force.  From known 
properties of the strong force, he was able to predict that this particle
should be about 200 times as massive as the electron,
or about a tenth the mass of a proton. He predicted that experimentalists would
find a particle with a mass in this range, and that it would interact strongly
when it collided with nuclei.

Partially because of the intervention of World War II, it took over
ten years for Yukawa's prediction to be vindicated.  After a 
famous false alarm (see Section~\ref{sec:generations}), 
it became clear by 1947 that a particle with the
expected properties had been found. It was called the \textbf{pion} and
it came in three varieties: one with positive charge, the $\pi^+$, one
neutral, the $\pi^0$, and one with negative charge, the $\pi^-$.

\begin{figure}[H]
\begin{center}
	\begin{tabular}{cc}
		\includegraphics[scale=0.75]{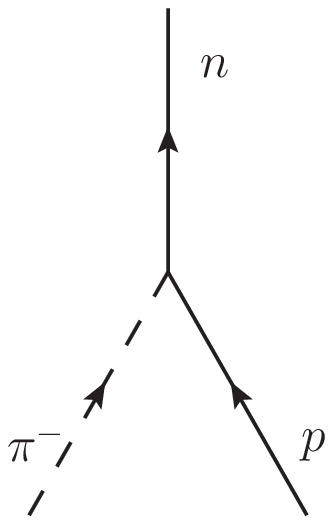} & \includegraphics[scale=0.75]{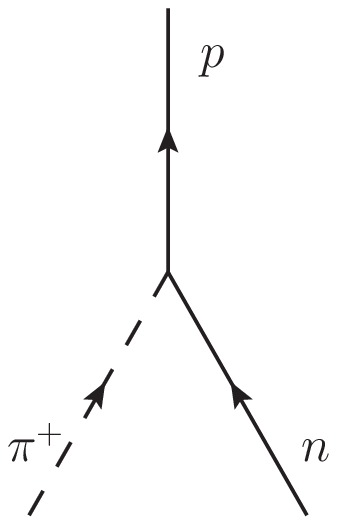} \\
		$\pi^- + p \to n$                             & $\pi^+ + n \to p$                             \\
		\\
		\\
		\includegraphics[scale=0.75]{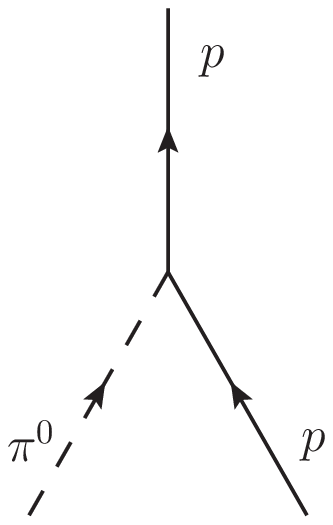} & \includegraphics[scale=0.75]{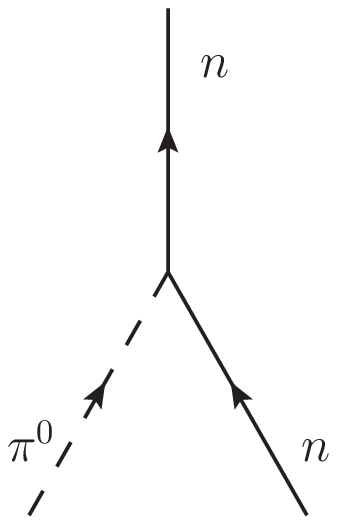} \\
		$\pi^0 + p \to p$                             & $\pi^0 + n \to n$                             \\
	\end{tabular}
	\caption{The nucleons absorbing pions.} \label{fig:piN_vertices}
\end{center}
\end{figure}

The pion proved to be the mechanism that can transform nucleons. To wit, we
observe processes such as those in Figure~\ref{fig:piN_vertices}, where we have
drawn the Feynman diagrams which depict the nucleons absorbing pions,
transforming where they are allowed to by charge conservation.

Because of isospin conservation, we can measure the $I_3$ of a pion by
looking at these interactions with the nucleons. It turns out that the
$I_3$ of a pion is the same as its charge:
\vskip 1em
\begin{center}
	\begin{tabular}{c|r}
		\hline
		Pion & $I_3$ \\
		\hline
		$\pi^+$ & $+1$ \\
		$\pi^0$ & 0 \\
		$\pi^-$ & $-1$ \\
		\hline
	\end{tabular}
\end{center}
Here we pause, because we can see the clearest example of a pattern that lies
at the heart of the Standard Model. It is the relationship between isospin
$I_3$ and charge $Q$. For the pion, isospin and charge are equal:
\[ Q(\pi) = I_3(\pi). \]
But they are also related for the nucleon, though in a subtler way:
\vskip 1em
\begin{center}
	\begin{tabular}{c|r|c}
		\hline
		Nucleon & $I_3$ & Charge \\
		\hline
		$p$ & $\half$  & 1 \\
		    &          &   \\
		$n$ & $-\half$ & 0 \\
		\hline
	\end{tabular}
\end{center}
\vskip 1em
The relationship for nucleons is
\[ Q(N) = I_3(N) + \frac{1}{2} . \]
This is nearly the most general relationship. It turns out that, for any given
family of particles that differ only by $I_3$, we have the
\textbf{Gell-Mann--Nishijima formula}:
\[ Q = I_3 + Y/2 \]
where the charge $Q$ and isospin $I_3$ depend on the particle, but a new
quantity, the \textbf{hypercharge} $Y$, depends only on the family. For example,
pions all have hypercharge $Y = 0$, while nucleons both have hypercharge $Y =
1$.

Mathematically, $Y$ being constant on `families' just means it is
constant on representations of the isospin symmetry group,
$\SU(2)$. The three pions, like the proton and neutron, are nearly
identical in terms of mass and their strong interactions.  In
Heisenberg's theory, the different pions are just different isospin
states of the same particle. Since there are three, they have to span
a three-dimensional representation of $\SU(2)$. 

Up to isomorphism, there is only one three-dimensional complex irrep 
of $\SU(2)$, which is $\Sym^2 \C^2$, the symmetric tensors of rank 2. 
In general, the unique $(n+1)$-dimensional irrep of $\SU(2)$ is given 
by $\Sym^n \C^2$.   Physicists call this the \textbf{\boldmath 
spin-$n/2$ representation} of $\SU(2)$, or in the present context, 
the `isospin-$n/2$ representation'.  This representation 
has a basis of vectors where $I_3$ ranges from $-n/2$ to $n/2$ 
in integer steps.  Nucleons lie in the isospin-$\half$ representation, 
while pions lie in the isospin-$1$ representation.

This sets up an interesting puzzle.   We know two ways to transform 
nucleons: the mathematical action of $\SU(2)$, and their physical 
interactions with pions. How are these related?

The answer lies in the representation theory. Just as the two nucleons
span the two-dimensional irrep of $\C^2$ of $\SU(2)$, the pions 
span the three-dimensional irrep $\Sym^2 \C^2$ of $\SU(2)$. But there
is another way to write this representation which sheds light on the
pions and the way they interact with nucleons: because $\SU(2)$ is
itself a three-dimensional \emph{real} manifold, its Lie algebra
$\su(2)$ is a three-dimensional \emph{real}
vector space. $\SU(2)$ acts on itself by conjugation, which fixes the
identity and thus induces linear transformations of $\su(2)$, giving a
representation of $\SU(2)$ on $\su(2)$ called the adjoint
representation.

For simple Lie groups such as $\SU(2)$, the adjoint representation is
irreducible.  Thus $\su(2)$ is a three-dimensional \emph{real} irrep
of $\SU(2)$.  This is different from the three-dimensional \emph{complex} 
irrep $\Sym^2 \C^2$, but very related. Indeed, $\Sym^2 \C^2$ is just the
complexification of $\su(2)$:
\[ \Sym^2 \C^2 \iso \C \otimes \su(2) \iso \sl(2,\C) . \]

The pions thus live in $\sl(2,\C)$, a complex Lie algebra, and
this acts on $\C^2$ because $\SU(2)$ does. To be precise, Lie group
representations induce Lie algebra representations, so the real Lie algebra
$\su(2)$ has a representation on $\C^2$.  This then extends
to a representation of the complex Lie algebra $\sl(2,\C)$. 
And this representation is even familiar---it is the fundamental 
representation of $\sl(2,\C)$ on $\C^2$.

Quite generally, whenever $\mathfrak{g}$ is the Lie algebra of a 
Lie group $G$, and $\rho \maps G \times V \to V$ is a representation of
$G$ on some finite-dimensional vector space $V$, we get
a representation of the Lie algebra $\mathfrak{g}$ on $V$, which
we can think of as a linear map
\[ d\rho \maps \mathfrak{g} \otimes V \to V .\]
And this map is actually an intertwining operator,
meaning that it commutes with the action of $G$: since $\mathfrak{g}$ 
and $V$ are both representations of $G$ this is a sensible thing to 
say, and it is easy to check.

Pions act on nucleons via precisely such an intertwining operator:
\[ \sl(2, \C) \otimes \C^2 \to \C^2 . \]
So, the interaction between pions and nucleons arises naturally from 
the action of $\SU(2)$ on $\C^2$ after we complexify the Lie algebra of 
this group!  

Physicists have invented a nice way to depict such
intertwining operators---\textbf{Feynman diagrams}:
\begin{figure}[H]
	\begin{center}
		\includegraphics[scale=0.75]{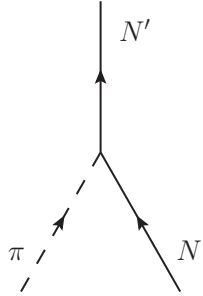}
	\end{center}
	\caption{A nucleon absorbs a pion.} \label{fig:piN_vertex}
\end{figure}

\noindent
Here we see a nucleon coming in, absorbing a pion, and leaving. That is, this
diagram depicts a basic \emph{interaction} between pions and nucleons.

Feynman diagrams are calculational tools in physics, though to actually
use them as such, we need quantum field theory. Then, instead of just
standing for intertwining operators between representations of compact
groups such as $\SU(2)$, they depict intertwining operators between
representations of the product of this group and the Poincar\'e group, 
which describes the symmetries of spacetime.  Unfortunately, the 
details are beyond the scope of this paper.  By ignoring the 
Poincar\'e group, we are, in the language of physics, restricting 
our attention to `internal degrees of freedom', and their `internal' 
(i.e., gauge) symmetries.

Nonetheless, we can put basic interactions like the one in
figure~\ref{fig:piN_vertex} together to form more complicated ones, like
this:

\begin{center}
	\includegraphics[scale=0.75]{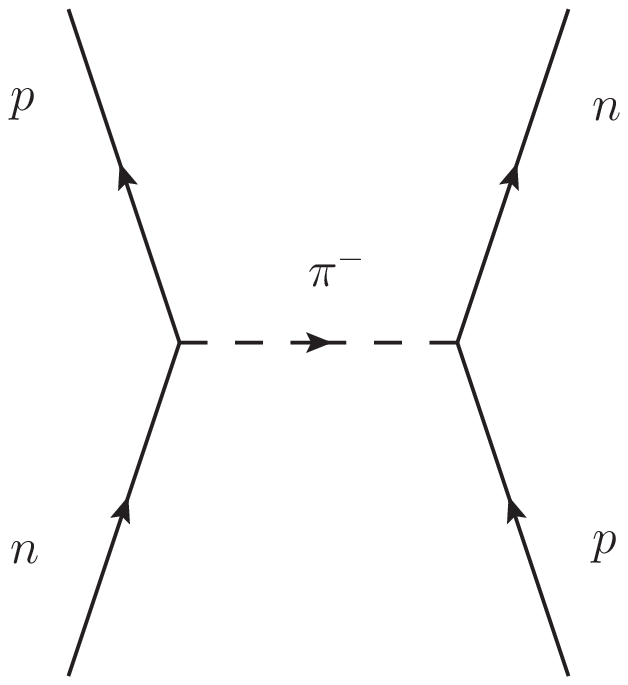}
\end{center}

Here, two nucleons interact by exchanging pions. This is the mechanism
for the strong force proposed by Yukawa, still considered
approximately right today.  Better, though, it depicts all the
representation-theoretic ingredients of a modern gauge theory in
physics. That is, it shows two nucleons, which live in a
representation $\C^2$ of the gauge group $\SU(2)$, interacting by the
exchange of a pion, which lives in the complexified adjoint rep,
$\C \otimes \su(2)$. In the coming sections we will see how these
ideas underlie the Standard Model.

\subsection{The Fundamental Fermions} \label{sec:fermions}

\subsubsection{Quarks} \label{sec:quarks}

In the last section, we learned how Heisenberg unified the proton and neutron
into the nucleon, and that Yukawa proposed that nucleons interact by exchanging
pions. This viewpoint turned out to be at least approximately true, but it was
based on the idea that the proton, neutron and pions were all fundamental
particles without internal structure, which was not ultimately supported by the
evidence.

Protons and neutrons are not fundamental. They are made of particles called
\textbf{quarks}. There are a number of different types of quarks, 
called \textbf{flavors}.  However, it takes only two
flavors to make protons and neutrons: the \textbf{up quark}, $u$, 
and the \textbf{down quark}, $d$. The proton consists of two up 
quarks and one down: 
\[ p = uud \]
while the neutron consists of one up quark and two down:
\[ n = udd \]
Protons have an electric charge of $+1$, exactly opposite the electron, while
neutrons are neutral, with $0$ charge. These two conditions are enough to
determine the charge of their constituents, which are fundamental fermions 
much like the electron:
\vskip 1em
\begin{center}
	\begin{tabular}{lccc}
		\hline
		\multicolumn{3}{|c|}{\bf{Fundamental Fermions (second try)}} \\
		\hline
		Name       & Symbol & Charge       \\
		\hline
		Electron   & $e^-$  & $-1$         \\
		\\
		Up quark   & $u \quad$ &  $+\twothirds$  \\
		\\
		Down quark & $d \quad$ &  $-\third$      \\
		\hline
	\end{tabular}
\end{center}
\vskip 1em
There are more quarks than these, but these are the lightest ones,
comprising the \textbf{first generation}.  They are all we need to make
protons and neutrons, and so, with the electron in tow, the above list
contains all the particles we need to make atoms.

Yet quarks, fundamental as they are, are never seen in isolation. They
are always bunched up into particles like the proton and neutron. This
phenomenon is called \textbf{confinement}.  It makes the long,
convoluted history of how we came to understand quarks, despite the
fact that they are never seen, all the more fascinating.
Unfortunately, we do not have space for this history here, but it can
be found in the books by Crease and Mann~\cite{CreaseMann:sc}, 
Segr\`e~\cite{segre:modern}, and Pais~\cite{pais:ib}.

It is especially impressive how physicists were able to discover
that each flavor of quark comes in three different states,
called \textbf{colors}: \textbf{red} $r$, \textbf{green} $g$, 
and \textbf{blue} $b$.  These `colors' have nothing to do with 
actual colors; they are just cute names---though as we shall see, 
the names are quite well chosen.  Mathematically, all that matters is
that the Hilbert space for a single quark is $\C^3$; we call the 
standard basis vectors $r, g$ and $b$.  The \textbf{color symmetry 
group} $\SU(3)$ acts on this Hilbert space in the obvious way, via 
its fundamental representation.

Since both up and down quarks come in three color states, there are 
really six kinds of quarks in the matter we see around us. 
Three up quarks, spanning a copy of $\C^3$:
\[ u^r, u^g, u^b \in \C^3 .\]
and three down quarks, spanning another copy of $\C^3$:
\[ d^r, d^g, d^b \in \C^3 .\]
The group $\SU(3)$ acts on each space.  
All six quarks taken together span this vector space:
\[ \C^3 \oplus \C^3 \iso \C^2 \otimes \C^3 \]
where $\C^2$ is spanned by the flavors $u$ and $d$.  Put another way,  a
first-generation quark comes in one of six flavor-color states.

How could physicists discover the concept of color, given that quarks
are confined?  In fact confinement was the key to this discovery!
Confinement amounts to the following decree: all observed
states must be {\bf white}, i.e., invariant under the action of
$\SU(3)$.  It turns out that this has many consequences.

For starters, this decree implies that we cannot see an individual
quark, because they all transform nontrivially under $\SU(3)$.  
Nor do we ever see a particle built from two quarks, since 
no unit vectors in $\C^3 \otimes \C^3$ are fixed by $\SU(3)$.  But 
we \emph{do} see particles made of three quarks: namely, nucleons!
This is because there \emph{are} unit vectors in 
\[ \C^3 \otimes \C^3 \otimes \C^3 \]
fixed by $\SU(3)$.  Indeed, as a representation of $\SU(3)$,
$\C^3 \otimes \C^3 \otimes \C^3$ contains precisely one copy of the trivial
representation: the antisymmetric rank three tensors, $\Ex^3 \C^3 \subseteq
\C^3 \otimes \C^3 \otimes \C^3$. This one dimensional vector space is spanned
by the wedge product of all three basis vectors:
\[ r \wedge b \wedge g \in \Ex^3 \C^3. \]
So, up to normalization, this \emph{must} be the color state of a nucleon. 
And now we see why the `color' terminology is well-chosen: an equal 
mixture of red, green and blue light is white.  This is just
a coincidence, but it is too cute to resist.

So: color is deeply related to confinement.  Flavor, on the other hand,
is deeply related to isospin.  Indeed, the flavor $\C^2$ is
suspiciously like the isospin $\C^2$ of the nucleon.  We even call the
quark flavors `up' and `down'.  This is no accident.  The proton and
neutron, which are the two isospin states of the nucleon, differ only
by their flavors, and only the flavor of one quark at that. If one
could interchange $u$ and $d$, one could interchange protons and
neutrons.

Indeed, we can use quarks to explain the isospin symmetry of
Section~\ref{sec:isospin}. Protons and neutrons are so similar, with
nearly the same mass and strong interactions, because $u$ and $d$
quarks are so similar, with nearly the same mass and truly identical
colors.

So as in Section~\ref{sec:isospin}, let $\SU(2)$ act on the flavor
states $\C^2$. By analogy with that section, we call this $\SU(2)$ the
isospin symmetries of the quark model. Unlike the color symmetries
$\SU(3)$, these symmetries are not exact, because $u$ and $d$ quarks
have different mass and charge.  Nevertheless, they are useful.

The isospin of the proton and neutron then arises from the isospin of
its quarks. Define $I_3(u) = \half$ and $I_3(d) = -\half$, making $u$
and $d$ the isospin up and down states at which their names hint. To
find the $I_3$ of a composite, like a proton or neutron, add the $I_3$
for its constituents.  This gives the proton and neutron the right
$I_3$:
\[
\begin{array}{ccccr}
	I_3( p ) & = & \half + \half - \half &=& \half \\
                                                       \\
	I_3( n ) & = & \half - \half - \half &=& -\half. \\
\end{array}
\]
Of course, having the right $I_3$ is not the whole story for
isospin. The states $p$ and $n$ must still span a copy of the fundamental rep
$\C^2$ of $\SU(2)$. Whether or not this happens depends on how the
constituent quark flavors transform under $\SU(2)$.

The states $u \otimes u \otimes d$ and $u \otimes d \otimes d$ 
do \emph{not} span a copy of the fundamental rep of $\SU(2)$ inside
$\C^2 \otimes \C^2 \otimes \C^2$.  So, as with color, the equations
\[ p = uud, \quad n = udd \]
fail to give us the whole story. For the proton, we actually need some linear
combination of the $I_3 = \half$ flavor states, which are made of two $u$'s and
one $d$:
\[ u \otimes u \otimes d, \quad u \otimes d \otimes u, \quad d \otimes u \otimes u  \quad \in \quad \C^2 \otimes \C^2 \otimes \C^2. \]
And for the neutron, we need some linear combination of the $I_3 = -\half$ 
flavor states, which are made of one $u$ and two $d$'s:
\[ u \otimes d \otimes d, \quad d \otimes u \otimes d, \quad d \otimes d \otimes u  \quad \in \quad \C^2 \otimes \C^2 \otimes \C^2.   \]
Writing $p = uud$ and $n = udd$ is just a sort of 
shorthand for saying that $p$ and $n$ are made from basis vectors 
with those quarks in them.

In physics, the linear combination required to make $p$ and $n$ work also
involves the `spin' of the quarks, which lies
outside of our scope. We will content ourselves with showing that it \emph{can}
be done. That is, we will show that $\C^2 \otimes \C^2 \otimes \C^2$ really
does contain a copy of the fundamental rep $\C^2$ of $\SU(2)$. To do this, we
use the fact that any rank 2 tensor can be decomposed into symmetric and
antisymmetric parts; for example,
\[ \C^2 \otimes \C^2 \iso \Sym^2 \C^2 \oplus \Ex^2 \C^2 \]
and this is actually how $\C^2 \otimes \C^2$ decomposes into irreps. $\Sym^2
\C^2$, as we noted in Section~\ref{sec:isospin}, is the unique 3-dimensional
irrep of $\SU(2)$; its othogonal complement $\Ex^2 \C^2$ in $\C^2 \otimes \C^2$
is thus also a subrepresentation, but this space is 1-dimensional, and must
therefore be the trivial irrep, $\Ex^2 \C^2 \iso \C$. In fact, for any
$\SU(n)$, the top exterior power of its fundamental rep, $\Ex^n \C^n$, is
trivial.

As a representation of $\SU(2)$, we thus have
\begin{eqnarray*}
	\C^2 \otimes \C^2 \otimes \C^2 
               & \iso & \C^2 \otimes ( \Sym^2 \C^2 \oplus \C) \\
               & \iso & \C^2 \otimes \Sym^2 \C^2 \quad \oplus \quad \C^2. 
\end{eqnarray*}
So indeed, $\C^2$ is a subrepresentation of $\C^2 \otimes \C^2 \otimes \C^2$.

As in the last section, there is no reason to have the full $\C^2$ of isospin
states for nucleons unless there is a way to change protons into neutrons.
There, we discussed how the pions provide this mechanism. The pions live in
$\sl(2, \C)$, the complexification of the adjoint representation of
$\SU(2)$, and this acts on $\C^2$:
\vskip 1em
\begin{center}
	\includegraphics[scale=0.75]{piN_vertex}
\end{center}
\vskip 1em
This Feynman diagram is a picture of the intertwining operator $\sl(2, \C)
\otimes \C^2 \to \C^2$ given by the representation of
$\sl(2, \C)$ on $\C^2$.  Now we know that nucleons are made of 
quarks and that isospin symmetry comes from their flavor symmetry. 
What about pions?

Pions also fit into this model, but they require more explanation, because they
are made of quarks and `antiquarks'. To every kind of particle, there is a
corresponding antiparticle, which is just like the original particle but with
opposite charge and isospin.   The antiparticle of a quark is called an
\textbf{antiquark}.

In terms of group representations, passing from a particle to its 
antiparticle corresponds to taking the dual representation.  
Since the quarks live in $\C^2 \otimes \C^3$, a representation 
of $\SU(2) \times \SU(3)$, the antiquarks live in the dual 
representation $\C^{2*} \otimes \C^{3*}$.  
Since $\C^2$ has basis vectors called {\bf up} and {\bf down}:
\[  
u = \left( \begin{array}{c} 1 \\ 0 \end{array} \right) \in \C^2  \qquad
d = \left( \begin{array}{c} 0 \\ 1 \end{array} \right) \in \C^2
\]
the space $\C^{2*}$ has a dual basis
\[	\ubar = \left( 1, 0 \right) \in \C^{2*}	 \qquad
        \dbar = \left( 0, 1 \right) \in \C^{2*}	\]
called \textbf{antiup} and \textbf{antidown}.  Similarly, since the
standard basis vectors for $\C^3$ are called red, green and blue, 
the dual basis vectors for $\C^{3*}$ are known as \textbf{anticolors}: namely
\textbf{antired} $\rbar$, \textbf{antigreen} $\gbar$, and \textbf{antiblue} 
$\bbar$.  When it comes to actual colors of light, antired is 
called `cyan': this is the color of light which blended with red gives
white.  Similarly, antigreen is magenta, and antiblue is yellow.  But 
few physicists dare speak of `magenta antiquarks'---apparently this
would be taking the joke too far.

All pions are made from one quark and one
antiquark.  The flavor state of the pions must therefore live in 
\[ \C^2 \otimes \C^{2*}. \]
We can use the fact that pions live in $\mathfrak{sl}(2, \C)$ to find out how
they decompose into quarks and antiquarks, since
\[ \mathfrak{sl}(2, \C) \subseteq \End(\C^2). \]
First, express the pions as matrices:
\[ 
\pi^+ = \left(\begin{array}{cc} 0 & 1 \\ 0 & 0 \end{array} \right) \quad  
\pi^0 = \left(\begin{array}{cc} 1 & 0 \\ 0 & -1 \end{array} \right) \quad
\pi^- = \left(\begin{array}{cc} 0 & 0 \\ 1 & 0 \end{array} \right) .\]
We know they have to be these matrices, up to normalization, because 
these act the right way on nucleons in $\C^2$:
\[	\pi^- + p \to n	\]
\[	\pi^+ + n \to p	\]
\[	\pi^0 + p \to p	\]
\[	\pi^0 + n \to n	.\]
Now, apply the standard isomorphism $\End (\C^2) \iso \C^2 \otimes \C^{2*}$ to
write these matrices as linear combinations of quarks and antiquarks:
\[ \pi^+ = u \otimes \dbar, \quad \pi^0 = u \otimes \ubar - d \otimes \dbar, \quad \pi^- = d \otimes \ubar .\]
Note these all have the right $I_3$, because isospins reverse for
antiparticles. For example, $I_3(\dbar) = +\half$, so $I_3(\pi^+) = 1$.

In writing these pions as quarks and antiquarks, we have once again neglected
to write the color, because this works the same way for all pions.  
As far as color goes, pions live in 
\[ \C^3 \otimes \C^{3*}. \]
Confinement says that pions need to be white, just like nucleons, and 
there is only a one-dimensional subspace of $\C^3 \otimes \C^{3*}$ on
which $\SU(3)$ acts trivially, spanned by
\[ r \otimes \rbar + 
g \otimes \gbar + b \otimes \bbar \in \C^3 \otimes \C^{3*} .\]
So, this must be the color state of all pions.

Finally, the Gell-Mann--Nishijima formula also still works for quarks, 
provided we define the hypercharge for both quarks to be $Y = \third$: 
\[
\begin{array}{ccccrcr}
	Q(u) &=& I_3(u) + Y/2 & = & \half + \frac{1}{6}  &=& \twothirds \\
                                                                        \\
	Q(d) &=& I_3(d) + Y/2 & = & -\half + \frac{1}{6} &=& -\third .  
\end{array}
\]
Since nucleons are made of three quarks, their total hypercharge is $Y = 1$,
just as before.

\subsubsection{Leptons} \label{sec:leptons}

With the quarks and electron, we have met all the fundamental fermions required
to make atoms, and almost all of the particles we need to discuss the Standard
Model.  Only one player remains to be introduced: the \textbf{neutrino}, 
$\nu$.  This particle completes the \textbf{first generation} of fundamental 
fermions:
\vskip 1em
\begin{center}
	\begin{tabular}{lcc}
		\hline
		\multicolumn{3}{|c|}{\bf{The First Generation of Fermions ---
Charge}}  \\
		\hline
		Name       & Symbol      & Charge          \\
		\hline
		\\
		Neutrino   & $\nu \quad$ & 0               \\
		\\
		Electron   & $e^-$       & $-1$            \\
		\\
		Up quark   & $u \quad$   &  $+\twothirds$  \\
		\\
		Down quark & $d \quad$   &  $-\third$      \\
                \\
		\hline
	\end{tabular}
\end{center}
\vskip 1em

Neutrinos are particles which show up in certain interactions, like the decay
of a neutron into a proton, an electron, and an antineutrino
\[ n \to p + e^- + \nubar .\]
Indeed, neutrinos $\nu$ have antiparticles $\nubar$, just like quarks and all
other particles. The electron's antiparticle, denoted $e^+$, was the first
discovered, so it wound up subject to an inconsistent naming convention: the
`antielectron' is called a \textbf{positron}.

Neutrinos carry no charge and no color. They interact very weakly with other
particles, so weakly that they were not observed until the 1950s, over 20 years
after they were hypothesized by Pauli. Collectively, neutrinos and electrons,
the fundamental fermions that do not feel the strong force, are called
\textbf{leptons}.

In fact, the neutrino only interacts via the \textbf{weak force}. Like the
electromagnetic force and the strong force, the weak force is a fundamental
force, hypothesized to explain the decay of the neutron, and eventually
required to explain other phenomena.

The weak force cares about the `handedness' of particles.  It seems
that every particle that we have discussed comes in left- and right-handed 
varieties, which (quite roughly speaking) spin in opposite ways.  There are
are left-handed leptons, which we denote as
\[ \nu_L \quad e^-_L \]
and left-handed quarks, which we denote as
\[ u_L \quad d_L \]
and similarly for right-handed fermions, which we will denote with a subscript
$R$. As the terminology suggests, looking in a mirror interchanges left and
right---in a mirror, the left-handed electron $e^-_L$ looks like a 
right-handed
electron, $e^-_R$, and vice versa. More precisely, applying any of the
reflections in the Poincar\'e group to the (infinite-dimensional)
representation we use to describe these fermions interchanges left and right.

Remarkably, the weak force interacts only with left-handed particles and
right-handed antiparticles. For example, when the neutron decays, we always
have
\[ n_L \to p_L + e^-_L + \nubar_R \]
and never 
\[ n_R \to p_R + e^-_R + \nubar_L. \]
This fact about the weak force, first noticed in the 1950s, left a deep
impression on physicists. No other physical law is asymmetric in left and
right. That is, no other physics, classical or quantum, looks different when
viewed in a mirror. Why the weak force, and only the weak force, exhibits this
behavior is a mystery.

Since neutrinos only feel the weak force, and the weak force only involves
left-handed particles, the right-handed neutrino $\nu_R$ has never been
observed directly. For a long time, physicists believed this particle did not 
even exist, but recent observations of neutrino oscillations suggest otherwise. In this paper, we will assume there are right-handed neutrinos, but the reader
should be aware that this is still open to some debate. In particular, even if
they do exist, we know very little about them. 

Note that isospin is not conserved in weak interactions. After all, we saw in
the last section that $I_3$ is all about counting the number of $u$ quarks over
the number of $d$ quarks. In a weak process such as neutron decay
\[ udd \to uud + e^- + \nubar, \]
the right-hand side has $I_3 = -\half$, while the left has $I_3 = \half$. 

Yet maybe we are not being sophisticated enough. Perhaps isospin can be
extended beyond quarks, and leptons can also carry $I_3$. Indeed, if we define
$I_3( \nu_L ) = \half$ and $I_3( e^- ) = -\half$, we get
\begin{center}
	\begin{tabular}{lccccccc}
		 & $n_L$    & $\to$ & $p_L$   & $+$  &  $e_L^-$ & $+$ & $\nubar_R$ \\
\\
		$I_3:$   & $-\half$ & $=$   & $\half$ & $-$  &  $\half$ & $-$ & $\half$ \\
	\end{tabular}
\end{center}
where we have used the rule that isospin reverses sign for antiparticles.

This extension of isospin is called \textbf{weak isospin} since it extends the
concept to weak interactions. Indeed, it turns out to be fundamental to the
theory of weak interactions. Unlike regular isospin symmetry, which is only
approximate, weak isospin symmetry turns out to be exact.

So from now on we shall discuss only weak isospin, and call it simply
\textbf{isospin}.  
Weak isospin is zero for right-handed particles, and $\pm \half$
for left-handed particles:
\vskip 1em
\begin{center}
	\begin{tabular}{lccc}
	\hline
	\multicolumn{4}{|c|}{{\bf The First Generation of Fermions ---
Charge and Isospin}} \\
	\hline
        \\   
	Name                    & Symbol  & Charge        & Isospin \\
	                        &         & $Q$           & $I_3$   \\
	\hline
	Left-handed neutrino    & $\nu_L$ & $0$           & $\half$ \\
	\\
	Left-handed electron    & $e^-_L$ & $-1$          & $-\half$ \\
	\\
	Left-handed up quark    & $u_L$   & $+\twothirds$ & $\half$ \\
	\\ 
	Left-handed down quark  & $d_L$   & $-\third$     & $-\half$ \\
	\\
	Right-handed neutrino   & $\nu_R$ & $0$           & $0$ \\
	\\
	Right-handed electron   & $e^-_R$ & $-1$          & $0$ \\
	\\
	Right-handed up quark   & $u_R$   & $+\twothirds$ & $0$ \\
	\\ 
	Right-handed down quark & $d_R$   & $-\third$     & $0$ \\  
        \\
	\hline
	\end{tabular}
\end{center}
\vskip 1em
The antiparticle of a left-handed particle is right-handed, and the
antiparticle of a right-handed particle is left-handed.  The isospins 
also change sign.  For example, $I_3( e^+_R ) = +\half$,
while $I_3( e^+_L ) = 0$.

In Section~\ref{sec:hypercharge}, we will see that the Gell-Mann--Nishijima
formula, when applied to weak isospin, defines a fundamental quantity, the
`weak hypercharge', that is vital to the Standard Model. But first, in
Section~\ref{sec:redux}, we discuss how to generalize the $\SU(2)$ symmetries
from isospin to weak isospin.

\subsection{The Fundamental Forces} \label{sec:forces}

\subsubsection{Isospin and {\rm{SU(2)}}, Redux} \label{sec:redux}

The tale we told of isospin in Section~\ref{sec:isospin} only concerned
the strong force, which binds nucleons together into nuclei. We learned
about an approximation in which nucleons live in the fundamental rep $\C^2$ 
of the isospin symmetry group $\SU(2)$, and that they interact by exchanging 
pions, which live in the complexified adjoint rep of this group,
namely $\sl(2, \C)$.

But this tale is mere prelude to the modern story, where \emph{weak} isospin, 
defined in Section~\ref{sec:leptons}, is the star of the show.  This story
is not about the strong force, but rather the weak force.   This story
parallels the old one, but it involves left-handed fermions instead of nucleons.
The left-handed fermions, with $I_3 = \pm \half$, are paired up into
fundamental representations of $\SU(2)$, the \textbf{weak isospin symmetry
group}.  There is one spanned by left-handed leptons:
\[ \nu_L, e^-_L \in \C^2, \]
and one spanned by each color of left-handed quarks:
\[ u^r_L, d^r_L \in \C^2, \quad u^g_L, d^g_L \in \C^2, \quad 
u^b_L, d^b_L \in \C^2. \]
The antiparticles of the left-handed fermions, the right-handed antifermions,
span the dual representation $\C^{2*}$.

Because these particles are paired up in the same $\SU(2)$ representation,
physicists often write them as \textbf{doublets}:
\[ \lep \qquad \quark \]
with the particle of higher $I_3$ written on top. Note that we have suppressed
color on the quarks. This is conventional, and is done because $\SU(2)$ acts
the same way on all colors.

The particles in these doublets then interact via the exchange of $W$ bosons,
which are the weak isospin analogues of the pions. Like the pions, there are
three $W$ bosons:
\[ W^+ = \left(\begin{array}{cc} 0 & 1 \\ 0 & 0 \end{array} \right), \quad  
W^0 = \left(\begin{array}{cc} 1 & 0 \\ 0 & -1 \end{array} \right), \quad
W^- = \left(\begin{array}{cc} 0 & 0 \\ 1 & 0 \end{array} \right). \]
They span the complexified adjoint rep of $\SU(2)$, $\sl(2, \C)$, and they act
on each of the doublets like the pions act on the nucleons, via the action of
$\sl(2,\C)$ on $\C^2$. For example,
\begin{center}
	\includegraphics[scale=0.75]{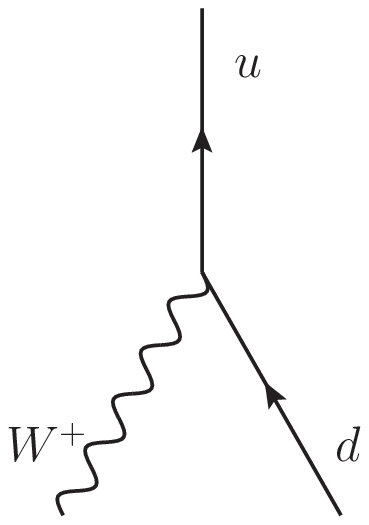}
\end{center}

Again, Feynman diagrams are the physicists' way of drawing
intertwining operators.  Since all the $\C^2$'s are acted on by the same
$\SU(2)$, they can interact with each other via $W$ boson exchange. For
example, quarks and leptons can interact via $W$'s:
\begin{center}
	\includegraphics[scale=0.75]{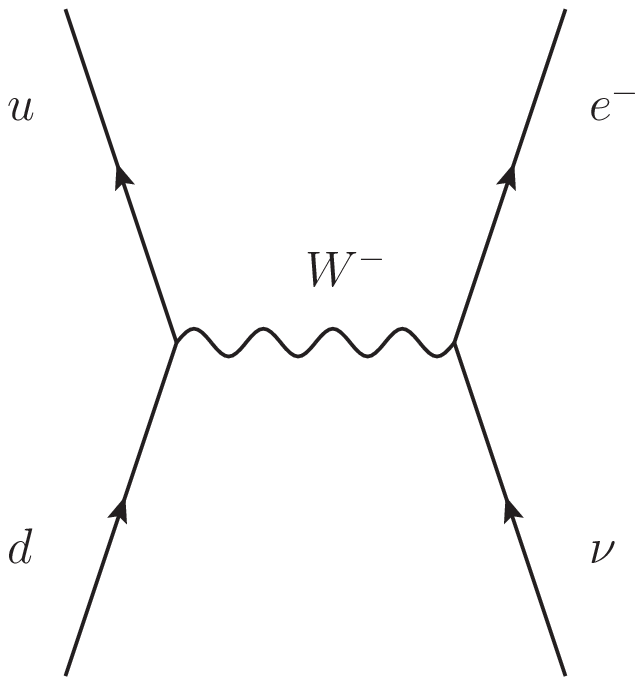}
\end{center}
This is in sharp contrast to the old isospin theory.
In the new theory, it is processes like these that are
responsible for the decay of the neutron:
\begin{center}
	\includegraphics[scale=0.75]{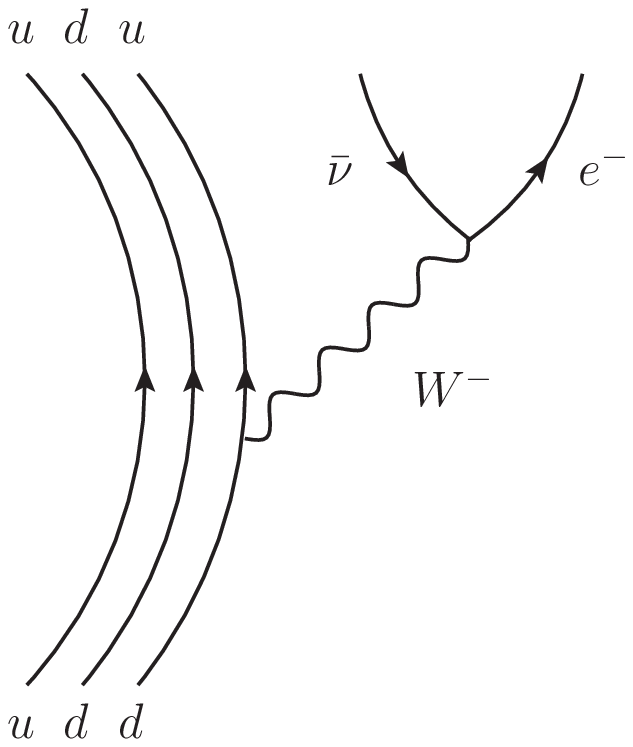}
\end{center}

The fact that only left-handed particles are combined into doublets reflects
the fact that only they take part in weak interactions. Every right-handed
fermion, on the other hand, is trivial under $\SU(2)$. Each one spans the
trivial rep, $\C$.  An example is the right-handed electron 
\[              e^-_R \in \C  .\]
Physicists call these particles \textbf{singlets}, meaning they are 
trivial under $\SU(2)$. This is just the representation theoretic way of 
saying the right-handed electron, $e^-_R$, does not participate in weak 
interactions.

In summary, left-handed fermions are grouped into doublets (nontrivial 
representations of $\SU(2)$ on $\C^2$), while right-handed fermions are 
singlets (trivial representations on $\C$).  So, the left-handed ones 
interact via the exchange of $W$ bosons, while the right-handed ones do not.
\vskip 1em
\begin{center}
	\begin{tabular}{lccc}
         \hline
         \multicolumn{4}{|c|}{\bf{The First Generation of Fermions ---
\boldmath $\SU(2)$ Representations}} \\
         \hline
          \\
         Name                    & Symbol   & Isospin    & $\SU(2)$ rep \\
         \hline
         Left-handed leptons     & $\lep$   & $\pm\half$ & $\C^2$ \\
         \\
         Left-handed quarks      & $\quark$ & $\pm\half$ & $\C^2$ \\
         \\
         Right-handed neutrino   & $\nu_R$  & $0$        & $\C$ \\
	 \\
         Right-handed electron   & $e^-_R$  & $0$        & $\C$ \\
	 \\
         Right-handed up quark   & $u_R$    & $0$        & $\C$ \\
	 \\
         Right-handed down quark & $d_R$    & $0$        & $\C$ \\
         \\  
         \hline
\end{tabular}
\end{center}

\subsubsection{Hypercharge and {\rm{U(1)}}} \label{sec:hypercharge}

In Section~\ref{sec:leptons}, we saw how to extend the notion of isospin to
weak isospin, which proved to be more fundamental, since we saw in
Section~\ref{sec:redux} how this gives rise to interactions among left-handed
fermions mediated via $W$ bosons. 

We grouped all the fermions into $\SU(2)$ representations. When we did this in
Section~\ref{sec:isospin}, we saw that the $\SU(2)$ representations of
particles were labeled by a quantity, the hypercharge $Y$, which relates the
isospin $I_3$ to the charge $Q$ via the Gell-Mann--Nishijima formula
\[ Q = I_3 + Y/2. \]

We can use this formula to extend the notion of hypercharge to \textbf{weak
hypercharge}, a quantity which labels the weak isospin representations. For
left-handed quarks, this notion, like weak isospin, coincides with the old
isospin and hypercharge.  We have weak hypercharge $Y = \third$ for these
particles:
\[ \begin{array}{ccccrcr}
	Q(u_L) &=& I_3(u_L) + Y/2 & = & \half + \frac{1}{6}  &=& \twothirds \\
                                                                           \\ 
	Q(d_L) &=& I_3(d_L) + Y/2 & = & -\half + \frac{1}{6}& =& -\third .  \\
\end{array}
\]
But just as weak isospin extended isospin to leptons, weak hypercharge extends
hypercharge to leptons. For left-handed leptons the
Gell-Mann--Nishijima formula holds if we set $Y = -1$:
\[ \begin{array}{ccccrcr}
	Q(\nu_L) &=& I_3(\nu_L) + Y/2 & = & \half - \half  &=& 0 \\
                                                                    \\
	Q(e^-_L) &=& I_3(e^-_L) + Y/2 & = & -\half - \half &=& -1 .  \\
\end{array}
\]
Note that the weak hypercharge of quarks comes in units one-third the size of
the weak hypercharge for leptons, a reflection of the fact that quark charges
come in units one-third the size of lepton charges. Indeed, thanks to the
Gell-Mann--Nishijima formula, these facts are equivalent.

For right-handed fermions, weak hypercharge is even simpler. Since $I_3 = 0$
for these particles, the Gell-Mann--Nishijima formula reduces to
\[ Q = Y/2. \]
So, the hypercharge of a right-handed fermion is twice its 
charge.  In summary, the fermions have these hypercharges:
\vskip 1em
\begin{center}
	\begin{tabular}{lcc}
         \hline
         \multicolumn{3}{|c|}{\textbf{The First Generation of Fermions ---
Hypercharge}} \\
         \hline
         \\  
         Name                    & Symbol   & Hypercharge  \\
	                         &          & $Y$          \\
         \hline
         Left-handed leptons     & $\lep$   & $-1$          \\
         \\
         Left-handed quarks      & $\quark$ & $\third$      \\
         \\
         Right-handed neutrino   & $\nu_R$  & $0$           \\
	 \\
         Right-handed electron   & $e^-_R$  & $-2$          \\
	 \\
         Right-handed up quark   & $u_R$    & $\fourthirds$ \\
	 \\
         Right-handed down quark & $d_R$    & $-\twothirds$ \\
         \\ 
         \hline
\end{tabular}
\end{center}
\vskip 1em

But what is the meaning of hypercharge?  We can start by reviewing
our answer for the quantity $I_3$.  This quantity, as we have seen, is 
related to how particles interact via $W$ bosons, because particles with 
$I_3 = \pm \half$ span the fundamental representation of $\SU(2)$, while
the $W$ bosons span the complexified adjoint representation, which 
acts on any other representation.  Yet there is a deeper connection. 

In quantum mechanics, observables such as $I_3$ correspond to self-adjoint
operators.  We will denote the operator corresponding to an observable with a
caret; for example, $\hat{I}_3$ is the operator corresponding to $I_3$. A state
of specific $I_3$, such as $\nu_L$, which has $I_3 = \half$, is an eigenvector,
\[ \hat{I}_3 \nu_L = \half \nu_L \]
with an eigenvalue that is the $I_3$ of the state. This makes it easy to write
$\hat{I}_3$ as a matrix when we let it act on the $\C^2$ with basis $\nu_L$ and
$e^-_L$, or any other doublet. We get
\[ \hat{I}_3 = \left(\begin{array}{cc} \half & 0 \\ 0 & -\half \end{array} \right). \]
Note that this is an element of $\su(2)$ divided by $i$.   So, it lies in
$\sl(2, \C)$, the complexified adjoint representation of $\SU(2)$.  
In fact it equals $\half W^0$, one of the gauge bosons. So, up to a constant
of proportionality, the observable $\hat{I}_3$ \emph{is} one of the gauge
bosons!

Similarly, corresponding to hypercharge $Y$ is an observable $\hat{Y}$. This is
also, up to proportionality, a gauge boson, though this gauge boson lives in
the complexified adjoint rep of $\U(1)$.

Here are the details. Particles with hypercharge $Y$ span irreps $\C_Y$ of $\U(1)$. Since
$\U(1)$ is abelian, all of its irreps are one-dimensional.  By $\C_Y$ we denote
the one-dimensional vector space $\C$ with action of $\U(1)$ given by 
\[ \alpha \cdot z = \alpha^{3Y} z. \]
The factor of $3$ takes care of the fact that $Y$ might not be an integer, 
but is only guaranteed to be an integral multiple of $\third$. 
For example, the left-handed leptons $\nu_L$ and $e^-_L$ both have hypercharge
$Y = -1$, so each one spans a copy of $\C_{-1}$:
\[ \nu_L \in \C_{-1}, \quad e^-_L \in \C_{-1} \]
or, more compactly, 
\[ \nu_L, e^-_L \in \C_{-1} \otimes \C^2 \]
where $\C^2$ is trivial under $\U(1)$. 

In summary, the fermions we have met thus far 
lie in these $\U(1)$ representations:
\vskip 1em
\begin{center}
	\begin{tabular}{lccc}
         \hline
         \multicolumn{3}{|c|}{\textbf{The First Generation of Fermions ---
\boldmath $\U(1)$ Representations}} \\
         \hline
         Name                    & Symbol     & $\U(1)$ rep \\
         \hline
	 \\
	 Left-handed leptons     & $\lep$      & $\C_{-1}$          \\
         \\
	 Left-handed quarks      & $\quark$    & $\C_{\third}$      \\
         \\
         Right-handed neutrino   & $\nu_R$    & $\C_{0}$           \\
	 \\
         Right-handed electron   & $e^-_R$    & $\C_{-2}$          \\
	 \\
         Right-handed up quark   & $u_R$      & $\C_{\fourthirds}$ \\
	 \\
         Right-handed down quark & $d_R$      & $\C_{-\twothirds}$ \\
	 \\
         \hline
\end{tabular}
\end{center}
\vskip 1em

Now, the adjoint representation $\u(1)$ of $\U(1)$ is just the tangent space to
the unit circle in $\C$ at 1. It is thus parallel to the imaginary axis, and
can be identified with $i\R$. It is generated by $i$. $i$ also generates the
complexification, $\C \otimes \u(1) \iso \C$, though this also has other
convenient generators, like 1.  Given a particle $\psi \in \C_Y$ of 
hypercharge $Y$, we can differentiate the action of $\U(1)$ on $\psi$
\[ e^{i\theta} \cdot \psi = e^{3iY\theta} \psi \]
and set $\theta = 0$ to find out how $\u(1)$ acts:
\[ i \cdot \psi = 3iY \psi .\]
Dividing by $i$ we obtain
\[ 1 \cdot \psi = 3Y \psi .\]
In other words, we have
\[ \hat{Y} = \third \in \C \]
as an element of the complexified adjoint rep of $\U(1)$.

Particles with hypercharge interact by exchange of a boson, called the
$B$ boson, which spans the complexified adjoint rep of $\U(1)$.
Of course, since $\C$ is one-dimensional, any
nonzero element spans it. Up to a constant of proportionality, the $B$ boson is just
$\hat{Y}$, and we might as well take it to be equal to $\hat{Y}$, but calling
it $B$ is standard in physics.

The $B$ boson is a lot like another, more familiar $\U(1)$ gauge boson---the
photon! The hypercharge force which the $B$ boson mediates is a lot like
electromagnetism, which is mediated by photons, but its strength is
proportional to hypercharge rather than charge.  As usual, we can
draw the $\U(1)$ intertwining operators as Feynman diagrams:
\begin{center}
	\includegraphics[scale=0.5]{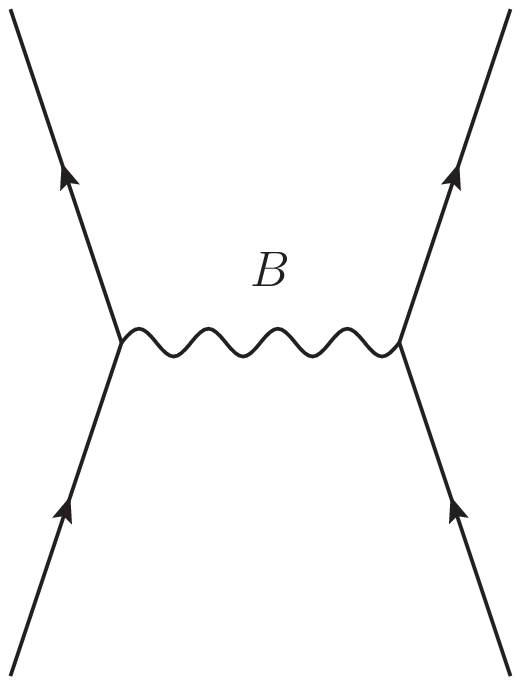}
\end{center}

\subsubsection{Electroweak Symmetry Breaking} \label{sec:electroweak}

In the Standard Model, electromagnetism and the weak force are unified into the
electroweak force. This is is a $\U(1) \times \SU(2)$ gauge theory, and without
saying so, we just told you all about it in sections~\ref{sec:redux} and
\ref{sec:hypercharge}. The fermions live in representations of hypercharge
$\U(1)$ and weak isospin $\SU(2)$, exactly as we described in those sections,
and we tensor these together to get representations of $\U(1) \times \SU(2)$:
\begin{center}
	\begin{tabular}{lcccc}
         \hline
         \multicolumn{5}{|c|}{\textbf{The First Generation of Fermions ---
\boldmath $\U(1) \times \SU(2)$ Representations}} \\
         \hline
         Name                    & Symbol   & Hypercharge   & Isospin    & $\U(1) \times \SU(2)$ rep \\
         \hline                              
	 \\
         Left-handed leptons     & $\lep$   & $-1$          & $\pm\half$ & $\C_{-1} \otimes \C^2$ \\
         \\                                                                                  
         Left-handed quarks      & $\quark$ & $\third$      & $\pm\half$ & $\C_{\third} \otimes \C^2$ \\
         \\                                                                                  
         Right-handed neutrino   & $\nu_R$  & $0$           & $0$        & $\C_{0} \otimes \C$ \\
	 \\                                                                                  
         Right-handed electron   & $e^-_R$  & $-2$          & $0$        & $\C_{-2} \otimes \C$ \\
	 \\                                                                                  
         Right-handed up quark   & $u_R$    & $\fourthirds$ & $0$        & $\C_{\fourthirds} \otimes \C$ \\
	 \\                                                                                  
         Right-handed down quark & $d_R$    & $-\twothirds$ & $0$        & $\C_{-\twothirds} \otimes \C$ \\
	 \\
         \hline                              
\end{tabular}                                
\end{center}

These fermions interact by exchanging $B$ and $W$ bosons, which span $\C
\oplus \sl(2, \C)$, the complexified adjoint representation of $\U(1) \times
\SU(2)$.  

Yet despite the electroweak unification, electromagnetism and the weak force
are very different at low energies, including most interactions in the everyday
world. Electromagnetism is a force of infinite range that we can describe by a
$\U(1)$ gauge theory, with the photon as gauge boson:
\begin{center}
	\includegraphics[scale=0.6]{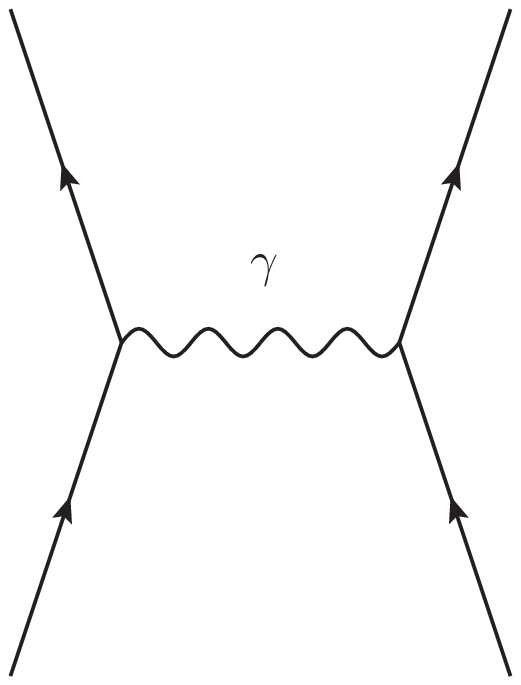}
\end{center}
The photon lives in $\C \oplus \sl(2, \C)$, alongside the $B$ and $W$ bosons.
It is given by a linear combination
\[ \gamma = W^0 + B/2 \]
that parallels the Gell-Mann--Nishijima formula, $Q = I_3 + Y/2$.

The weak force is of very short range and mediated by the $W$ and $Z$ bosons:
\begin{center}
	\begin{tabular}{ccc}
		\includegraphics[scale=0.6]{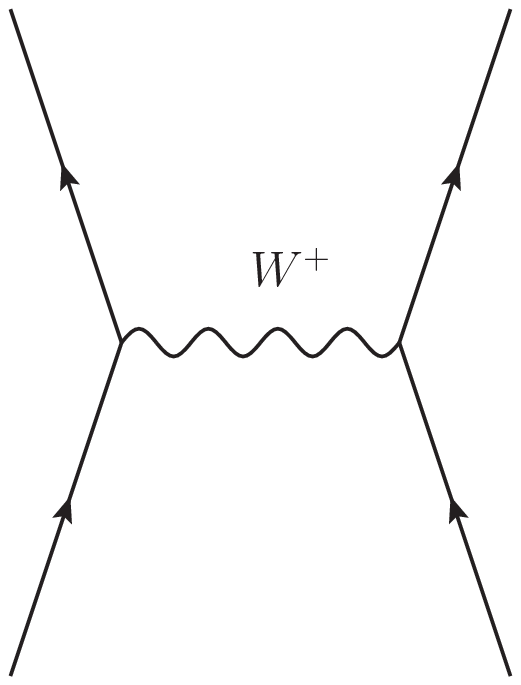} &
                \includegraphics[scale=0.6]{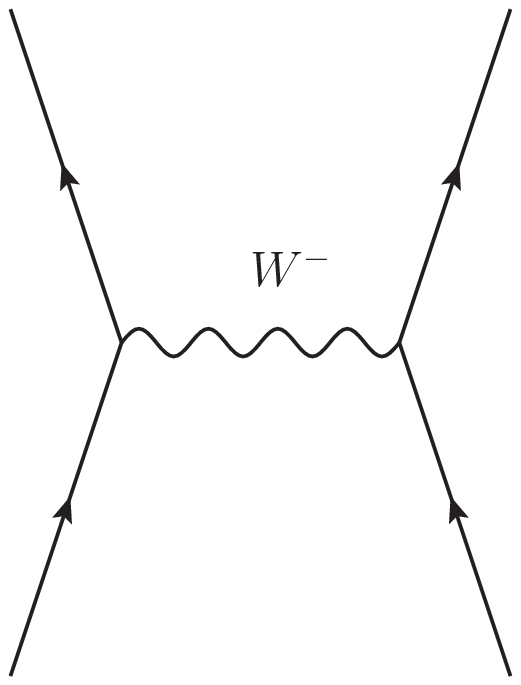} & 
                \includegraphics[scale=0.6]{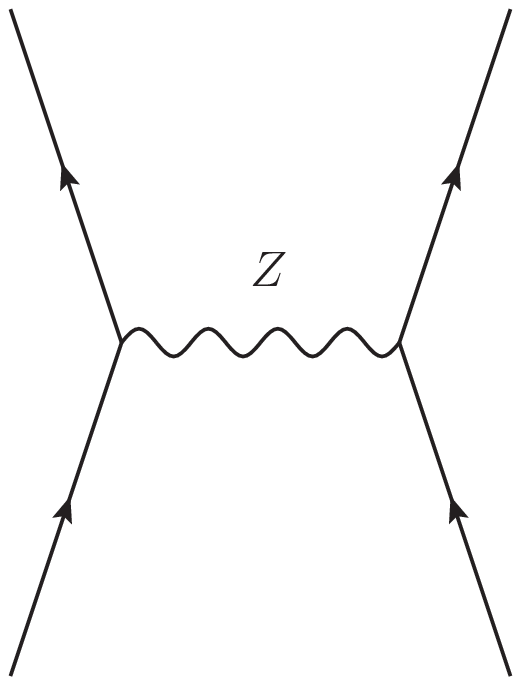} \\
	\end{tabular}
\end{center}
The $Z$ boson lives in $\C \oplus \sl(2, \C)$, and is given by the linear
combination
\[ Z = W^0 - B/2 \]
which is in some sense `perpendicular' to the photon. So, we can
expand our chart of gauge bosons to include a basis for all of $\C
\oplus \sl(2, \C)$ as follows:
\vskip 1em
\begin{center}
	\begin{tabular}{llc}
		\hline
		\multicolumn{3}{|c|}{\bf{Gauge Bosons (second try)}} \\
		\hline
		Force & Gauge boson & Symbol \\
		\hline
		Electromagnetism & Photon & $\gamma$ \\
		\\
		Weak force & $W$ and $Z$ bosons & $W^+$, $W^-$ and $Z$ \\
		\hline
	\end{tabular}
\end{center}
\vskip 1em

What makes the photon (and electromagnetism) so different from the $W$ and
$Z$ bosons (and the weak force)? It is symmetry breaking. Symmetry breaking
allows the full electroweak $\U(1) \times \SU(2)$ symmetry group to be hidden
away at high energy, replaced with the electomagnetic subgroup $\U(1)$ at lower
energies. This electromagnetic $\U(1)$ is not the obvious factor of $\U(1)$
given by $\U(1) \times 1$. It is another copy, one which wraps around inside
$\U(1) \times \SU(2)$ in a manner given by the Gell-Mann--Nishijima formula.

The dynamics behind symmetry breaking are beyond the scope of this paper. We
will just mention that, in the Standard Model, electroweak symmetry breaking is
believed to be due to the `Higgs mechanism'.  In this mechanism, all
particles in the Standard Model, including the photon and the $W$ and $Z$
bosons, interact with a particle called the `Higgs boson', and it is their
differing interactions with this particle that makes them appear so different
at low energies. 

The Higgs boson has yet to be observed, and remains one of the most mysterious
parts of the Standard Model. As of this writing, the Large Hadron Collider at
CERN is beginning operations; searching for the Higgs boson is one of its
primary aims.

For the details on symmetry breaking and the Higgs mechanism, which is
essential to understanding the Standard Model, see Huang~\cite{huang:qlgf}. For
a quick overview, see Zee~\cite{zee:nutshell}.

\subsubsection{Color and {\rm{SU(3)}}} \label{sec:color}

There is one more fundamental force in the Standard Model: the
\textbf{strong force}. We have already met this force, as the force that
keeps the nucleus together, but we discussed it before we knew that
protons and neutrons are made of quarks.  Now we need a force to keep
quarks together inside the nucleons, and quark confinement tells us it
must be a very strong force indeed.  It is this force that, in modern
parlance, is called the strong force and is considered fundamental. The
force between nucleons is a side effect of these more fundamental
interactions among quarks.

Like all three forces in the Standard Model, the strong force is
explained by a gauge theory, this time with gauge group $\SU(3)$, the
color symmetry group of the quarks.  The picture is simpler than that
of electromagnetism and the weak force, however, because this symmetry
is `unbroken'.  

By now you can guess how this goes. Every kind of quark spans the
fundamental representation $\C^3$ of $\SU(3)$. For example, the
left-handed up quark, with its three colors, lives in
\[ u^r_L, u^g_L, u^b_L \in \C^3 \]
and the left-handed down quark, with its three colors, spans another
copy of $\C^3$,
\[ d^r_L, d^g_L, d^b_L \in \C^3 \]
Together, these span the $\SU(3)$ representation
\[ \C^2 \otimes \C^3 \]
where $\C^2$ is trivial under $\SU(3)$.

The quarks interact by the exchange of \textbf{gluons}, the gauge bosons
of the strong force. These gauge bosons live in $\C \otimes \su(3)
\iso \sl(3, \C)$, the complexified adjoint representation of
$\SU(3)$. The interactions are drawn as Feynman diagrams, which now
depict intertwining operators between representations of $\SU(3)$:
\begin{center}
	\includegraphics[scale=0.6]{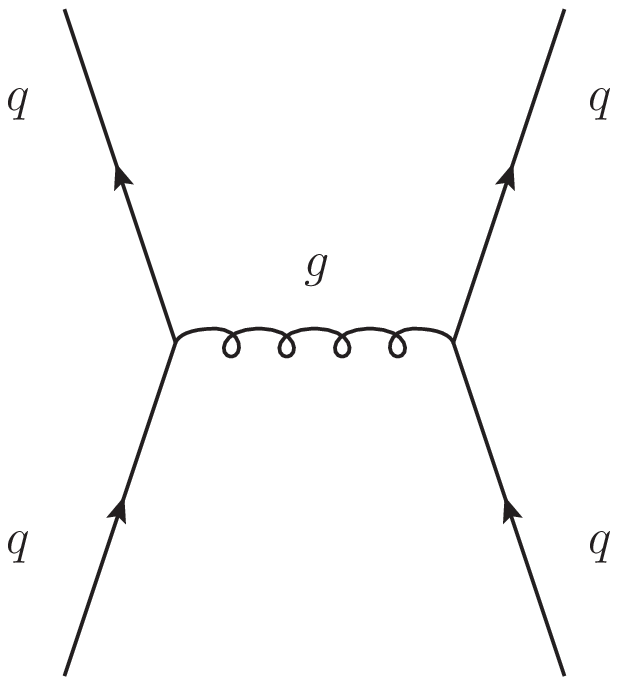}
\end{center}
The gluons are fundamental particles, gauge
bosons of the strong force, and they complete our table of gauge bosons:
\vskip 1em
\begin{center}
	\begin{tabular}{llc}
		\hline
		\multicolumn{3}{|c|}{\bf{Gauge Bosons}} \\
		\hline
		Force & Gauge Boson & Symbol \\
		\hline
		Electromagnetism & Photon & $\gamma$ \\
		\\
		Weak force & $W$ and $Z$ bosons & $W^+$, $W^-$ and $Z$ \\
		\\
		Strong force & Gluons & $g$ \\  
                \\ 
		\hline
	\end{tabular}
\end{center}
\vskip 1em

On the other hand, the leptons are `white': they transform trivially
under $\SU(3)$.  So, they do not exchange gluons.  In other words, they
are not affected by the strong force.
We can capture all of this information in a table, where we give the $\SU(3)$
representations in which all our fermions live.

\vskip 1em
\begin{center}
	\begin{tabular}{lccc}
	\hline
	\multicolumn{4}{|c|}{\textbf{The First Generation of Fermions ---
\boldmath $\SU(3)$ Representations}} \\
	\hline
        \\
	Name                     & Symbol                  & Colors    & $\SU(3)$ rep \\
	\hline
	Left-handed neutrino     & $\nu_L$                 & white     & $\C$   \\
	\\
	Left-handed electron     & $e^-_L$                 & white     & $\C$   \\
	\\
	Left-handed up quarks    & $u^r_L, u^g_L, u^b_L$   & $r, g, b$ & $\C^3$ \\
	\\ 
	Left-handed down quarks  & $d^r_L, d^g_L, d^b_L$   & $r, g, b$ & $\C^3$ \\
	\\
	Right-handed electron    & $e^-_R$                 & white     & $\C$   \\
	\\
	Right-handed neutrino    & $\nu_R$                 & white     & $\C$   \\
	\\
	Right-handed up quarks   & $u^r_R, u^g_R, u^b_R$   & $r, g, b$ & $\C^3$ \\
	\\ 
	Right-handed down quarks & $d^r_R, d^g_R, d^b_R$   & $r, g, b$ & $\C^3$ \\
\\
	\hline
	\end{tabular}
\end{center}
\vskip 1em

\subsection{The Standard Model Representation} \label{sec:smrep}

We are now in a position to put the entire Standard Model together in a single
picture, much as we combined the isospin $\SU(2)$ and hypercharge $\U(1)$
into the electroweak gauge group, $\U(1) \times \SU(2)$, in
Section~\ref{sec:electroweak}. We then tensored the hypercharge $\U(1)$
representations with the isospin $\SU(2)$ representations to get the
electroweak representations.

Now let us take this process one step further, by bringing in a factor of
$\SU(3)$, for the color symmetry, and tensoring the representations of $\U(1)
\times \SU(2)$ with the representations of $\SU(3)$. Doing this, we get the
Standard Model. The Standard Model has this gauge group:
\[ G_{\mbox{SM}} = \U(1) \times \SU(2) \times \SU(3). \]
The fundamental fermions described by the Standard Model combine to form
representations of this group. We know what these are, and describe all of 
them in Table~\ref{tab:smrep}.

\begin{table}[H]
	\renewcommand{\arraystretch}{0.8}
\begin{center}
	\begin{tabular}{lcc}
         \hline
	 \multicolumn{3}{|c|}{\bf{The Standard Model Representation}} \\
         \hline
         Name                     & Symbol                 & $\U(1) \times \SU(2) \times \SU(3)$ rep \\
         \hline                              
	 \\
         Left-handed leptons      & $\lep$                 & $\C_{-1} \otimes \C^2 \otimes \C$ \\
         \\                                                               
         Left-handed quarks       & $\quarkwithcolor$      & $\C_{\third} \otimes \C^2 \otimes \C^3$ \\
         \\                                                               
         Right-handed neutrino    & $\nu_R$                & $\C_{0} \otimes \C \otimes \C$ \\
	 \\                                                               
         Right-handed electron    & $e^-_R$                & $\C_{-2} \otimes \C \otimes \C$ \\
	 \\                                                               
         Right-handed up quarks   & $(u^r_R, u^g_R, u^b_R)$  &  $\C_{\fourthirds} \otimes \C \otimes \C^3$ \\
	 \\                                                               
         Right-handed down quarks & $(d^r_R, d^g_R, d^b_R)$  & $\C_{-\twothirds} \otimes \C \otimes \C^3$ \\
	 \\
         \hline                              
	\end{tabular}
	\vspace{-5pt}
\caption{Fundamental fermions as representations of
$\GSM = \U(1) \times \SU(2) \times \SU(3)$} \label{tab:smrep}
\end{center}
	\vspace{-10pt}
	\renewcommand{\arraystretch}{1}
\end{table}

\noindent
All of the representations of $\GSM$ in the left-hand column are irreducible,
since they are made by tensoring irreps of this group's three factors,
$\U(1)$, $\SU(2)$ and $\SU(3)$.  This is a general
fact: if $V$ is an irrep of $G$, and $W$ is an irrep of $H$, then $V \otimes W$
is an irrep of $G \times H$. Moreover, all irreps of $G \times H$ arise in this
way.

On the other hand, if we take the direct sum of all these irreps,
\[ F = (\C_{-1} \otimes \C^2 \otimes \C) \quad \oplus \quad \cdots \quad 
\oplus \quad (\C_{-\twothirds} \otimes \C \otimes \C^3), \]
we get a reducible representation containing all the first-generation fermions
in the Standard Model.  We call $F$ the {\bf fermion representation}.
If we take the dual of $F$, we get a representation describing all the 
antifermions in the first generation.  And taking the direct sum of these
spaces:
\[ F \oplus F^* \]
we get a representation of $\GSM$ that we will call the \textbf{Standard Model
representation}. It contains all the first-generation elementary particles in
the Standard Model. It does not contain the gauge bosons or the mysterious
Higgs. 

The fermions living in the Standard Model representation interact by exchanging
gauge bosons that live in the complexified adjoint representation of $\GSM$. We
have already met all of these, and we collect them in
Table~\ref{tab:gaugebosons}.

\begin{table}[H]
\begin{center}
	\begin{tabular}{llc}
		\hline
		\multicolumn{3}{|c|}{\bf{Gauge Bosons}} \\
		\hline
		Force & Gauge Boson & Symbol \\
		\hline
                \\ 
		Electromagnetism & Photon & $\gamma$ \\
		\\
		Weak force & $W$ and $Z$ bosons & $W^+$, $W^-$ and $Z$ \\
		\\
		Strong force & Gluons & $g$ \\  \\
		\hline 
	\end{tabular}
	\caption{Gauge bosons} \label{tab:gaugebosons}
\end{center}
\end{table}

Of all the particles and antiparticles in $F \oplus F^*$, exactly two of
them are fixed by the action of $\GSM$. These are the right-handed neutrino
\[ \nu_R \in \C_0 \otimes \C \otimes \C \]
and its antiparticle
\[ \nubar_L \in (\C_0 \otimes \C \otimes \C)^*, \]
both of which are trivial representations of $\GSM$; they thus do not
participate in any forces mediated by the gauge bosons of the Standard
Model. They might interact with the Higgs boson, but very little about
right-handed neutrinos is known with certainty at this time.

\subsection{Generations} \label{sec:generations}

Our description of the Standard Model is almost at an end. We have told you
about its gauge group, $\GSM$, its representation $F \oplus F^*$ on the
first generation of fermions and antifermions, and a bit about how
these fermions interact by exchanging gauge bosons, which live in the
complexified adjoint rep of $\GSM$.  For the grand unified theories we
are about to discuss, that is all we need.  The stage is set.

Yet we would be derelict in our duty if we did not mention the
\emph{second} and \emph{third} generation of fermions.  The first
evidence for these came in the 1930s, when a charged particle 207
times as heavy as the electron was found.  At first researchers
thought it was the particle predicted by Yukawa---the one that
mediates the strong force between nucleons.  But then it 
turned out the newly discovered particle was \emph{not} affected by
the strong force.  This came as a complete surprise.  
As the physicist Rabi quipped at the time: ``Who ordered that?''

Dubbed the \textbf{muon} and denoted $\mu^-$, this new particle turned
out to act like an overweight electron.  Like the electron, it
feels only the electromagnetic and weak force---and
like the electron, it has its own neutrino!  So, the neutrino we have
been discussing so far is now called the \textbf{electron neutrino}, $\nu_e$,
to distinguish it from the \textbf{muon neutrino}, $\nu_\mu$.  Together, the
muon and the muon neutrino comprise the second generation of leptons.
The muon decays via the weak force
into an electron, a muon neutrino, and an electron antineutrino:
\[ \mu^- \to e^- + \nu_{\mu} + \nubar_e .\]

Much later, in the 1970s, physicists realized there was also a second
generation of quarks: the \textbf{charm quark}, $c$, and the
\textbf{strange quark}, $s$. This was evidence of another pattern in the
Standard Model: there are as many flavors of quark as there are
leptons. In Section~\ref{sec:g(2,2,4)}, we will learn about the
Pati--Salam model, which explains this pattern by unifying quarks and
leptons.

Today, we know about three generations of fermions. Three of quarks:
\vskip 1em 
\begin{center}
	\begin{tabular}{lclclc}
        \hline
	\multicolumn{6}{|c|}{\textbf{Quarks by Generation}} \\
	\hline
	              \multicolumn{2}{|c|}{1st Generation} & \multicolumn{2}{|c|}{2nd Generation} & \multicolumn{2}{|c|}{3rd Generation} \\
        \hline
	   Name & Symbol                        & Name    & Symbol                     & Name & Symbol \\
	\hline
        \\ 
	  Up   & $u$                           & Charm   & $c$                        & Top    & $t$ \\
	\\
	 Down & $d$                           & Strange & $s$                        & Bottom & $b$ \\  
        \\
	\hline
	\end{tabular}
 \end{center}
 \vskip 1em
and three of leptons:
\vskip 1em 
\begin{center}
	\begin{tabular}{lclclc}
        \hline
	\multicolumn{6}{|c|}{\textbf{Leptons by Generation}} \\
	\hline
	        \multicolumn{2}{|c|}{1st Generation} & \multicolumn{2}{|c|}{2nd Generation} & \multicolumn{2}{|c|}{3rd Generation} \\
        \hline
	 Name              & Symbol           & Name          & Symbol               & Name & Symbol \\
	\hline
        \\ 
	Electron & $\nu_e$          & Muon & $\nu_{\mu}$          & Tau & $\nu_{\tau}$ \\
        neutrino &                  & neutrino &                  & neutrino &
       \\
       \\
	Electron          & $e^-$            & Muon          & $\mu^-$              & Tau        & $\tau^-$ \\
        \\  
	\hline
	\end{tabular}
\end{center}
\vskip 1em
The second and third generations of quarks and \emph{charged} leptons
differ from the first by being more massive and able to decay into
particles of the earlier generations. The various neutrinos do not
decay, and for a long time it was thought they were massless, but now
it is known that some and perhaps all of them are massive.  This allows 
them to change back and forth from one type to another, a phenomenon 
called `neutrino oscillation'.

The Standard Model explains all of this by something called the Higgs
mechanism.  Apart from how they interact with the Higgs boson, the
generations are identical. For instance, as representations of $\GSM$,
each generation spans another copy of $F$.  Each generation of
fermions has corresponding antifermions, spanning a copy of $F^*$.

All told, we thus have three copies of the Standard Model
representation, $F \oplus F^*$.  We will only need to discuss one
generation, so we find it convenient to speak as if $F \oplus F^*$
contains particles of the first generation.  No one knows why the
Standard Model is this redundant, with three sets of very similar
particles.  It remains a mystery.

\section{Grand Unified Theories} \label{sec:guts}

Not all of the symmetries of $\GSM$, the gauge group of the Standard
Model, are actually seen in ordinary life.  This is because some of
the symmetries are `spontaneously broken'.  This means that while they
are symmetries of the laws of physics, they are not symmetries of the
vacuum.  To see these symmetries we need to do experiments at
very high energies, where the asymmetry of the vacuum has less effect.
So, the behavior of particles at lower energies is like a shadow of
the fundamental laws of physics, cast down from on high: a cryptic
clue we must struggle to interpret.

It is reasonable to ask if this process continues.  Could the
symmetries of the Standard Model be just a subset of all the
symmetries in nature? Could they be the low energy shadows of laws
still more symmetric?

A grand unified theory, or GUT, constitutes a guess at what these
`more symmetric' laws might be. It is a theory with more symmetry than
the Standard Model, which reduces to the Standard Model at lower
energies. It is also, therefore, an attempt to describe the physics at
higher energies.

GUTs are speculative physics.  The Standard Model has been tested in
countless experiments.  There is a lot of evidence that it is an
incomplete theory, and some vague clues about what the next theory
might be like, but so far there is no empirical evidence that any GUT
is correct---and even some empirical evidence that some GUTs, like
$\SU(5)$, are incorrect.

Nonetheless, GUTs are interesting to theoretical physicists, because
they allow us to explore some very definite ideas about how to extend
the Standard Model. And because they are based almost entirely on the
representation theory of compact Lie groups, the underlying physical
ideas provide a marvelous playground for this beautiful area of
mathematics.

Amazingly, this beauty then becomes a part of the physics.  The
representation of $\GSM$ used in the Standard Model seems ad hoc.  Why
this one?  Why all those seemingly arbitrary hypercharges floating
around, mucking up some otherwise simple representations? Why do both
leptons and quarks come in left- and right-handed varieties, which
transform so differently?  Why do quarks come in charges which are in
units $\third$ times an electron's charge? Why are there the same
number of quarks and leptons?  GUTs can shed light on these questions,
using only group representation theory.

\subsection{The {\rm{SU(5)}} GUT} \label{sec:su(5)}

The $\SU(5)$ grand unified theory appeared in a 1974 paper by Howard
Georgi and Sheldon Glashow \cite{GeorgiGlashow:su(5)}. It was the
first grand unified theory, and is still considered the prototypical
example.  As such, there are many accounts of it in the physics
literature.  The textbooks by Ross \cite{ross:gut} and Mohapatra
\cite{mohapatra:us} both devote an entire chapter to the $\SU(5)$
theory, and a lucid summary can be found in a review article by
Witten~\cite{witten:grandunification}, which also
discusses the supersymmetric generalization of this theory.

In this section, we will limit our attention to the nonsupersymmetric version
of $\SU(5)$ theory, which is how it was originally proposed. Unfortunately,
this theory has since been ruled out by experiment; it predicts that protons
will decay faster than the current lower bound on proton
lifetime~\cite{pati:decay}.  Nevertheless, because of its prototypical
status and intrinsic interest, we simply must talk about the $\SU(5)$ theory.

The core idea behind the $\SU(5)$ grand unified theory is that because
the Standard Model representation $F \oplus F^*$ is 32-dimensional,
each particle or antiparticle in the first generation of fermions can
be named by a 5-bit code.  Roughly speaking, these bits are the answers 
to five yes-or-no questions:
\begin{itemize}
\item Is the particle isospin up?
\item Is it isospin down?
\item Is it red?
\item Is it green?
\item Is it blue?
\end{itemize}
There are subtleties involved when we answer `yes' to both the
first two questions, or `yes' to more than one of the last three, but
let us start with an example where these issues do not arise:
the bit string $01100$.  This names a particle that is down and red.
So, it refers to a red quark whose isospin is down, meaning $-\half$.
Glancing at Table \ref{tab:smrep}, we see just one particle meeting
this description: the red left-handed down quark, $d^r_L$.

We can flesh out this scheme by demanding that the operation of taking
antiparticles correspond to switching 0's for 1's in the code.  So the
code for the antiparticle of $d^r_L$, the `antired right-handed
antidown antiquark', is $10011$.  This is cute: it means that being
antidown is the same as being up, while being antired is the same as
being both green and blue.

Furthermore, in this scheme all antileptons are `black' (the particles
with no color, ending in 000), while leptons are `white' (the particles
with every color, ending in 111).  Quarks have exactly one color, and
antiquarks have exactly two.

We are slowly working our way to the $\SU(5)$ theory.  Next let us bring
Hilbert spaces into the game.  We can take the basic properties of
being up, down, red, green or blue, and treat them as basis vectors for
$\C^5$.  Let us call these vectors $u,d,r,g,b$.  The exterior algebra
$\Lambda \C^5$ has a basis given by wedge products of these 5 vectors.
This exterior algebra is 32-dimensional, and it has a basis labelled by
5-bit strings.  For example, the bit string $01100$ corresponds to the
basis vector $d \wedge r$, while the bit string $10011$ corresponds to
$u \wedge g \wedge b$.

Next we bring in representation theory.  The group $\SU(5)$ has an
obvious representation on $\C^5$.  And since the operation of taking
exterior algebras is functorial, this group also has a representation
on $\Lambda \C^5$.  In the $\SU(5)$ grand unified theory, this is the
representation we use to describe a single generation of fermions and
their antiparticles.

Just by our wording, though, we are picking out a splitting of $\C^5$
into $\C^2 \oplus \C^3$: the isospin and color parts,
respectively. Choosing such a splitting of $\C^5$ picks out a subgroup
of $\SU(5)$, the set of all group elements that preserve this
splitting.  This subgroup consists of block diagonal matrices with a $2
\times 2$ block and a $3 \times 3$ block, both unitary, such that the
determinant of the whole matrix is 1.  Let us denote this subgroup as
$\S(\U(2) \times \U(3))$.

Now for the miracle: the subgroup $\S(\U(2) \times \U(3))$ is isomorphic
to the Standard Model gauge group (at least modulo a finite subgroup).
And, when we restrict the representation of $\SU(5)$ on $\Ex
\C^5$ to $\S(\U(2) \times \U(3))$, we get the Standard Model representation!

There are two great things about this.  The first is that it gives a
concise and mathematically elegant description of the Standard Model
representation.  The second is that the seemingly ad hoc hypercharges
in the Standard Model \emph{must be exactly what they are} for this
description to work.  So, physicists say the $\SU(5)$ theory explains
the fractional charges of quarks: the fact that quark charges come in
units $\third$ the size of the electron charge pops right out of this
theory.

With this foretaste of the fruits the $\SU(5)$ theory will bear, let
us get to work and sow the seeds.  Our work will have two
parts.  First we need to check that
\[ \S(\U(2) \times \U(3)) \iso \GSM/N \]
where $N$ is some finite normal subgroup that acts trivially on 
$F \oplus F^*$.  Then we need to check that indeed
\[ \Ex \C^5 \iso F \oplus F^* \]
as representations of $\S(\U(2) \times \U(3))$. 

First, the group isomorphism.  Naively, one might seek to build
the $\SU(5)$ theory by including $\GSM$ as a subgroup of $\SU(5)$. 
Can this be done?
Clearly, we can include $\SU(2) \times \SU(3)$ as block diagonal 
matrices in $\SU(5)$:
\[ \begin{array}{ccc}
   \SU(2) \times \SU(3) &\to& \SU(5) \\
   (g, h) &\longmapsto& 
\left( 
\begin{array}{cc}
g & 0 \\
0 & h
\end{array}
\right).
\end{array}
\]
but this is not enough, because $\GSM$ also has that pesky factor of
$\U(1)$, related to the hypercharge.  How can we fit that in?

The first clue is that elements of $\U(1)$ must commute with the
elements of $\SU(2) \times \SU(3)$.  But the only elements of $\SU(5)$
that commute with everybody in the $\SU(2) \times \SU(3)$ subgroup are
diagonal, since they must separately commute with $\SU(2) \times 1$
and $1 \times \SU(3)$, and the only elements doing so are
diagonal. Moreover, they must be scalars on each block.  So, they have
to look like this:
\[
\left( 
\begin{array}{c c}
\alpha & 0 \\
0      & \beta
\end{array}
\right)
\]
where $\alpha$ stands for the $2 \times 2$ identity matrix times
the complex number $\alpha \in \U(1)$, and similarly for $\beta$ in the
$3 \times 3$ block.  For the above matrix to lie in $\SU(5)$, it
must have determinant 1, so $\alpha^2 \beta^3 = 1$.  This condition 
cuts the group of such matrices from $\U(1) \times \U(1)$ down to 
$\U(1)$. In fact, all such matrices are of the form
\[
\left( 
\begin{array}{c c}
\alpha^3 & 0 \\
0        & \alpha^{-2}
\end{array}
\right)
\]
where $\alpha$ runs over $\U(1)$.

So if we throw in elements of this form, do we get $\U(1) \times
\SU(2) \times \SU(3)$?  More precisely, does this map:
\[ \begin{array}{rccc}
  \phi \maps & \GSM &\to& \SU(5) \\   
             & (\alpha, g, h) &\longmapsto& 
\left( 
\begin{array}{c c}
\alpha^{3}g & 0 \\
0           & \alpha^{-2}h
\end{array}
\right)
\end{array}
\]
give an isomorphism between $\GSM$ and $\S(\U(2) \times \U(3))$?
It is clearly a homomorphism.  It clearly maps 
$\GSM$ \emph{into} the subgroup $\S(\U(2) \times \U(3))$, and 
it is easy to check that it maps $\GSM$ \emph{onto} this subgroup.
But is it one-to-one?

The answer is \emph{no}: the map $\phi$ has a kernel, $\Z_6$.  
The kernel is the set of all elements of the form
\[ (\alpha, \alpha^{-3}, \alpha^2) \in \U(1) \times \SU(2) \times \SU(3) \]
and this is $\Z_6$, because scalar matrices $\alpha^{-3}$ and $\alpha^2$ live
in $\SU(2)$ and $\SU(3)$, respectively, if and only if $\alpha$ is a sixth root
of unity.  So, all we get is
\[  \GSM/\Z_6 \cong \S(\U(2) \times \U(3)) \inclusion \SU(5). \]

This sets up a nerve-wracking test that the $\SU(5)$ theory must pass
for it to have any chance of success.  After all, not all
representations of $\GSM$ factor through $\GSM/\Z_6$, but all those
coming from representations of $\SU(5)$ must do so.  A representation
of $\GSM$ will factor through $\GSM/\Z_6$ only if the $\Z_6$ subgroup acts
trivially.

In short: the $\SU(5)$ GUT is doomed unless $\Z_6$ acts trivially on
every fermion.  (And antifermion, but that amounts to the same thing.)
For this to be true, some nontrivial relations between hypercharge, 
isospin and color must hold.

For example, consider the left-handed electron
\[ e^-_L \in \C_{-1} \otimes \C^2 \otimes \C .\]
For any sixth root of unity $\alpha$, we need
\[ (\alpha, \alpha^{-3}, \alpha^2) \in \U(1) \times \SU(2) \times \SU(3) \]
to act trivially on this particle.  Let us see how it acts.  Note that:
\begin{itemize}
\item
$\alpha \in \U(1)$ acts on $\C_{-1}$ as multiplication by $\alpha^{-3}$;
\item 
$\alpha^{-3} \in \SU(2)$ acts on $\C^2$ as multiplication by $\alpha^{-3}$;
\item 
$\alpha^2 \in \SU(3)$ acts trivially on $\C$.
\end{itemize}
So, we have 
\[ (\alpha, \alpha^{-3}, \alpha^2) \cdot e^-_L = 
\alpha^{-3} \alpha^{-3} e^-_L = e^-_L .\]
The action is indeed trivial---precisely because 
$\alpha$ is a sixth root of unity.

Or, consider the right-handed $d$ quark:
\[ d_R \in \C_{-\twothirds} \otimes \C \otimes \C^3. \]
How does $(\alpha, \alpha^{-3}, \alpha^2)$ act on this?
We note:
\begin{itemize}
\item
$\alpha \in \U(1)$ acts on $\C_{-\twothirds}$ as multiplication by 
$\alpha^{-2}$;
\item 
$\alpha^{-3} \in \SU(2)$ acts trivially on the trivial representation
$\C$;
\item 
$\alpha^2 \in \SU(3)$ acts on $\C^3$ as multiplication by $\alpha^2$.
\end{itemize}
So, we have
\[ (\alpha, \alpha^{-3}, \alpha^2) \cdot d_R 
= \alpha^{-2} \alpha^2 d_R = d_R .\]
Again, the action is trivial.

For $\SU(5)$ to work, though, $\Z_6$ has to act trivially on \emph{every}
fermion.  There are 16 cases to check, and it is an awful lot to demand that
hypercharge, the most erratic part of the Standard Model 
representation, satisfies 16 relations.

Or is it? In general, for a fermion with hypercharge $Y$, there are four
distinct possibilities:
\vskip 1em
\begin{center}
\begin{tabular}{lcccc}
       	\hline
       	\multicolumn{5}{|c|}{\bf{Hypercharge relations}} \\
       	\hline
       	Case & & Representation & & Relation \\
       	\hline
       	Nontrivial $\SU(2)$, nontrivial $\SU(3)$ & $\Rightarrow$ & $\C_Y \otimes \C^2 \otimes \C^3$ & $\Rightarrow$ & $\alpha^{3Y - 3 + 2} = 1$ \\
	\hline
       	Nontrivial $\SU(2)$, trivial $\SU(3)$    & $\Rightarrow$ & $\C_Y \otimes \C^2 \otimes \C$   & $\Rightarrow$ & $\alpha^{3Y - 3} = 1$ \\
	\hline
       	Trivial $\SU(2)$, nontrivial $\SU(3)$    & $\Rightarrow$ & $\C_Y \otimes \C \otimes \C^3$   & $\Rightarrow$ & $\alpha^{3Y + 2} = 1$ \\
	\hline
       	Trivial $\SU(2)$, trivial $\SU(3)$       & $\Rightarrow$ & $\C_Y \otimes \C \otimes \C$     & $\Rightarrow$ & $\alpha^{3Y} = 1$ \\
       	\hline
\end{tabular}
\end{center}
Better yet, say it like a physicist!
\vskip 1em
\begin{center}
\begin{tabular}{lcccc}
       	\hline
       	\multicolumn{5}{|c|}{\bf{Hypercharge relations}} \\
       	\hline
       	Case & & Representation & & Relation \\
       	\hline
       	Left-handed quark   & $\Rightarrow$ & $\C_Y \otimes \C^2 \otimes \C^3$ & $\Rightarrow$ & $\alpha^{3Y - 3 + 2} = 1$ \\
	\hline
       	Left-handed lepton  & $\Rightarrow$ & $\C_Y \otimes \C^2 \otimes \C$   & $\Rightarrow$ & $\alpha^{3Y - 3} = 1$ \\
	\hline
       	Right-handed quark  & $\Rightarrow$ & $\C_Y \otimes \C \otimes \C^3$   & $\Rightarrow$ & $\alpha^{3Y + 2} = 1$ \\
	\hline
       	Right-handed lepton & $\Rightarrow$ & $\C_Y \otimes \C \otimes \C$     & $\Rightarrow$ & $\alpha^{3Y} = 1$\\
       	\hline
\end{tabular}
\end{center}
But $\alpha$ is sixth root of unity, so all this really says is that those
exponents are multiples of six:
\vskip 1em
\begin{center}
\begin{tabular}{lcc}
       	\hline
       	\multicolumn{3}{|c|}{\bf{Hypercharge relations}} \\
       	\hline
       	Case & & Relation \\
       	\hline
       	Left-handed quark   & $\Rightarrow$ & $3Y - 3 + 2 \in 6\Z$ \\
	\hline
       	Left-handed lepton  & $\Rightarrow$ & $3Y - 3 \in 6\Z$ \\
	\hline
       	Right-handed quark  & $\Rightarrow$ & $3Y + 2 \in 6\Z$ \\
	\hline
       	Right-handed lepton & $\Rightarrow$ & $3Y \in 6\Z$ \\
       	\hline
\end{tabular}
\end{center}
Dividing by 3 and doing a little work, it is easy to see these are just saying:
\vskip 1em
\begin{table}[H]
\begin{center}
	\begin{tabular}{ll}
		\hline
		\multicolumn{2}{|c|}{\bf{Hypercharge relations}} \\
		\hline
		Case & Hypercharge \\
		\hline
		Left-handed quark   & $Y$ = even integer $+ \third$ \\
		\hline
		Left-handed lepton  & $Y$ = odd integer \\
		\hline
		Right-handed quark  & $Y$ = odd integer $+ \third$ \\
		\hline
		Right-handed lepton & $Y$ = even integer \\
		\hline
\end{tabular}
	\caption{Hypercharge relations} \label{tab:hypercharge}
\end{center}
\end{table}
\noindent
Now it is easy to check that this indeed holds for every fermion in the standard
model.  $\SU(5)$ passes the test, not despite the bizarre pattern followed by
hypercharges, but \emph{because of it!}

By this analysis, we have shown that $\Z_6$ acts trivially on the Standard Model
rep, so it is contained in the kernel of this rep. It is better than just a
containment though: $\Z_6$ is the entire kernel. Because of this, we could say
that $\GSM/\Z_6$ is the `true' gauge group of the Standard Model.  And because
we now know that
\[ \GSM/\Z_6 \iso \S(\U(2) \times \U(3)) \inclusion \SU(5), \]
it is almost as though this $\Z_6$ kernel, lurking inside $\GSM$ this whole
time, was a cryptic hint to try the $\SU(5)$ theory.

Of course, we still need to find a representation of $\SU(5)$ that extends the 
Standard Model representation.   Luckily, there is a very beautiful choice
that works: the exterior algebra $\Ex \C^5$.
Since $\SU(5)$ acts on $\C^5$, it has a representation on $\Ex \C^5$.
Our next goal is to check that pulling back this representation from
$\SU(5)$ to $\GSM$ using $\phi$, we obtain the Standard Model
representation $F \oplus F^*.$

As we do this, we will see another fruit of the $\SU(5)$ theory ripen. The
triviality of $\Z_6$ already imposed some structure on hypercharges,
as outlined above in Table~\ref{tab:hypercharge}. As we fit the
fermions into $\Ex \C^5$, we will see this is no accident: the
hypercharges have to be \emph{exactly what they are} for the
$\SU(5)$ theory to work.

To get started, our strategy will be to use the fact that, being
representations of compact Lie groups, both the fermions $F \oplus
F^*$ and the exterior algebra $\Ex \C^5$ are completely reducible, 
so they can be written as a direct sum of irreps. We will then match up
these irreps one at a time.

The fermions are already written as a direct sum of irreps, so we need
to work on $\Ex \C^5$. Now, any element $g \in \SU(5)$ acts as an automorphism
of the exterior algebra $\Ex \C^5$:
\[ g(v \wedge w) = gv \wedge gw \]
where $v,w \in \Ex \C^5$. Since we know how $g$ acts on the vectors in $\C^5$,
and these generate $\Ex \C^5$, this rule is enough to tell us how $g$ acts on
all of $\Ex \C^5$. This action respects grades in $\Ex \C^5$, so each
exterior power in
\[ \Ex \C^5 \iso \Ex^0 \C^5 \oplus \Ex^1 \C^5 \oplus \Ex^2 \C^5 \oplus \Ex^3 \C^5 \oplus \Ex^4 \C^5 \oplus \Ex^5 \C^5    \]
is a subrepresentation. In fact, these are all irreducible, so this is
how $\Ex \C^5$ breaks up into irreps of $\SU(5)$.  Upon restriction to $\GSM$,
some of these summands break apart further into irreps of $\GSM$.

Let us see how this works, starting with the easiest cases.  $\Ex^0
\C^5$ and $\Ex^5 \C^5$ are both trivial irreps of $\GSM$.   There
are two trivial irreps in the Standard Model representation, namely 
$\langle \nu_R \rangle$ and its dual $\langle \nubar_L \rangle$,
where we use angle brackets to stand for the Hilbert space spanned by 
a vector or collection of vectors.
So, we could select $\Ex^0 \C^5 = \langle \nubar_L \rangle$ and $\Ex^5
\C^5 = \langle \nu_R \rangle$, or vice versa. At this juncture, we
have no reason to prefer one choice to the other.

Next let us chew on the next piece: the first exterior power, $\Ex^1
\C^5$.  We have
\[ \Ex^1 \C^5 \iso \C^5 \]
as vector spaces, and as representations of $\GSM$.  But what is $\C^5$ as
a representation of $\GSM$?  The Standard Model gauge group acts on 
$\C^5$ via the map
\[ \phi \maps (\alpha, g, h) \longmapsto 
\left( 
\begin{array}{c c}
\alpha^{3}g & 0 \\
0 & \alpha^{-2}h
\end{array}
\right) .
\]
Clearly, this action preserves the splitting into the `isospin part'
and the `color part' of $\C^5$:
\[ \C^5 \iso \C^2 \oplus \C^3 .\]
So, let us examine these two subrepresentations in turn:
\begin{itemize}
	\item The $\C^2$ part transforms in the hypercharge 1 rep of $\U(1)$:
		that is, $\alpha$ acts as multiplication by $\alpha^3$.  It
		transforms according to the fundamental representation of
		$\SU(2)$, and the trivial representation of $\SU(3)$.  This
		seems to describe a left-handed lepton with hypercharge 1.

	\item The $\C^3$ part transforms in the hypercharge $-\twothirds$ rep
		of $\U(1)$: that is, $\alpha$ acts as multiplication by
		$\alpha^{-2}$.  It transforms trivially under $\SU(2)$, and
		according to the fundamental representation of $\SU(3)$.  Table
		\ref{tab:smrep} shows that these are the features of a
		right-handed down quark.
\end{itemize}

In short, as a rep of $\GSM$, we have
\[ \C^5 \quad \iso \quad 
\C_1 \otimes \C^2 \otimes \C \quad \oplus 
\quad \C_{-\twothirds} \otimes \C \otimes \C^3 \]
and we have already guessed which particles these correspond to. 
The first summand looks like a left-handed lepton with hypercharge 1, 
while the second is a right-handed quark with hypercharge $-\twothirds$.

Now this is problematic, because another glance at Table \ref{tab:smrep}
reveals that there is no left-handed lepton with 
hypercharge 1.  The only particles with hypercharge 1 are the right-handed 
antileptons, which span the representation
\[ \angantilep \iso \C_1 \otimes \C^{2*} \otimes \C. \]
But wait! $\SU(2)$ is unique among the $\SU(n)$'s in that its fundamental rep
$\C^2$ is self-dual:
\[ \C^2 \iso \C^{2*}. \]
This saves the day. As a rep of $\GSM$, $\C^5$ becomes
\[ \C^5 \quad \iso \quad
\C_1 \otimes \C^{2*} \otimes \C \quad \oplus \quad 
\C_{-\twothirds} \otimes \C \otimes \C^3 \]
so it describes the right-handed antileptons with hypercharge 1 
and the right-handed quarks with hypercharge $-\twothirds$. 
In other words:
\[ \Ex^1 \C^5 \iso \C^5 \iso \angantilep \oplus \langle d_R \rangle \]
where we have omitted the color label on $d_R$ to save space. Take heed of 
this: $\langle d_R \rangle$ is short for the vector space
$\langle d^r_R, d^g_R, d^b_R \rangle$, and it is three-dimensional.

Now we can use our knowledge of the first exterior power to
compute the second exterior power, by applying the formula
\[ \Ex^2 (V \oplus W) \quad \iso \quad 
\Ex^2 V \; \oplus \; (V \otimes W) \; \oplus \; \Ex^2 W. \]

So, let us calculate!   As reps of $\GSM$ we have
\[  
\Ex^2 \C^5   \quad \iso \quad
\Ex^2 ( \C_1 \otimes \C^2 \otimes \C \quad 
\oplus \quad \C_{-\twothirds} \otimes \C^2 \otimes \C^3 )  
\]
\[
 \iso  \quad
\Ex^2 ( \C_1 \otimes \C^2 \otimes \C ) \quad \oplus 
     \quad (\C_1 \otimes \C^2 \otimes \C) \otimes 
(\C_{-\twothirds} \otimes \C \otimes \C^3) \quad \oplus 
    \quad \Ex^2 ( \C_{-\twothirds} \otimes \C \otimes \C^3 ).
\]

Consider the first summand, $\Ex^2 ( \C_1 \otimes
\C^2 \otimes \C)$.  As a rep of $\SU(2)$ this space is just $\Ex^2 \C^2$, 
which is the one-dimensional trivial rep, $\C$. 
As a rep of $\SU(3)$ it is also trivial.  But as a rep of
$\U(1)$, it is nontrivial.  Inside it we are juxtaposing 
two particles with hypercharge 1. Hypercharges add, just like charges, 
so the composite particle, which consists of one particle \textsl{and} 
the other, has hypercharge 2.  So, as a representation of the Standard
Model gauge group we have
\[ \Ex^2 (\C_1 \otimes \C^2 \otimes \C) \iso \C_2 \otimes \C \otimes \C  .\]
Glancing at Table \ref{tab:smrep} we see this matches the
left-handed positron, $e^+_L$.  Note that the hypercharges are becoming 
useful now, since they uniquely identify all the fermion and antifermion 
representations, except for neutrinos.

Next consider the second summand:
\[  (\C_1 \otimes \C^2 \otimes \C) \otimes
(\C_{-\twothirds} \otimes \C \otimes \C^3) . \]
Again, we can add hypercharges, so this representation of $\GSM$ is 
isomorphic to
\[ \C_\third \otimes \C^2 \otimes \C^3. \]
This is the space for left-handed quarks of hypercharge $\third$, which
from Table \ref{tab:smrep} is:
\[ \angquark \]
where once again we have suppressed the label for colors.

Finally, the third summand in $\Ex^2 \C^5$ is 
\[ \Ex^2 ( \C_{-\twothirds} \otimes \C \otimes \C^3 ) .\]
This has isospin $-\fourthirds$, so by Table \ref{tab:smrep}
it had better correspond to the left-handed antiup
antiquark, which lives in the representation
\[ \C_{-\fourthirds} \otimes \C \otimes \C^{3*}. \]
Let us check. 
The rep $\Ex^2 ( \C_{-\twothirds} \otimes \C \otimes \C^3 )$ 
is trivial under $\SU(2)$.  As a rep of $\SU(3)$ it is the same as $\Ex^2 \C^3$.  
But because $\SU(3)$ preserves the volume form on $\C^3$, taking Hodge duals 
gives an isomorphism 
\[ \Ex^p \C^3 \iso (\Ex^{3 - p} \C^3)^* \]
so we have
\[ \Ex^2 \C^3 \iso (\Ex^1 \C^3)^* \iso \C^{3*} \]
which is just what we need to show
\[ \Ex^2 ( \C_{-\twothirds} \otimes \C^3 ) \iso 
 \C_{-\fourthirds} \otimes \C \otimes \C^{3*} \iso 
\langle \ubar_L \rangle .\]

In summary, the following pieces of the Standard Model rep
sit inside $\Ex^2 \C^5$:
\[ \Ex^2 \C^5 \iso \langle e^+_L \rangle \oplus \angquark \oplus 
\langle \ubar_L \rangle \]

We are almost done. Because $\SU(5)$ preserves the canonical volume form on
$\C^5$, taking Hodge duals gives an isomorphism between
\[ \Ex^p \C^5 \iso (\Ex^{5 - p} \C^5)^* \]
as representations of $\SU(5)$.  Thus given our results so far:
\begin{eqnarray*}
\Ex^0 \C^5 & \iso & \langle \nubar_L \rangle \\
\Ex^1 \C^5 & \iso & \angantilep \oplus \langle d_R \rangle \\
\Ex^2 \C^5 & \iso & \langle e^+_L \rangle \oplus \angquark \oplus 
\langle \ubar_L \rangle
\end{eqnarray*}
we automatically get the antiparticles of these upon taking Hodge duals,
\begin{eqnarray*}
\Ex^3 \C^5 & \iso & \langle e^-_R \rangle \oplus \angantiquark \oplus 
\langle u_R \rangle \\
\Ex^4 \C^5 & \iso & \anglep \oplus \langle \dbar_L \rangle \\
\Ex^5 \C^5 & \iso & \langle \nu_R \rangle.
\end{eqnarray*}
So $\Ex \C^5 \iso F \oplus F^*$, as desired.

How does all this look in terms of the promised binary code? 
Remember, a 5-bit code is short for a wedge product of
basis vectors $u,d,r,g,b \in \C^5$.  For example, 01101 corresponds
to $d \wedge r \wedge b$.  And now that we have found an isomorphism 
$\Ex \C^5 \iso F \oplus F^*$, each of these wedge products corresponds
to a fermion or antifermion.  How does this correspondence go, exactly?
 
First consider the grade-one part $\Ex^1 \C^5 \iso \C^2 \oplus \C^3$.
This has basis vectors called $u,d,r,g,$ and $b$.  We have seen that
the subspace $\C^2$, spanned by $u$ and $d$, corresponds to 
\[   \angantilep.   \]
The top particle here has isospin up, while the bottom one has
isospin down, so we must have $e^+_R = u$ and $\nubar_R = d$.
Likewise, the subspace $\C^3$ spanned by $r,g$ and $b$ corresponds to
\[     \langle d_R \rangle = \langle d_R^r, d_R^g, d_R^b \rangle . \] 
Thus we must have $d^c_R = c$, where $c$ runs over the colors $r,g,b$.

Next consider the grade-two part:
\[        \Ex^2 \C^5  \iso  \langle e^+_L \rangle \oplus \angquark \oplus 
\langle \ubar_L \rangle .\]
Here $e^+_L$ lives in the one-dimensional $\Ex^2 \C^2$ rep
of $\SU(2)$, which is spanned by the vector $u \wedge d$. 
Thus, $e^+_L = u \wedge d$.  The left-handed quarks 
live in the $\C^2 \otimes \C^3$ rep of $\SU(2) \times \SU(3)$,
which is spanned by vectors that consist of one isospin and one color. We must
have $u^c_L = u \wedge c$ and $d^c_L = d \wedge c$, where again $c$ runs 
over all the colors $r,g,b$.  And now for the tricky part: the $\ubar_L$ 
quarks live in the $\Ex^2 \C^3$ rep of $\SU(3)$, but this is isomorphic 
to the fundamental representation of $\SU(3)$ on $\C^{3*}$, which is
spanned by antired, antired and antiblue:
\[ \rbar = g \wedge b , \qquad
   \gbar = b \wedge r , \qquad
   \bbar = r \wedge g. \]
These vectors form the basis of $\Ex^2 \C^3$ that is dual to $r$, $g$, 
and $b$ under Hodge duality in $\Ex \C^3$.  So we must have
\[ u^\cbar_L = \cbar \]
where $\cbar$ can be any anticolor.  Take heed of the fact that $\cbar$ 
is grade 2, even though it may look like grade 1.

To work out the other grades, note that Hodge duality corresponds to switching 
0's and 1's in our binary code.  For instance, the dual of 01101 is 
10010: or written in terms of basis vectors, the dual of 
$d \wedge r \wedge b$ is $u \wedge g$.  Thus given the binary codes
for the first few exterior powers:
\vskip 1em
\begin{center}
	\begin{tabular}{lll}
		\hline
		$\Ex^0 \C^5$   & $\Ex^1 \C^5$    & $\Ex^2 \C^5$            \\
		\hline
		$\nubar_L = 1$ & $e^+_R = u$     & $e^+_L = u \wedge d$    \\
		               & $\nubar_R = d$  & $u^c_L = u \wedge c$    \\
			       & $d^c_R = c$     & $d^c_L = d \wedge c$    \\
			       &                 & $\ubar^\cbar_L = \cbar$ \\
		\hline
	\end{tabular}
\end{center}
\vskip 1em
\noindent taking Hodge duals gives the binary codes for the rest: 
\vskip 1em
\begin{center}
	\begin{tabular}{lll}
		\hline
		$\Ex^3 \C^5$             & $\Ex^4 \C^4$              & $\Ex^5 \C^5$ \\
		\hline
	$e^-_R = r \wedge g \wedge b$         & $e^-_L = d \wedge r \wedge g \wedge b$            & $\nu_R = u \wedge d \wedge r \wedge g \wedge b$ \\
$\ubar^\cbar_R = d \wedge \cbar$ & $\nu_L = u \wedge r \wedge g \wedge b$ 
    & \\
$\dbar^\cbar_R = u \wedge \cbar$ & 
$\dbar^\cbar_L = u \wedge d \wedge \cbar$ & \\
$u^c_R = u \wedge d \wedge c$            &                           & \\
		\hline
	\end{tabular}
\end{center}
\vskip 1em
Putting these together, we get the binary code for every particle
and antiparticle in the first generation of fermions.  To save space,
let us omit the wedge product symbols:
\vskip 1em
	\begin{table}[H]
\begin{center} 
	\begin{tabular}{cccccc} 
		\hline
		\multicolumn{6}{|c|}{\bf{The Binary Code for \boldmath $\SU(5)$}} \\
		\hline
		$\Ex^0 \C^5$   & $\Ex^1 \C^5$    & $\Ex^2 \C^5$            & $\Ex^3 \C^5$             & $\Ex^4 \C^4$              & $\Ex^5 \C^5$ \\ 
		\hline
\\
		$\nubar_L = 1$ & $e^+_R = u$     & $e^+_L = ud$            & $e^-_R = rgb$            & $e^-_L = drgb$            & $\nu_R = udrgb$ \\  \\
		               & $\nubar_R = d$  & $u^c_L = uc$            & $\ubar^\cbar_R = d\cbar$ & $\nu_L = urgb$            & \\  \\
			       & $d^c_R = c$     & $d^c_L = dc$            & $\dbar^\cbar_R = u\cbar$ & $\dbar^\cbar_L = ud\cbar$ & \\   \\
			       &                 & $\ubar^\cbar_L = \cbar$ & $u^c_R = udc$            &                           & \\   \\
		\hline
	\vspace{-10pt}
	\end{tabular}
		\caption{Binary code for first-generation fermions, where
      $c= r, g, b$ and $\cbar = gb, br, rg$} \label{tab:su5code}
\end{center}
\end{table}
\vspace{-1em}

Now we can see a good, though not decisive, reason to choose $\Ex^0 \C^5 \iso
\nubar_L$. With this choice, and not the other, we get left-handed particles in
the even grades, and right-handed particles in the odd grades. We \emph{choose}
to have this pattern now, but later on we need it.

Table~\ref{tab:su5code} defines a linear isomorphism $f \maps F \oplus F^* \to
\Ex \C^5$ in terms of the basis vectors, so the equations in this table
are a bit of an exaggeration.  When we write say, $e^+_R = u$, we really
mean $f(e^+_R) = u$.  This map $f$ is an isomorphism between 
representations of $\GSM$. It tells us how these representations are 
the `same'.

More precisely, we mean these representations are the same when we 
identify $\S(\U(2) \times \U(3))$ with $\GSM/\Z_6$ using the isomorphism 
induced by $\phi$. In general,
we can think of a unitary representation as a Lie group homomorphism
\[ G \to \U(V) \]
where $V$ is a finite-dimensional Hilbert space and $\U(V)$ is the 
Lie group of unitary operators on $V$.  
In this section we have been comparing two unitary representations:
an ugly, complicated representation of $\GSM$:
\[ \GSM \to \U(F \oplus F^*) \]
and a nice, beautiful representation of $\SU(5)$:
\[ \SU(5) \to \U(\Ex \C^5). \]
We built a homomorphism
\[ \phi \maps \GSM \to \SU(5), \]
so it is natural to wonder if is there a fourth homomorphism
\[ \U(F \oplus F^*) \to \U(\Ex \C^5) \]
such that this square:
\[
\xymatrix{
\GSM \ar[r]^\phi \ar[d] & \SU(5) \ar[d] \\
\U(F \oplus F^*) \ar[r] & \U(\Ex \C^5)
} 
\]
commutes.

Indeed, we just showed this!  We have seen that there exists a
unitary operator from the Standard Model rep to $\Ex \C^5$, say
\[ f \maps F \oplus F^* \stackrel{\sim}{\to} \Ex \C^5 , \]
such that the induced isomorphism of the unitary groups, 
\[ \U(f) \maps \U(F \oplus F^*) \stackrel{\sim}{\to} \U(\Ex \C^5), \] 
makes the above square commute.  So, let us summarize this result as a theorem:

\begin{thm}\et
\label{thm:su(5)}
	The following square commutes:
	\[
	\xymatrix{
	\GSM \ar[r]^\phi \ar[d] & \SU(5) \ar[d] \\
	\U(F \oplus F^*) \ar[r]^-{\U(f)} & \U(\Ex \C^5)
	}
	\]
where the left vertical arrow is the Standard Model representation
and the right one is the natural representation of $\SU(5)$ on
the exterior algebra of $\C^5$.
\end{thm}

\subsection{The Spin(10) Theory} \label{sec:so(10)}

We now turn our attention to another grand unified theory.  Physicists
call it the `$\SO(10)$ theory', but we shall call it the
$\Spin(10)$ theory, because the Lie group involved is really
$\Spin(10)$, the double cover of $\SO(10)$.  This theory appeared in a 
1974 paper by Georgi~\cite{georgi:so(10)}, shortly after his paper with
Glashow on the $\SU(5)$ theory.  However, Georgi has said that he conceived 
of the $\Spin(10)$ theory first.  See Zee~\cite{zee:nutshell}, Chapter VII.7, 
for a concise and readable account. 

The $\SU(5)$ GUT has helped us explain the pattern of hypercharges in the
Standard Model, and thanks to the use of the exterior algebra, $\Ex \C^5$, we
can interpret it in terms of a binary code. This binary code explains another
curious fact about the Standard Model. Specifically, why is the number of
fermions a power of 2? There are 16 fermions, and 16 antifermions, which 
makes the
Standard Model rep have dimension
\[ \dim (F \oplus F^*) = 2^5 = 32. \]
With the binary code interpretation, it could not be any other way.

In actuality, however, the existence of a right-handed neutrino (or its
antiparticle, the left-handed antineutrino) has been controversial. Because it
transforms trivially in the Standard Model, it does not interact
with anything except perhaps the Higgs.

The right-handed neutrino certainly improves the aesthetics of the
$\SU(5)$ theory.  When we include this particle (and its antiparticle),
we obtain the rep
\[ \Ex^0 \C^5 \oplus \Ex^1 \C^5 \oplus \Ex^2 \C^5 \oplus \Ex^3 \C^5 \oplus \Ex^4 \C^5 \oplus \Ex^5 \C^5    \]
which is all of $\Ex \C^5$, whereas without this particle
we would just have
\[ \Ex^1 \C^5 \oplus \Ex^2 \C^5 \oplus \Ex^3 \C^5 \oplus \Ex^4 \C^5  \]
which is much less appealing---it \emph{wants} to be $\Ex \C^5$, but
it comes up short.   

More importantly, there is increasing indirect evidence from
experimental particle physics that right-handed neutrinos \emph{do}
exist.   For details, see Pati \cite{pati:probing}.  
If this is true, the number of fermions really could be 16, and we have a
ready-made explanation for that number in the binary code.

However, this creates a new mystery.  The
$\SU(5)$ works nicely with the representation $\Ex \C^5$, but $\SU(5)$ does 
not \emph{require} this.  It works just fine with the smaller rep
\[ \Ex^1 \C^5 \oplus \Ex^2 \C^5 \oplus \Ex^3 \C^5 \oplus \Ex^4 \C^5 .\]
It would be nicer to have a theory that \emph{required} us
to use all of $\Lambda \C^5$.  Better yet, if our new GUT were an
\emph{extension} of $\SU(5)$, the beautiful explanation of hypercharges
would live on in our new theory.  With luck, we might even get away with 
using the same underlying vector space, $\Ex \C^5$.
Could it be that the $\SU(5)$ GUT is only the beginning of the story? Could
unification go on, with a grand unified theory that extends $\SU(5)$ just as
$\SU(5)$ extended the Standard Model?

Let us look for a group that extends $\SU(5)$ and has an irrep whose
dimension is some power of 2.  The dimension is a big clue.  What
representations have dimensions that are powers of 2? \emph{Spinors}.

What are spinors? They are certain representations of $\Spin(n)$,
the double cover of the rotation group in $n$ dimensions, which do not factor
through the quotient $\SO(n)$. Their dimensions are always a power of two.
We build them by exhibiting $\Spin(n)$ as a subgroup of a 
Clifford algebra.  Recall that the {\bf Clifford algebra} $\Cliff_n$ is
the associative algebra freely generated by $\R^n$ with relations
\[	v w + w v = -2\langle v, w \rangle .	\]
If we take products of pairs of unit vectors in $\R^n$ inside
this algebra, these generate the group $\Spin(n)$: multiplication 
in this group coincides with multiplication in the Clifford algebra.
Using this fact, we can get representations of $\Spin(2n)$ 
from modules of $\Cliff_n$.  

We can use this method to get a rep of $\Spin(10)$ on $\Lambda \C^5$ 
that extends the rep of $\SU(5)$ on this space.
In fact, quite generally $\Cliff_{2n}$ acts on $\Ex \C^n$.  Then, because
\[ \Spin(2n) \inclusion \Cliff_{2n}, \]
$\Ex \C^n$ becomes a representation of $\Spin(2n)$, called the
\textbf{Dirac spinor representation}.  

To see this, we use operators on $\Ex \C^n$ 
called `creation and annihilation operators'.  Let $e_1, \ldots, e_n$ be the 
standard basis for $\C^n$.  Each of these gives a {\bf creation operator}:
\[ 
\begin{array}{rcl}
     a_j^*  \maps \Ex \C^n &\to& \Ex \C^n    \\
                       \psi &\mapsto& e_j \wedge \psi .
\end{array}\]
We use the notation $a_j^*$ because $\Ex \C^n$ is a Hilbert space,
so $a_j^*$ is the adjoint of some other operator
\[         a_j  \maps \Ex \C^n \to \Ex \C^n  ,   \]
which is called an {\bf annihilation operator}.   

In physics, we can think of the basis vectors $e_j$ as particles.
For example, in the binary code approach to the $\SU(5)$
theory we imagine five particles from which the observed
particles in the Standard Model are composed: {\em{up, down, red, green}}
and {\em blue}.  Taking the wedge product with $e_j$ `creates a particle' 
of type $j$, while the adjoint `annihilates a particle' of type $j$.  

It may seem odd that creation is the adjoint of annihilation, 
rather than its inverse. One reason for this is that the creation 
operator, $a_j^*$, has no inverse.  In some sense, its adjoint $a_j$ 
is the best substitute.

This adjoint \emph{does} do what want, which is to delete any particle of type
$j$.  Explicitly, it deletes the `first' occurrence of $e_j$ from any basis
element, bringing out any minus signs we need to make this respect the
antisymmetry of the wedge product:
\[	a_j ( e_{i_1} \wedge \cdots \wedge e_{i_p} ) = 
( -1 )^{k+1} e_{i_1} \wedge \cdots \wedge e_{i_{k-1}} \wedge e_{i_{k+1}} \cdots \wedge e_{i_p}, \mbox{ if }j=i_k.	\]
And if no particle of type $j$ appears, we get zero.

Now, whenever we have an inner product space like $\C^n$, we get an inner
product on $\Ex \C^n$. The fastest, if not most elegant, route to this inner
product is to remember that, given an orthonormal basis $e_1, \ldots, e_n$ for
$\C^n$, the induced basis, consisting of elements of the form $e_{i_1} \wedge
\cdots \wedge e_{i_p}$, should be orthonormal in $\Ex \C^n$. But choosing an
orthonormal basis defines an inner product, and in this case it defines an
inner product on the whole exterior algebra, one that reduces to the usual one
for the grade one elements, $\Ex^1 \C^n \iso \C^n$.

It is with respect to this inner product on $\Ex \C^n$ that $a_j$ and $a_j^*$
are adjoint. That is, they satisfy
\[ \langle v, a_jw \rangle = \langle a_j^*v, w \rangle \]
for any elements $v,w \in \Ex \C^n$. Showing this from the definitions we have
given is a straightforward calculation, which we leave to the reader.

These operators satisfy the following relations:
\begin{eqnarray*} 
	\{ a_j, a_k \} & = & 0 \\
	\{ a_j^*, a_k^* \} & = & 0 \\
	\{ a_j, a_k^* \} & = & \delta_{jk}
\end{eqnarray*}
where curly brackets denote the \textbf{anticommutator} of two linear
operators, namely $\{a, b\} = ab + ba$.

As an algebra, $\Cliff_{2n}$ is generated by the standard basis vectors of
$\R^{2n}$.  Let us call the elements of $\Cliff_{2n}$ corresponding to these
basis vectors $\gamma_1, \ldots, \gamma_{2n}$.  From the definition of
the Clifford algebra, it is easy to check that
\[      \{   \gamma_k , \gamma_\ell \} = -2\delta_{k\ell}  .\]  
In other words, the elements
$\gamma_k$ are anticommuting square roots of $-1$.  So, we can turn 
$\Ex \C^n$ into a $\Cliff_{2n}$-module by finding $2n$ linear operators 
on $\Ex \C^n$ that anticommute and square to $-1$.  We build these from 
the raw material provided by $a_j$ and $a_j^*$.  Indeed, it is easy to
see that 
\begin{eqnarray*}
	\phi_j & = & i(a_j + a_j^*) \\
	\pi_j  & = & a_j - a_j^*
\end{eqnarray*}
do the trick.  Now we can map $\gamma_1, \ldots, \gamma_{2n}$ to these 
operators, in any order, and $\Ex \C^n$ becomes a $\Cliff_{2n}$-module as 
promised.  

Now for $n > 1$ we may define $\Spin(2n)$ to be the universal cover of 
$\SO(2n)$, with group structure making the covering map
\[
\xymatrix{
	\Spin(2n) \ar[d]^p \\
	\SO(2n)
}
\]
into a homomorphism.  This universal cover is a double cover, because
the fundamental group of $\SO(2n)$ is $\Z_2$ for $n$ in this range.

This construction of $\Spin(2n)$ is fairly abstract.  Luckily, we can realize
$\Spin(2n)$ as the multiplicative group in $\Cliff_{2n}$ generated by products
of pairs of unit vectors.  This gives us the inclusion
\[ \Spin(2n) \inclusion \Cliff_{2n} \]
we need to make $\Ex \C^n$ into a representation of $\Spin(2n)$. 
From this, one can show that the Lie algebra $\so(2n)$ is 
generated by the commutators of the $\gamma_j$.  Because we know
how to map each $\gamma_j$ to an operator on $\Ex \C^n$, 
this gives us an explicit formula for the action of $\so(2n)$ on $\Ex \C^n$. 
Each $\gamma_j$
changes the parity of the grades, and their commutators do this twice,
restoring grade parity.  Thus, $\so(2n)$ preserves the parity of the grading on
$\Ex \C^n$, and $\Spin(2n)$ does the same.  This breaks $\Ex \C^n$ into
two subrepresentations:
\[	\Ex \C^n = \Exev \C^n  \oplus \Exodd \C^n	\]
where $\Exev \C^n$ is the direct sum of the even-graded parts:
\[	\Exev \C^n = \Ex^0 \C^n \oplus \Ex^2 \C^n \oplus \cdots	\]
while $\Exodd \C^n$ is the sum of the odd-graded parts:
\[	\Exodd \C^n = \Ex^1 \C^n \oplus \Ex^3 \C^n \oplus \cdots. \]

In fact, both these representations of $\Spin(2n)$ are irreducible, and
$\Spin(2n)$ acts faithfully on their direct sum $\Ex \C^n$. Elements of these
two irreps of $\Spin(2n)$ are called \define{left- and right-handed 
Weyl spinors}, respectively, while elements of $\Ex \C^n$ are called 
\define{Dirac spinors}.

All this works for any $n$, but we are especially interested in the 
case $n=5$.  The big question is: does the Dirac spinor representation of 
$\Spin(10)$ extend the obvious representation of $\SU(5)$ on $\Ex \C^5$?
Or, more generally, does the Dirac spinor representation of $\Spin(2n)$ 
extend the representation of $\SU(n)$ on $\Ex \C^n$?

Remember, we can think of a unitary representation as a group homomorphism
\[ G \to \U(V) \]
where $V$ is the Hilbert space on which $G$ acts as unitary operators.
Here we are concerned with two representations. One of them is the familiar
representation of $\SU(n)$ on $\Ex \C^n$:
\[ \rho \maps \SU(n) \to \U(\Ex \C^n), \]
which acts as the fundamental rep 
on $\Ex^1 \C^n \iso \C^n$ and respects wedge
products. The other is the representation of $\Spin(2n)$ on
the Dirac spinors, which happen to form the same vector space $\Ex \C^n$:
\[ \rho' \maps \Spin(2n) \to \U(\Ex \C^n). \]
Our big question is answered affirmatively by this theorem, which
can be found in a classic paper by Atiyah, Bott and Shapiro
\cite{ABS}:

\begin{thm}\et
\label{thm:spinor}
        There exists a Lie group homomorphism $\psi$ that makes this 
triangle commute:
	\[
	\xymatrix{
	\SU(n) \ar[r]^{\psi} \ar[dr]_\rho & \Spin(2n) \ar[d]^{\rho'} \\
	                    & \U(\Ex \C^n)
	}
	\]
\end{thm}

\emph{Proof.}  The complex vector space $\C^n$ has an underlying
real vector space of dimension $2n$, and the real part of the 
usual inner product on $\C^n$ gives an inner product on this 
underlying real vector space, so we have an inclusion $\U(n) 
\hookrightarrow \O(2n)$.  The connected component of the identity
in $\O(2n)$ is $\SO(2n)$, and $\U(n)$ is connected, so this gives
an inclusion $\U(n) \hookrightarrow \SO(2n)$ and thus 
$\SU(n) \hookrightarrow \SO(2n)$.  Passing to Lie algebras,
we obtain an inclusion $\su(n) \hookrightarrow \so(2n)$.  
A homomorphism of Lie algebras gives a homomorphism of the
corresponding simply-connected Lie groups, so this in turn gives
the desired map $\psi \maps \SU(n) \to \Spin(2n)$.

Next we must check that $\psi$ makes the above triangle commute.
Since all the groups involved are connected, it suffices to check
that this diagram
\[
\xymatrix{
\su(n) \ar[r]^{d\psi} \ar[dr]_{d\rho} & \so(2n) \ar[d]^{d\rho'} \\
                    & \u(\Ex \C^n)
}
\]
commutes.  Since the Dirac representation
$d\rho'$ is defined in terms of creation and annihilations operators, 
we should try to express $d\rho$ this way.  To do so, we will
need a good basis for $\su(n)$. Remember,
\[ \su(n) = \{n \times n \mbox{ traceless skew-adjoint matrices over } \C \}. \]
If $E_{jk}$ denotes the matrix with 1 in the $jk$th entry and 0 everywhere
else, then the traceless skew-adjoint matrices have this basis:
\[ 
\begin{array}{ll}
E_{jk} - E_{kj} &  j>k  \\
i(E_{jk} + E_{kj}) & j>k  \\
i(E_{jj} - E_{j+1,j+1}) & j=1, \ldots, n-1. 
\end{array} \] 
For example, $\su(2)$ has the basis
\[
\left(
\begin{array}{rr}
	0 & -1 \\
	1 & 0
\end{array}
\right)
\quad
\left(
\begin{array}{rr}
	0 & i \\
	i & 0
\end{array}
\right)
\quad
\left(
\begin{array}{rr}
	i & 0 \\
	0 & -i
\end{array}
\right)
\]
and our basis for $\su(n)$ simply generalizes this.

Now, it is easy to guess a formula for $d\rho$ in terms of 
creation and annihilation operators.  After all, the 
elementary matrix $E_{jk}$ satisfies
\[
E_{jk} (e_\ell) =
\left\{ \begin{array}{cl}
      e_j   & \textrm{if} \; \ell = k \\
       0    & \textrm{if} \; \ell \ne k 
\end{array} \right.
\]
and $a^*_j a_k$ acts the same way on $\Ex^1 \C^n \subseteq \Ex \C^n$. 
So, we certainly have
\[
\begin{array}{rcl}
d\rho \left(E_{jk} - E_{kj}\right) & = & a^*_j a_k - a^*_k a_j \\
d\rho \left(i(E_{jk} + E_{kj})\right) & = & i(a^*_j a_k + a^*_k a_j) \\
d\rho \left(i(E_{jj} - E_{j+1,j+1})\right)	& = & 
i(a^*_j a_j - a^*_{j+1} a_{j+1}) \\
\end{array}
\]
on the subspace $\Ex^1 \C^n$. But do these operators agree
on the rest of $\Ex \C^n$?  Remember, $\rho$ preserves
wedge products:
\[ \rho(x)(v \wedge w) = \rho(x)v \wedge \rho(x)w \]
for all $x \in \SU(n)$. Differentiating this condition, we see that $\su(n)$
must act as derivations:
\[ d\rho(X)(v \wedge w) = d\rho(X)v \wedge w + v \wedge d\rho(X)w\]
for all $X \in \su(n)$.  Derivations of $\Ex \C^n$ are determined by
their action on $\Ex^1 \C^n$.  So, $d\rho$ will be given on all
of $\Ex \C^n$ by the above formulas if we can show that
\[
a^*_j a_k - a^*_k a_j , 
\quad
i(a^*_j a_k + a^*_k a_j), 
\quad
\textrm{and} \quad
i(a^*_j a_j - a^*_{j+1} a_{j+1})
\]
are derivations.

Now, the annihilation operators are a lot like derivations: they are
\textbf{antiderivations}. That is, if $v \in \Ex^p \C^n$ and 
$w \in \Ex^q \C^n$, then
\[ a_j(v \wedge w) = a_j v \wedge w + (-1)^p v \wedge a_j w .\]
However, the creation operators are nothing like derivations. 
They satisfy
\[ a^*_j (v \wedge w) = a^*_j v \wedge w = (-1)^p v \wedge a^*_j w, \]
because $a^*_j$ acts by wedging with $e_j$, and moving this through $v$
introduces $p$ minus signs.  Luckily, this relation combines with the previous 
one to make the composites $a^*_j a_k$  into derivations for every $j$ and 
$k$.  We leave this for the reader to check.

So, $d\rho$ can really be expressed in terms of annihilation and
creation operators as above.  Checking that 
\[
\xymatrix{
\su(n) \ar[r]^{d\psi} \ar[dr]_{d\rho} & \so(2n) \ar[d]^{d\rho'} \\
                    & \u(\Ex \C^n)
}
\]
commutes is now a straightforward but somewhat tedious job, 
which we leave to the dedicated reader.  \qed

This theorem had a counterpart for the $\SU(5)$ GUT---namely,
Theorem \ref{thm:su(5)}.  There we saw a homomorphism $\phi$ that 
showed us how to extend the Standard Model group $\GSM$ to $\SU(5)$, and 
made this square commute:
\[
\xymatrix{
\GSM \ar[r]^\phi \ar[d] & \SU(5) \ar[d] \\
\U(F \oplus F^*) \ar[r]^-{\U(f)} & \U(\Ex \C^5)
}
\]
Now $\psi$ says how to extend $\SU(5)$ further to $\Spin(10)$, and 
makes this square commute:
\[
\xymatrix{
\SU(5) \ar[r]^{\psi} \ar[d]_\rho & \Spin(10) \ar[d]^{\rho'} \\
\U(\Ex \C^5) \ar[r]^1 & \U(\Ex \C^5)
}
\]
We can put these squares together to get this commutative diagram:
\[
\xymatrix{
\GSM \ar[r]^{\psi \phi} \ar[d] & \Spin(10) \ar[d] \\
\U(F \oplus F^*) \ar[r]^-{\U(f)} & \U(\Ex \C^5) 
}
\]
This diagram simply says that $\Spin(10)$ is a GUT: it extends the Standard
Model group $\GSM$ in a way that is compatible with the Standard Model
representation, $F \oplus F^*$. In Section~\ref{sec:su(5)}, all the hard work
lay in showing the representations $F \oplus F^*$ and $\Ex \C^5$ of $\GSM$ were
the same.  Here, we do not have to do that.  We just showed that $\Spin(10)$
extends $\SU(5)$. Since $\SU(5)$ already extended $\GSM$, $\Spin(10)$ extends
that, too.

\subsection{The Pati--Salam Model} \label{sec:g(2,2,4)}

Next we turn to a unified theory that is not so `grand':
its gauge group is not a simple Lie group, as it was for the
$\SU(5)$ and $\Spin(10)$ theories.  This theory is called the
Pati--Salam model, after its inventors~\cite{PatiSalam:model}; it has gauge
group $\SU(2) \times \SU(2) \times \SU(4)$, which is merely semisimple.

We might imagine the $\SU(5)$ theory as an answer to this question:
\begin{quote}
	Why are the hypercharges in the Standard Model what they are?
\end{quote}
The answer it provides is something like this:
\begin{quote}
	Because $\SU(5)$ is the actual gauge group of the world, acting on the
	representation $\Ex \C^5$.   
\end{quote}
But there are other intriguing patterns in the Standard Model that $\SU(5)$
does \emph{not} explain---and these lead us in different directions. 

First, there is a strange similarity between quarks and leptons. Each
generation of fermions in the Standard Model has two quarks and two
leptons.  For example, in the first generation we have the quarks $u$
and $d$, and the leptons $\nu$ and $e^-$.  The quarks come in three
`colors': this is a picturesque way of saying that they transform in
the fundamental representation of $\SU(3)$ on $\C^3$.  The leptons, on
the other hand, are `white': they transform in the trivial
representation of $\SU(3)$ on $\C$.
\vskip 1em
\begin{center}
	\begin{tabular}{lc}
		\hline
		\multicolumn{2}{|c|}{\bf{Representations of \boldmath $\SU(3)$}} \\
		\hline
		Particle & Representation \\
		\hline
		Quark & $\C^3$ \\
		Lepton & $\C$ \\
		\hline
	\end{tabular}
\end{center}
\vskip 1em
\noindent
Could the lepton secretly be a fourth color of quark? Maybe it could in a
theory where the $\SU(3)$ color symmetry of the Standard Model is extended  
to $\SU(4)$.  Of course this larger symmetry would need to be broken to
explain the very real \textit{difference} between leptons and quarks. 

Second, there is a strange difference between left- and right-handed
fermions. The left-handed ones participate in the weak force governed by
$\SU(2)$, while the right-handed ones do not.  Mathematically speaking, the
left-handed ones live in a nontrivial representation of $\SU(2)$, while the
right-handed ones live in a trivial one. The nontrivial one is $\C^2$, while
the trivial one is $\C \oplus \C$:
\vskip 1em
\begin{center}
	\begin{tabular}{lc}
		\hline
		\multicolumn{2}{|c|}{\bf{Representations of \boldmath $\SU(2)$}} \\
		\hline
		Particle & Representation \\
		\hline
		Left-handed fermion & $\C^2$ \\
		Right-handed fermion & $\C \oplus \C$ \\
		\hline
	\end{tabular}
\end{center}
\vskip 1em
But there is a suspicious similarity between $\C^2$ and $\C \oplus \C$. Could
there be another copy of $\SU(2)$ that acts on the right-handed particles?
Again, this `right-handed' $\SU(2)$ would need to be broken, to explain why
we do not see a `right-handed' version of the weak force that acts
on right-handed particles.

Following Pati and Salam, let us try to sculpt a theory that makes
these ideas precise.  In the last two sections, we saw some of the
ingredients we need to make a grand unified theory: we need to extend
the symmetry group $\GSM$ to a larger group $G$ using an inclusion
\[ \GSM \inclusion G \]
(up to some discrete kernel), and we need a representation $V$ of $G$ which
reduces to the Standard Model representation when restricted to $\GSM$:
\[ F \oplus F^* \iso V. \]
We can put all these ingredients together into a diagram
\[
\xymatrix{
	\GSM \ar[r] \ar[d] & G \ar[d] \\
	\U(F \oplus F^*) \ar[r]^-\sim    & \U(V)   \\
}
\]
which commutes only when our $G$ theory works out.

We now use the same methods to chip away at our current challenge.  We
asked if leptons correspond to a fourth color.  We already know that
every quark comes in three colors, $r$, $g$, and $b$, which form a basis 
for the vector space $\C^3$.  This is the fundamental representation 
of $\SU(3)$, the color symmetry group of the Standard Model.  If leptons 
correspond to a fourth color, say `white', then we should use the colors 
$r$, $g$, $b$ and $w$, as a basis for the vector space $\C^4$.  This is the 
fundamental representation of $\SU(4)$, so let us take that group to
describe color symmetries in our new GUT.

Now $\SU(3)$ has an obvious inclusion into $\SU(4)$, using block 
diagonal matrices:
\[ g \mapsto 
\left(
\begin{array}{cc}
	g & 0 \\
	0 & 1
\end{array}
\right)
\]
When restricted to this subgroup, the fundamental 
representation $\C^4$ breaks into a direct sum of irreps:
\[ \C^4 \iso \C^3 \oplus \C. \]
These are precisely the irreps of $\SU(3)$ that describe
quarks and leptons.  For antiquarks and antileptons we can use
\[ \C^{4*} \iso \C^{3*} \oplus \C. \]
It looks like we are on the right track.

We can do even better if we \emph{start} with the splitting 
\[     \C^4 \iso \C^3 \oplus \C.   \]
Remember that when we studied $\SU(5)$, the splitting 
\[ \C^5 \iso \C^2 \oplus \C^3 \]
had the remarkable effect of introducing $\U(1)$, and thus hypercharge, into
$\SU(5)$ theory.  This was because the subgroup of 
$\SU(5)$ that preserves this splitting is larger than $\SU(2) \times
\SU(3)$, roughly by a factor of $\U(1)$:
\[ (\U(1) \times \SU(2) \times \SU(3)) / \Z_6 \iso \S(\U(2) \times \U(3)) \]
It was this factor of $\U(1)$ that made $\SU(5)$ theory so fruitful.

So, if we choose a splitting $\C^4 \iso \C^3 \oplus \C,$ 
we should again look at the subgroup that preserves this splitting. 
Namely:
\[ \S(\U(3) \times \U(1)) \subseteq \SU(4) .\]
Just as in the $\SU(5)$ case, this group is bigger than $\SU(3) \times
\SU(1)$, roughly by a factor of $\U(1)$.  And again, this factor of $\U(1)$ 
is related to hypercharge!

This works very much as it did for $\SU(5)$.  We want a map
\[ \U(1) \times \SU(3) \to \SU(4) \]
and we already have one that works for the $\SU(3)$ part:
\[
\begin{array}{ccc}
 \SU(3) &\to & \SU(4) \\
 h &\mapsto & 
\left(
\begin{array}{cc}
	h & 0 \\
	0 & 1
\end{array}
\right)
\end{array}
\]
So, we just need to include a factor of $\U(1)$ that commutes
with everything in the $\SU(3)$ subgroup. 
Elements of $\SU(4)$ that do this are of the form
\[
\left( 
\begin{array}{c c}
\alpha & 0 \\
0 & \beta
\end{array}
\right)
\]
where $\alpha$ stands for
the $3 \times 3$ identity matrix times
the complex number $\alpha \in \U(1)$, and similarly for $\beta$ in the
$1 \times 1$ block.  For the above matrix to lie in $\SU(4)$, it
must have determinant 1, so $\alpha^3 \beta = 1$.  Thus we
must include $\U(1)$ using matrices of this form:
\[
\left( 
\begin{array}{c c}
\alpha & 0 \\
0 & \alpha^{-3}
\end{array}
\right).
\]
This gives our map:
\[
\begin{array}{ccc}
 \U(1) \times \SU(3) &\to& \SU(4)  \\
( \alpha, h) &\mapsto &
\left( 
\begin{array}{c c}
\alpha h & 0 \\
0 & \alpha^{-3}
\end{array}
\right) .
\end{array}
\]

If we let $\U(1) \times \SU(3)$
act on $\C^4 \iso \C^3 \oplus \C$ via this map, the `quark part' 
$\C^3$ transforms as though it has
hypercharge $\third$: that is, it gets multiplied by a factor of $\alpha$.
Meanwhile, the `lepton part' $\C$ transforms as though it has 
hypercharge $-1$, getting multiplied by a factor of $\alpha^{-3}$.  
So, as a representation of $\U(1) \times \SU(3)$, we have
\[ \C^4 \quad \iso \quad
\C_{\third} \otimes \C^3 \quad \oplus \quad \C_{-1} \otimes \C .\]
A peek at Table \ref{tab:smrep} reveals something nice.
This is exactly how the left-handed quarks and leptons 
in the Standard Model transform under $\U(1) \times \SU(3)$!  

The right-handed leptons do not work this way.  That is a problem
we need to address.  But this brings us to our second question, 
which was about the strange difference between left- and right-handed 
particles. 

Remember that in the Standard Model, the left-handed particles live 
in the fundamental rep of $\SU(2)$ on $\C^2$, while the right-handed 
ones live in the trivial rep on $\C \oplus \C$. Physicists write this 
by grouping left-handed particles into `doublets', while leaving the
right-handed particles as `singlets':
\[ \lep \quad \quad 
\begin{array}{c}
	\nu_R \\
	e^-_R 
\end{array}.
\]
But there is a suspicious similarity between $\C^2$ and $\C \oplus \C$. Could
there be another copy of $\SU(2)$ that acts on the right-handed particles?
Physically speaking, this would mean that the left- and right-handed particles 
both form doublets:
\[ \lep \quad \quad 
\left( \!
\begin{array}{c}
	\nu_R \\
	e^-_R
\end{array}
\! \right)
\]
but under the actions of different $\SU(2)$'s.  Mathematically, this 
would amount to extending the representations of the `left-handed' $\SU(2)$:
\[ \C^2 \quad \quad \C \oplus \C \]
to representations of $\SU(2) \times \SU(2)$:
\[ \C^2 \otimes \C \quad \quad \C \otimes \C^2 \]
where the first copy of $\SU(2)$ acts on
the first factor in these tensor products, while the second copy acts on
the second factor.  The first copy of $\SU(2)$ is the `left-handed'
one familiar from the Standard Model.  The second copy is a new
`right-handed' one.

If we restrict these representations to the `left-handed' $\SU(2)$
subgroup, we obtain:
\begin{eqnarray*} 
	\C^2 \otimes \C & \iso & \C^2 \\ 
	\C \otimes \C^2 & \iso & \C \oplus \C .
\end{eqnarray*}
These are exactly the representations of $\SU(2)$ that appear in 
the Standard Model. It looks like we are on the right track!

Now let us try to combine these ideas into a theory with symmetry group 
$\SU(2) \times \SU(2) \times \SU(4)$.   We have seen that letting 
$\SU(4)$ act on $\C^4$ is a good way to unify our treatment
of color for all the left-handed fermions.  Similarly, the dual representation 
on $\C^{4*}$ is good for their antiparticles.   So, we will tackle color
by letting $\SU(4)$ act on the direct sum $\C^4 \oplus \C^{4*}$.   This 
space is 8-dimensional.  We have also seen that letting $\SU(2) \times \SU(2)$ 
act on $\C^2 \otimes \C \oplus \C \otimes \C^2$ is a good way to unify our 
treatment of isospin for left- and right-handed fermions.  This space is 
4-dimensional. 

Since $8 \times 4 = 32$, and the Standard Model representation is 
32-dimensional, let us take the tensor product
\[     V =   \big((\C^2 \otimes \C) \; \oplus \; (\C \otimes \C^2)\big)
\; \otimes \; \big(\C^4 \; \oplus \; \C^{4*}\big). \]
This becomes a representation of $\SU(2) \times \SU(2) \times \SU(4)$,
which we call the {\bf Pati--Salam representation}.
To obtain a theory that extends the Standard Model, 
we also need a way to map $\GSM$ to $\SU(2) \times \SU(2) \times 
\SU(4)$, such that pulling back $V$ to a representation of $\GSM$ 
gives the Standard model representation.

How can we map $\GSM$ to $\SU(2) \times \SU(2) \times \SU(4)$?  There
are several possibilities.  Our work so far suggests this option:
\[ \begin{array}{ccl}
\U(1) \times \SU(2) \times \SU(3) &\to& \SU(2) \times \SU(2) \times \SU(4) \\
  (\alpha, g, h)  &\mapsto&  
\left( g, \; 1, \;
\left(
\begin{array}{c c}
	\alpha h & 0 \\
	0 & \alpha^{-3}
\end{array}
\right) 
\right)
\end{array}
\]

Let us see what this gives.  The Pati--Salam representation
of $\SU(2) \times \SU(2) \times \SU(4)$ is a direct sum
of four irreducibles:
\[   V \quad \iso \quad 
\C^2 \otimes \C \otimes \C^4 \quad \oplus \quad
\C \otimes \C^2 \otimes \C^4 \quad \oplus \quad
\C^2 \otimes \C \otimes \C^{4*} \quad \oplus \quad
\C \otimes \C^2 \otimes \C^{4*}.
\]
We hope the first two will describe left- and right-handed
fermions, so let us give them names that suggest this:
\[         F_L = \C^2 \otimes \C \otimes \C^4,  \]
\[         F_R = \C \otimes \C^2 \otimes \C^4 . \]
The other two are the duals of the first two, since the 2-dimensional
irrep of $\SU(2)$ is its own dual:
\[         F_L^* = \C^2 \otimes \C \otimes \C^{4*},   \]
\[         F_R^* = \C \otimes \C^2 \otimes \C^{4*} . \]

Given our chosen map from $\GSM$ to $\SU(2) \times \SU(2) \times \SU(4)$,
we can work out which representations of the $\GSM$ these four spaces give.
For example, consider $F_L$.   We have already seen that under our
chosen map, 
\[ \C^4 \quad \iso \quad
\C_{\third} \otimes \C^3 \quad \oplus \quad \C_{-1} \otimes \C \]
as representations of $\U(1) \times \SU(3)$, while
\[  \C^2 \otimes \C \iso \C^2 \] 
as representations of the left-handed $\SU(2)$. 
So, as representations of $\GSM$ we have
\[         F_L \quad \iso 
\quad \C_{\third} \otimes \C^2 \otimes \C^3  \quad \oplus 
\quad \C_{-1} \otimes \C^2 \otimes \C .\]
Table \ref{tab:smrep} shows that these indeed match the left-handed
fermions.

If we go ahead and do the other four cases, we see that everything
works \emph{except for the hypercharges of the right-handed particles}---and 
their antiparticles.  Here we just show results for the particles:

\begin{table}[H]
	\renewcommand{\arraystretch}{0.8}
\begin{center}
	\begin{tabular}{ccc}
         \hline
	 \multicolumn{3}{|c|} {\bf{The Pati--Salam Model --- First Try}} \\
         \hline
 Particle   & Hypercharge: predicted  & Hypercharge: actual \\  
\hline                              
\\
 $\lep$     & $-1$ & $-1$
\\    \\                                                               
 $\quark$   & $\third$ & $\third$ 
\\     \\                                                               
 $\nu_R$   &  $-1$ & $0$
\\     \\                                                               
 $e^-_R$   & $-1$ & $-2$
\\     \\                                                               
 $u_R$     & $\third$ & $\fourthirds$
\\     \\                                                               
 $d_R$     & $\third$ &  $-\twothirds$          
\\    \\
         \hline                              
	\end{tabular}
	\vspace{-10pt}
\end{center}
	\renewcommand{\arraystretch}{1}
\end{table}

The problem is that the right-handed particles are getting the
same hypercharges as their left-handed brethren.  
To fix this problem, we need a more clever map from 
$\GSM$ to $\SU(2) \times \SU(2) \times \SU(4)$.  This map must behave
differently on the $\U(1)$ factor of $\GSM$, so the hypercharges
come out differently.  And it must take advantage of the 
right-handed copy of $\SU(2)$, which acts nontrivially only on the
right-handed particles.  For example, we can try this map:
\[ \begin{array}{ccc}
\U(1) \times \SU(2) \times \SU(3) &\to& \SU(2) \times \SU(2) \times \SU(4) \\
  (\alpha, g, h)  &\mapsto&  
\left( g, \;
\left(
\begin{array}{c c}
	\alpha^k & 0 \\
	0 & \alpha^{-k}
\end{array}
\right), \;
\left(
\begin{array}{c c}
	\alpha h & 0 \\
	0 & \alpha^{-3}
\end{array}
\right) 
\right)
\end{array}
\]
for any integer $k$.  This will not affect the above table
except for the hypercharges of right-handed particles.  It will
add $k/3$ to the hypercharges of the `up' particles in right-handed
doublets ($\nu_R$ and $u_R$), and subtract $k/3$ from the `down'
ones ($e^-_R$ and $d_R$).  So, we obtain these results:

\begin{table}[H]
	\renewcommand{\arraystretch}{0.8}
\begin{center}
	\begin{tabular}{ccc}
         \hline
	 \multicolumn{3}{|c|}{\bf{The Pati--Salam Model --- Second Try}} \\
         \hline
 Particle   & Hypercharge: predicted  & Hypercharge: actual \\  
\hline                              
\\
 $\lep$     & $-1$ & $-1$
\\    \\                                                               
 $\quark$   & $\third$ & $\third$  
\\     \\                                                               
 $\nu_R$    & $-1 + \frac{k}{3}$ & $0$
\\     \\                                                               
 $e^-_R$    & $-1 - \frac{k}{3}$  & $-2$    
\\     \\                                                               
 $u_R$      & $\third + \frac{k}{3}$ & $\fourthirds$
\\     \\                                                               
 $d_R$      & $\third - \frac{k}{3}$ &  $-\twothirds$          
\\    \\
         \hline                              
	\end{tabular}
	\vspace{-10pt}
\end{center}
	\renewcommand{\arraystretch}{1}
\end{table}

\noindent
Miraculously, all the hypercharges match if we choose $k = 3$.
So, let us use this map:
\[ \begin{array}{ccc}
\beta \maps \U(1) \times \SU(2) \times \SU(3) 
&\to& \SU(2) \times \SU(2) \times \SU(4) \\  \\
  (\alpha, g, h)  &\mapsto&  
\left( g, \;
\left(
\begin{array}{c c}
	\alpha^3 & 0 \\
	0 & \alpha^{-3}
\end{array}
\right), \;
\left(
\begin{array}{c c}
	\alpha h & 0 \\
	0 & \alpha^{-3}
\end{array}
\right) 
\right).
\end{array}
\]
When we take the Pati--Salam representation of 
$\SU(2) \times \SU(2) \times \SU(4)$
and pull it back along this map $\beta$, we obtain the Standard Model
representation. As in Section~\ref{sec:su(5)}, we use complete reducibility to
see this, but we can be more concrete. We saw in Table~\ref{tab:su5code} how we
can specify the intertwining map precisely by using a specific basis, which for
$\Lambda \C^5$ results in the binary code.

Similarly, we can create a kind of `Pati--Salam code' to specify an 
isomorphism of Hilbert spaces
\[  \ell \maps F
\oplus F^* \to \big((\C^2 \otimes \C) \; \oplus \; (\C \otimes \C^2)\big) \;
\otimes \; \big(\C^4 \; \oplus \; \C^{4*}\big), \]
and doing this provides a nice
summary of the ideas behind the Pati--Salam model. We take the space
$\C^2 \otimes \C$ to be spanned by $u_L$ and $d_L$, the \define{left-isospin
up} and \define{left-isospin down} states. Similarly, the space $\C \otimes
\C^2$ has basis $u_R$ and $d_R$, called \define{right-isospin up} and
\define{right-isospin down}. Take care not to confuse these with the similarly
named quarks. These have no color, and only correspond to isospin.

The color comes from $\C^4$ of course, which we already decreed to be spanned
by $r$, $g$, $b$ and $w$. For antiparticles, we also require anticolors, which
we take to be the dual basis $\rbar$, $\gbar$, $\bbar$ and $\wbar$, spanning
$\C^{4*}$.

It is now easy, with our knowledge of how the Pati--Salam model is to work, to
construct this code. Naturally, the left-handed quark doublet corresponds to
the left-isospin up and down states, which come in all three colors
$c = r,g,b$:
\[ u^c_L = u_L \otimes c \quad \quad d^c_L = d_L \otimes c. \]
The corresponding doublet of left-handed leptons is just the `white'
version of this:
\[ \nu_L = u_L \otimes w \quad \quad e^-_L = d_L \otimes w .\]
The right-handed fermions are the same, but with $R$'s instead of $L$'s. Thus
we get the Pati--Salam code for the fermions:
\vskip 1em 
\begin{center}
	\begin{tabular}{cc}
		\hline
		$F_L$                   & $F_R$ \\
		\hline
		$\nu_L = u_L \otimes w$ & $\nu_R = u_R \otimes w$ \\
		$e^-_L = d_L \otimes w$ & $e^-_R = d_R \otimes w$ \\
		$u^c_L = u_L \otimes c$ & $u^c_R = u_R \otimes c$ \\
		$d^c_L = d_L \otimes c$ & $d^c_R = d_R \otimes c$ \\
		\hline
	\end{tabular}
\end{center}
\vskip 1em
The result is very similar for the antifermions in $F^*_L$ and $F^*_R$, but
watch out: taking antiparticles \emph{swaps up and down}, and also \emph{swaps
left and right}, so the particles in $F^*_L$ are right-handed, despite
the subscript $L$, while those in $F^*_R$ are left-handed. This is because 
it is the right-handed antiparticles that feel the weak force, 
which in terms of representation theory means they are nontrivial under the 
left $\SU(2)$.  So, the Pati--Salam code for the antifermions is this:
\vskip 1em 
\begin{center}
	\begin{tabular}{cc}
		\hline
		$F^*_L$                             & $F^*_R$ \\
		\hline
		$e^+_R         = u_L \otimes \wbar$ & $e^+_L         = u_R \otimes \wbar$ \\
		$\nubar_R      = d_L \otimes \wbar$ & $\nubar_L      = d_R \otimes \wbar$ \\
		$\dbar^\cbar_R = u_L \otimes \cbar$ & $\dbar^\cbar_L = u_R \otimes \cbar$ \\
		$\ubar^\cbar_R = d_L \otimes \cbar$ & $\ubar^\cbar_L = d_R \otimes \cbar$ \\
		\hline
	\end{tabular}
\end{center}
\vskip 1em
Putting these together we get the full Pati--Salam code:
\vskip 1em
	\begin{table}[H]
\begin{center}
	\begin{tabular}{cccc}
		\hline
		\multicolumn{4}{|c|}{\bf{The Pati--Salam Code}} \\
		\hline
		$F_L$                   & $F_R$                   & $F^*_L$                             & $F^*_R$ \\
		\hline
		$\nu_L = u_L \otimes w$ & $\nu_R = u_R \otimes w$ & $e^+_R         = u_L \otimes \wbar$ & $e^+_L         = u_R \otimes \wbar$ \\
		$e^-_L = d_L \otimes w$ & $e^-_R = d_R \otimes w$ & $\nubar_R      = d_L \otimes \wbar$ & $\nubar_L      = d_R \otimes \wbar$ \\
		$u^c_L = u_L \otimes c$ & $u^c_R = u_R \otimes c$ & $\dbar^\cbar_R = u_L \otimes \cbar$ & $\dbar^\cbar_L = u_R \otimes \cbar$ \\
		$d^c_L = d_L \otimes c$ & $d^c_R = d_R \otimes c$ & $\ubar^\cbar_R = d_L \otimes \cbar$ & $\ubar^\cbar_L = d_R \otimes \cbar$ \\
		\hline
	\end{tabular}
	\caption{Pati--Salam code for first-generation fermions, where $c = r, g, b$ and $\cbar = \rbar, \bbar, \gbar$.} \label{tab:pscode1}
\end{center}
\end{table}
\vskip -1em

This table defines an isomorphism of Hilbert spaces
\[  \ell \maps F
\oplus F^* \to \big((\C^2 \otimes \C) \; \oplus \; (\C \otimes \C^2)\big) \;
\otimes \; \big(\C^4 \; \oplus \; \C^{4*}\big) \]
so where it says, for example, $\nu_L = u_L \otimes w$, that is just
short for $\ell(\nu_L) = u_L \otimes w$.   This map $\ell$
is also an isomorphism between representations of $\GSM$. It tells us 
how these representations are the `same', just as the map $h$ did for 
$F \oplus F^*$ and $\Ex \C^5$ at the end of
Section~\ref{sec:su(5)}.

As with $\SU(5)$ and $\Spin(10)$, we can summarize all the results
of this section in a commutative square:

\begin{thm}\et
\label{thm:Pati--Salam}
The following square commutes:
\[
\xymatrix{
\GSM \ar[r]^-{\beta} \ar[d]  & \SU(2) \times \SU(2) \times \SU(4) \ar[d] \\
\U( F \oplus F^* ) \ar[r]^-{U(\ell)} & 
\U\left(\big((\C^2 \otimes \C) \oplus (\C \otimes \C^2)\big)
\otimes \big(\C^4 \oplus \C^{4*}\big)\right) \\
}
\]
where the left vertical arrow is the Standard Model
representation and the right one is the Pati--Salam representation.
\end{thm}

The Pati--Salam representation and especially the homomorphism
$\beta$ look less natural than the representation of $\SU(5)$ on 
$\Lambda \C^5$ and the homomorphism $\phi \maps \GSM \to \SU(5)$.  
But appearances can be deceiving: in the next section we shall see 
a more elegant way to describe them.

\subsection{The Route to Spin(10) via Pati--Salam} \label{sec:route}

In the last section, we showed how the Pati--Salam model answers two questions
about the Standard Model:

\begin{quote}
	Why are quarks and leptons so similar?  Why are left and right so
	different?
\end{quote}

We were able to describe leptons as a fourth color of quark, `white', and
treat right-handed and left-handed particles on a more equal footing.
Neither of these ideas worked on its own, but together, they made
a full-fledged extension of the Standard Model, much like $\SU(5)$ and
$\Spin(10)$, but based on seemingly different principles.

Yet thinking of leptons as `white' should be strangely familiar, not
just from the Pati--Salam perspective, but
from the binary code that underlies both the $\SU(5)$ and the
$\Spin(10)$ theories. There, leptons were indeed white: they all have
color $r \wedge g \wedge b \in \Ex \C^5$.

Alas, while $\SU(5)$ hints that leptons might be a fourth color, it does not
deliver on this. The quark colors 
\[ r, g, b \in \Ex^1 \C^5 \]
lie in a different irrep of $\SU(5)$ than does $r \wedge g \wedge b \in \Ex^3
\C^5$.  So, leptons in the $\SU(5)$ theory are white, but unlike
the Pati--Salam model, this theory
does not unify leptons with quarks.

Yet $\SU(5)$ theory is not the only game in town when it comes to the binary
code. We also have $\Spin(10)$, which acts on the same vector space as
$\SU(5)$. As a representation of $\Spin(10)$, $\Ex \C^5$ breaks up into just
two irreps: the even grades, $\Exev \C^5$, which contain the left-handed
particles and antiparticles:
\[ \Exev \C^5 \iso \langle \nubar_L \rangle \oplus \langle e^+_L \rangle \oplus \angquark \oplus \langle \ubar_L \rangle \oplus \anglep \oplus \langle \dbar_L \rangle \]
and the odd grades $\Exodd \C^5$, which contain the right-handed particles and
antiparticles:
\[ \Exodd \C^5 \iso \langle \nu_R \rangle \oplus \langle e^-_R \rangle \oplus \angantiquark \oplus \langle u_R \rangle \oplus \angantilep \oplus \langle d_R \rangle.
\]
Unlike $\SU(5)$, the $\Spin(10)$ GUT really does unify 
$r \wedge g \wedge b$ with the colors $r$, $g$ and $b$, because they 
both live in the irrep $\Exodd \C^5$. 

In short, it seems that the $\Spin(10)$ GUT, which we built as an 
extension of the $\SU(5)$ GUT, somehow managed to pick up
this feature of the Pati--Salam model. 
How does $\Spin(10)$ relate to Pati--Salam's gauge group $\SU(2) \times \SU(2) \times \SU(4)$, 
exactly? In general, we only know there is a map $\SU(n) \to \Spin(2n)$, 
but in low dimensions, there is much more, because some groups coincide:
\begin{eqnarray*}
	\Spin(3) & \iso & \SU(2) \\
	\Spin(4) & \iso & \SU(2) \times \SU(2) \\
	\Spin(5) & \iso & \rm{Sp}(2) \\
	\Spin(6) & \iso & \SU(4) 
\end{eqnarray*}
What really stands out is this: 
\[ \SU(2) \times \SU(2) \times \SU(4) \iso \Spin(4) \times \Spin(6) .\]

This brings out an obvious relationship between the Pati--Salam model and the
$\Spin(10)$ theory, because the inclusion $\SO(4) \times \SO(6) \inclusion
\SO(10)$ lifts to the universal covers, so we get a homomorphism 
\[ \eta \maps \Spin(4) \times \Spin(6) \to \Spin(10). \]
A word of caution is needed here. While $\eta$ is the lift of an inclusion, it
is not an inclusion itself: it is two-to-one.  This is because the universal 
cover $\Spin(4) \times \Spin(6)$ of $\SO(4) \times \SO(6)$ is a four-fold 
cover, being a double cover on each factor. 

So we can try to extend the symmetries $\SU(2) \times \SU(2) \times \SU(4)$ to
$\Spin(10)$, though this can only work if the kernel of $\eta$ acts
trivially on the Pati--Salam representation.
What is this representation like?  There is an obvious
representation of $\Spin(4) \times \Spin(6)$ that extends to a $\Spin(10)$ rep.
Both $\Spin(4)$ and $\Spin(6)$ have Dirac spinor representations, so their
product $\Spin(4) \times \Spin(6)$ has a representation on 
$\Ex \C^2 \otimes \Ex \C^3$.  And in fact, the obvious map 
\[ g \maps \Ex \C^2 \otimes \Ex \C^3 \to \Ex \C^5 \]
given by 
\[ v \otimes w \mapsto v \wedge w \]
is an isomorphism compatible with the actions of $\Spin(4) \times \Spin(6)$
on these two spaces.  More concisely, this square:
\[
\xymatrix{
\Spin(4) \times \Spin(6) \ar[r]^-\eta \ar[d] & \Spin(10) \ar[d] \\
\U(\Ex \C^2 \otimes \Ex \C^3) \ar[r]^-{\U(g)} & \U(\Ex \C^5) \\
}
\]
commutes.

We will prove this in a moment. First though, we must check that
this representation of $\Spin(4) \times \Spin(6)$ is secretly
just another name for the Pati--Salam representation of 
$\SU(2) \times \SU(2) \times \SU(4)$ on the space we discussed in
Section~\ref{sec:g(2,2,4)}:
\[           \big((\C^2 \otimes \C) \; \oplus \; (\C \otimes \C^2)\big)
\; \otimes \; \big(\C^4 \; \oplus \; \C^{4*}\big). \]
Checking this involves choosing an isomorphism between 
$\SU(2) \times \SU(2) \times \SU(4)$ and $\Spin(4) \times \Spin(6)$.
Luckily, we can choose one that works: 

\begin{thm}\et
\label{thm:Pati--Salam2}
There exists an isomorphism of Lie groups 
\[ \alpha \maps \SU(2) \times \SU(2) \times \SU(4) \to 
\Spin(4) \times \Spin(6) \]
and a unitary operator
\[ 
k \maps \big((\C^2 \otimes \C) \oplus (\C \otimes \C^2)\big) 
\otimes  \big(\C^4 \oplus \C^{4*}\big)
\;\;  \to \;\; \Ex \C^2 \otimes \Ex \C^3 \]
that make this square commute:
\[
\xymatrix{
\SU(2) \times \SU(2) \times \SU(4) \ar[r]^-{\alpha} \ar[d] 
& \Spin(4) \times \Spin(6) \ar[d] \\
\U\left(\left((\C^2 \otimes \C) \oplus (\C \otimes \C^2)\right) 
\otimes \left(\C^4  \oplus  \C^{4*}\right)\right) 
\ar[r]^-{\U(k)} & \U(\Ex \C^2 \otimes \Ex \C^3) \\
}
\]
where the left vertical arrow is the Pati--Salam representation
and the right one is a tensor product of Dirac spinor representations.
\end{thm}

\emph{Proof.} We can prove this in pieces, by separately finding 
a unitary operator
\[  (\C^2 \otimes \C) \oplus (\C \otimes \C^2) \iso \Ex \C^2 \]
that makes this square commute:
\[
\xymatrix{
\SU(2) \times \SU(2) \ar[r]^-\sim \ar[d] & \Spin(4) \ar[d] \\
\U(\C^2 \otimes \C \oplus \C \otimes \C^2) \ar[r]^-\sim & \U(\Ex \C^2) \\
}
\]
and a unitary operator
\[          \C^4 \oplus \C^{4*} \iso \Ex \C^3 \]
that make this square:
\[
\xymatrix{
\SU(4) \ar[r]^\sim \ar[d] & \Spin(6) \ar[d] \\
\U(\C^4 \oplus \C^{4*}) \ar[r]^-\sim & \U(\Ex \C^3) \\
}
\]
commute.

First, the $\Spin(6)$ piece.  It suffices to show that the Dirac spinor rep 
of $\Spin(6) \iso \SU(4)$ on $\Ex \C^3$ 
is isomorphic to $\C^4 \oplus \C^{4*}$ as a rep of
$\SU(4)$. We start with the action of $\Spin(6)$ on $\Ex \C^3$. This breaks up
into irreps:
\[	\Ex \C^3 \iso \Exev \C^3 \oplus \Exodd \C^3	\]
called the {\bf left-handed}
and {\bf right-handed Weyl spinors}, and these are dual to each
other because 6 = 2 mod 4, by a theorem that can be found
in Adams' lectures~\cite{adams:lelg}. Call these representations
\[ \rev \maps \Spin(6) \to \U( \Exev \C^3 ), \qquad 
   \rodd \maps \Spin(6) \to \U( \Exodd \C^3 ). \]
Since these reps are dual, it suffices just to consider one of them, say 
$\rodd$. 

Passing to Lie algebras, we have a homomorphism
\[ d\rodd \maps \so(6) \to \u(\Exodd \C^3) \iso \u(4)
\iso \u(1) \oplus \su(4). \]
Homomorphic images of semisimple Lie algebras are semisimple, so the image of
$\so(6)$ must lie entirely in $\su(4)$.  In fact $\so(6)$ is simple, so this
nontrivial map must be an injection
\[ d\rodd \maps \so(6) \to \su(4), \]
and because the dimension is 15 on both sides, this map is also
onto. Thus $d\rodd$ is an isomorphism of Lie algebras, so
$\rodd$ is an isomorphism of the simply connected Lie 
groups $\Spin(6)$ and $\SU(4)$:
\[ \rodd \maps \Spin(6) \to \SU(\Exodd \C^3) \iso \SU(4). \]
Furthermore, under this isomorphism
\[ \Exodd \C^3 \iso \C^4 \]
as a representation of $\SU(\Exodd \C^3) \iso \SU(4)$. 
Taking duals, we obtain an isomorphism
\[ \Exev \C^3 \iso \C^{4*} \]
Putting these together, we get an isomorphism $\C^{4} \oplus \C^{4*} \iso
\Ex \C^3$ that makes this square commute:
\[
\xymatrix{
\Spin(6) \ar[d] \ar[r]^\rodd & \SU(4) \ar[d] \\
\U(\Ex \C^3) \ar[r] & \U(\C^4 \oplus \C^{4*})
}
\] 
which completes the proof for $\Spin(6)$.

Next, the $\Spin(4)$ piece.  It suffices 
to show that the spinor rep $\Ex \C^2$ of $\Spin(4) \iso \SU(2) \times \SU(2)$
is isomorphic to $\C^2 \otimes \C \; \oplus \; \C \otimes \C^2$ as a rep of $\SU(2)
\times \SU(2)$. We start with the action of $\Spin(4)$ on $\Ex \C^2$. This
again breaks up into irreps:
\[	\Ex \C^2 \iso \Exodd \C^2 \oplus \Exev \C^2. \]
Again we call these representations 
\[ \rev \maps \Spin(4) \to \U( \Exev \C^2 ) , \qquad
 \rodd \maps \Spin(4) \to \U( \Exodd \C^2 ) .\]

First consider $\rev$.  Passing to Lie algebras, this gives
a homomorphism
\[ d\rev \maps \so(4) \to \u(\Exev \C^2) \iso \u(2) \iso \u(1) \oplus \su(2) \]
Homomorphic images of semisimple Lie algebras are semisimple, so the image of
$\so(4)$ must lie entirely in $\su(2)$.  Similarly, $d\rodd$ also takes 
$\so(4)$ to $\su(2)$:
\[ d\rodd \maps \so(4) \to \su(2) \]
and we can combine these maps to get
\[ d\rodd \oplus d\rev \maps \so(4) \to \su(2) \oplus \su(2), \]
which is just the derivative of $\Spin(4)$'s representation on $\Ex \C^2$.
Since this representation is faithful, the map $d\rodd \oplus d\rev$ of Lie
algebras is injective. But the dimensions of $\so(4)$ and $\su(2) \oplus
\su(2)$ agree, so $d\rodd \oplus d\rev$ is also onto. Thus it is an isomorphism
of Lie algebras.  This implies that $\rodd \oplus \rev$ is an isomorphism of
the simply connected Lie groups $\Spin(4)$ and $\SU(2) \times \SU(2)$
\[ \rodd \oplus \rev \maps \Spin(4) \to \SU(\Exodd \C^2) \times \SU(\Exev \C^2) \iso \SU(2) \times \SU(2) \]
under which $\SU(2) \times \SU(2)$ acts on $\Exodd \C^2 \oplus \Exev \C^2$. The
left factor of $\SU(2)$ acts irreducibly on $\Exodd \C^2$, which the second
factor is trivial on.  Thus $\Exodd \C^2 \iso
\C^2 \otimes \C$ as a rep of $\SU(2) \times \SU(2)$. Similarly, $\Exev \C^2
\iso \C \otimes \C^2$ as a rep of this group.  Putting these together,
we get an isomorphism $\C^2 \otimes \C \oplus \C \otimes \C^2 \iso
\Ex \C^2$ that makes this square commute:
\[
\xymatrix{
\Spin(4) \ar[d] \ar[r]^-{\rev \oplus \rodd} & \SU(2) \times \SU(2) \ar[d] \\
\U(\Ex \C^2) \ar[r] & \U(\C^2 \otimes \C \oplus \C \otimes \C^2)
}
\] 
which completes the proof for $\Spin(4)$.  \qed

In the proof of the preceding theorem, we merely showed that there 
\emph{exists} an isomorphism 
\[   k \maps \big((\C^2 \otimes \C) \oplus
(\C \otimes \C^2)\big) \otimes \big(\C^4 \oplus \C^{4*}\big) \to \Ex \C^2
\otimes \Ex \C^3 \]
making the square commute.  We did not say exactly what $k$ was.  
In the proof, we built it after quietly choosing
three unitary operators, giving these isomorphisms:
\[ \C^2 \otimes \C \iso \Exodd \C^2, \quad \C \otimes \C^2 \iso \Exev \C^2, \quad \C^4 \iso \Exodd \C^3. \]
Since the remaining map $\Exev \C^3 \iso \C^{4*}$ is determined by 
duality, these three operators determine $k$, and they also 
determine the Lie group isomorphism $\alpha$ 
via the construction in our proof.

There is, however, a specific choice for these unitary operators that we 
prefer, because this choice makes the particles in the Pati--Salam 
representation $\Ex \C^2 \otimes \Ex \C^3$ look almost exactly like 
those in the $\SU(5)$ representation $\Ex \C^5$.

First, since $\C^2 = \Ex^1 \C^2 = \Exodd \C^2$ is spanned by $u$ and
$d$, the (left-handed) isospin states of the Standard Model, we really ought
to identify the left-isospin states $u_L$ and $d_L$ of the Pati--Salam model
with these.  So, we should use this unitary operator:
\[          
\begin{array}{ccl}
 \C^2 \otimes \C & \stackrel{\sim}{\longrightarrow} &  \Exodd \C^2  \\
             u_L & \mapsto & u   \\
             d_L & \mapsto & d .
\end{array}
\]

Next, we should use this unitary operator for right-isospin states:
\[          
\begin{array}{ccl}
 \C \otimes \C^2 & \stackrel{\sim}{\longrightarrow} &  \Exev \C^2  \\
             u_R & \mapsto & u \wedge d   \\
             d_R & \mapsto & 1 .
\end{array}
\]
Why?  Because the right-isospin up particle is the right-handed neutrino 
$\nu_R$, which corresponds to $u \wedge d \wedge r \wedge g \wedge b$ 
in the $\SU(5)$ theory, but $u_R \otimes w$ 
in the Pati--Salam model.  This suggests that $u \wedge d$ and $u_R$
should be identified.

Finally, because $\C^3$ is spanned by the colors $r$, $g$ and $b$, while
$\C^4$ is spanned by the colors $r$, $g$, $b$ and $w$, we really ought
to use this unitary operator:
\[          
\begin{array}{ccl}
 \C^4 & \stackrel{\sim}{\longrightarrow} &  \Exodd \C^3  \\
    r & \mapsto & r \\ 
    g & \mapsto & g \\ 
    b & \mapsto & b \\ 
    w & \mapsto & r \wedge g \wedge b 
\end{array}
\]
Dualizing this, we get the unitary operator
\[          
\begin{array}{ccl}
 \C^{4*} & \stackrel{\sim}{\longrightarrow} &  \Exev \C^3  \\
    \overline{r} & \mapsto & g \wedge b \\ 
    \overline{g} & \mapsto & b \wedge r \\ 
    \overline{b} & \mapsto & r \wedge g \\ 
    \overline{w} & \mapsto & 1
\end{array}
\]

These choices determine the unitary operator
\[         k \maps \big((\C^2 \otimes \C)
\oplus (\C \otimes \C^2)\big) \otimes \big(\C^4 \oplus \C^{4*}\big)
\to \Ex \C^2 \otimes \Ex \C^3  .\]
With this specific choice of $k$, we can combine the commutative squares
built in Theorems \ref{thm:Pati--Salam} and
\ref{thm:Pati--Salam2}: 
\[
\xymatrix{
\GSM \ar[r]^-{\beta} \ar[d] 
& \SU(2) \times \SU(2) \times \SU(4) \ar[r]^-{\alpha} \ar[d] 
& \Spin(4) \times \Spin(6) \ar[d]
\\
\U( F \oplus F^* ) \ar[r]^-{U(\ell)} 
& \U\left(\big((\C^2 \otimes \C) \oplus (\C \otimes \C^2)\big)
\otimes \big(\C^4 \oplus \C^{4*}\big)\right) \ar[r]^-{\U(k)} 
& \U(\Lambda \C^5) 
}
\]
to obtain the following result:

\begin{thm}\et
\label{thm:Pati--Salam3}
Taking $\theta = \alpha \beta$ and $h = k \ell$, the
following square commutes:
	\[ 
		\xymatrix{
		\GSM \ar[r]^-\theta \ar[d]       & \Spin(4) \times \Spin(6) \ar[d] \\
		\U(F \oplus F^*) \ar[r]^-{\U(h)} & \U(\Ex \C^2 \otimes \Ex \C^3) \\
		}
	\]
where the left vertical arrow is the Standard Model representation and
the right one is a tensor product of Dirac spinor representations.
\end{thm}

The map $h$, which tells us how to identify $F \oplus F^*$ and $\Ex \C^2
\otimes \Ex \C^3$, is given by applying $k$ to the `Pati--Salam code' in
Table~\ref{tab:pscode1}. This gives a binary code for the 
Pati--Salam model:
\vskip 1em
	\begin{table}[H]
\begin{center}
	\begin{tabular}{cccc}
		\hline
		\multicolumn{4}{|c|}{\bf{The Binary Code for Pati--Salam}} \\
		\hline
		$\Exodd \C^2 \otimes \Exodd \C^3$ & $\Exev \C^2 \otimes \Exodd \C^3$ & $\Exodd \C^2 \otimes \Exev \C^3$  & $\Exev \C^2 \otimes \Exev \C^3$ \\
		\hline
		$\nu_L = u \otimes rgb$           & $\nu_R = ud \otimes rgb$         & $e^+_R         = u \otimes 1$     & $e^+_L         = ud \otimes 1$ \\
		$e^-_L = d \otimes rgb$           & $e^-_R = 1 \otimes rgb$          & $\nubar_R      = d \otimes 1$     & $\nubar_L      = 1 \otimes 1$ \\
		$u^c_L = u \otimes c$             & $u^c_R = ud \otimes c$           & $\dbar^\cbar_R = u \otimes \cbar$ & $\dbar^\cbar_L = ud \otimes \cbar$ \\
		$d^c_L = d \otimes c$             & $d^c_R = 1 \otimes c$            & $\ubar^\cbar_R = d \otimes \cbar$ & $\ubar^\cbar_L = 1 \otimes \cbar$ \\
		\hline
	\end{tabular}
	\caption{Pati--Salam binary code for first-generation fermions, where $c = r, g, b$ and $\cbar = gb, br, rg$.} \label{tab:pscode2}
\end{center}
\end{table}
\vskip -1em
\noindent
We have omitted wedge product symbols to save space. Note that if we apply the
obvious isomorphism
\[ g \maps \Ex \C^2 \otimes \Ex \C^3 \to \Ex \C^5 \]
given by 
\[ v \otimes w \mapsto v \wedge w \]
then the above table does more than 
merely resemble Table~\ref{tab:su5code}, which gives the
binary code for the $\SU(5)$ theory.  The two tables become 
\emph{identical!}

This fact is quite intriguing.   We will explore its meaning
in the next section.  But first, let us start by relating the 
Pati--Salam model to the $\Spin(10)$ theory:

\begin{thm}\et
\label{thm:Spin(10)}
The following square commutes:
\[
\xymatrix{
\Spin(4) \times \Spin(6) \ar[r]^-\eta \ar[d] & \Spin(10) \ar[d] \\
\U(\Ex \C^2 \otimes \Ex \C^3) \ar[r]^-{\U(g)} & \U(\Ex \C^5)
}
\]
where the right vertical arrow is the Dirac spinor representation,
the left one is the tensor product of Dirac spinor representations, 
and 
\[  \eta \maps \Spin(4) \times \Spin(6) \to \Spin(10)  \] 
is the homomorphism lifting the inclusion of $\SO(4) \times \SO(6)$ in
$\SO(10)$.
\end{thm}

\emph{Proof}. At the Lie algebra level, we have the inclusion
\[ \so(4) \oplus \so(6) \inclusion \so(10) \]
by block diagonals, which is also just the differential of the inclusion 
$\SO(4) \times \SO(6) \inclusion \SO(10)$ at the Lie group level. Given how the
spinor reps are defined in terms of creation and annihilation operators, it is
easy to see that
\[
\xymatrix{
\so(4) \oplus \so(6) \ar@{^{(}->}[r] \ar[d] & \so(10) \ar[d] \\
\gl(\Ex \C^2 \otimes \Ex \C^3) \ar[r]^-{\gl(g)}  & \gl(\Ex \C^5) \\
}
\]
commutes, because $g$ is an
intertwining operator between representations
of $\so(4) \oplus \so(6)$.
That is because the $\so(4)$ part only acts on $\Ex \C^2$, while the $\so(6)$
part only acts on $\Ex \C^3$. 

But these Lie algebras act by skew-adjoint operators, so really
\[
\xymatrix{
\so(4) \oplus \so(6) \ar@{^{(}->}[r] \ar[d] & \so(10) \ar[d] \\
\u(\Ex \C^2 \otimes \Ex \C^3) \ar[r]^-{\u(g)}  & \u(\Ex \C^5) \\
}
\]
commutes. Since the $\so(n)$'s and their direct sums are semisimple, so are
their images. Therefore, their images live in the semisimple part of the
unitary Lie algebras, which is just another way of saying the special unitary
Lie algebras. We get that
\[
\xymatrix{
\so(4) \oplus \so(6) \ar@{^{(}->}[r] \ar[d] & \so(10) \ar[d] \\
\su(\Ex \C^2 \otimes \Ex \C^3) \ar[r]^-{\su(g)}  & \su(\Ex \C^5) \\
}
\]
commutes, and this gives a commutative square in the world of
simply connected Lie groups:
\[
\xymatrix{
\Spin(4) \times \Spin(6) \ar[r]^-\eta \ar[d] & \Spin(10) \ar[d] \\
\SU(\Ex \C^2 \otimes \Ex \C^3) \ar[r]^-{\SU(g)} & \SU(\Ex \C^5)
}
\]
This completes the proof.  \qed

This result shows us how to reach the $\Spin(10)$ theory, not 
through the $\SU(5)$ theory, but through the Pati--Salam model.
For physics texts that treat this issue, see for example 
Zee~\cite{zee:nutshell} and Ross~\cite{ross:gut}.

\subsection{The Question of Compatibility} \label{sec:compatibility}

We now have two routes to the $\Spin(10)$ theory. In Section~\ref{sec:so(10)}
we saw how to reach it via the $\SU(5)$ theory:

\vbox{
\[
\xymatrix{
\GSM \ar[r]^-\phi \ar[d] & \SU(5) \ar[r]^-\psi \ar[d] & \Spin(10) \ar[d] \\  
\U(F \oplus F^*) \ar[r]^-{\U(f)} & \U(\Ex \C^5) \ar[r]^-1 & \U(\Ex \C^5) \\
}
\]
\[ \xymatrix{ & \ar@{~>}[r]_{\txt{More Unification}} & } \]
}

\noindent
Our work in that section and in Section~\ref{sec:su(5)} 
showed that this diagram commutes, which is a way of saying that
the $\Spin(10)$ theory extends the Standard Model.

In Section \ref{sec:route} we saw another route to the
$\Spin(10)$ theory, which goes through $\Spin(4) \times \Spin(6)$:
\[
\xymatrix{
\GSM \ar[r]^-\theta \ar[d]       & \Spin(4) \times \Spin(6) \ar[r]^-\eta \ar[d]  & \Spin(10) \ar[d] \\  
\U(F \oplus F^*) \ar[r]^-{\U(h)} & \U(\Ex \C^2 \otimes \Ex \C^3) \ar[r]^-{\U(g)} & \U(\Ex \C^5) \\
}
\]
\[ \xymatrix{ & \ar@{~>}[r]_{\txt{More Unification}} & } \]
Our work in that section and Section~\ref{sec:g(2,2,4)} 
showed that this diagram commutes as well.  So, we have
\emph{another} way to extend the Standard Model and get 
the $\Spin(10)$ theory.

Drawing these two routes to $\Spin(10)$ together gives us a cube:
\[
\xymatrix{
& \GSM \ar[rr]^\phi \ar[dl]_\theta \ar'[d] [ddd] & & \SU(5) \ar[dl]_\psi \ar[ddd] \\
\Spin(4) \times \Spin(6) \ar[rr]^\eta \ar[ddd] & & \Spin(10) \ar[ddd] & \\
& & & \\
& \U(F \oplus F^*) \ar'[r]^(0.8){\U(f)}[rr] \ar[dl]_-{\U(h)} & & \U(\Ex \C^5) \ar[dl]_-1 \\
\U(\Ex \C^2 \otimes \Ex \C^3) \ar[rr]^-{\U(g)}	& & \U(\Ex \C^5) & \\
}
\] 
Are these two routes to $\Spin(10)$ theory the same? That is, does the
cube commute? 

\begin{thm}\et
\label{thm:cube}
	The cube commutes.
\end{thm}

\emph{Proof}. 
We have already seen in Sections \ref{sec:su(5)}-\ref{sec:route} that the
vertical faces commute.   So, we are left with two questions involving the
horizontal faces.  First: does the top face of the cube
\[
\xymatrix{
& \GSM \ar[rr]^\phi \ar[dl]_\theta  & & \SU(5) \ar[dl]_\psi \\
\Spin(4) \times \Spin(6) \ar[rr]^\eta & & \Spin(10) & \\
}
\]
commute?  In other words: does a symmetry in $\GSM$ go to the same place in
$\Spin(10)$ no matter how we take it there? 
And second: does the bottom face of the cube commute?  In other words:
does this triangle:
\[
\xymatrix{
F \oplus F^* \ar[r]^f \ar[d]_h & \Ex \C^5 \\
\Ex \C^2 \otimes \Ex \C^3 \ar[ru]_g \\
}
\]
commute?

In fact they both do, and we can use our affirmative answer to the 
second question to settle the first.
As we remarked in Section~\ref{sec:route}, applying the map $g$
to the Pati--Salam binary code given in Table~\ref{tab:pscode2}, we get the
$\SU(5)$ binary code given in Table~\ref{tab:su5code}. Thus, the linear maps
$f$ and $gh$ agree on a basis, so this triangle commutes:
\[
\xymatrix{
F \oplus F^* \ar[r]^f \ar[d]_h & \Ex \C^5 \\
\Ex \C^2 \otimes \Ex \C^3 \ar[ru]_g \\
}
\]
This in turn implies that the bottom face of the cube commutes, from which 
we see that the two maps from $\GSM$ to $\U(\Ex \C^5)$ going around the 
bottom face are equal:
\[
\xymatrix{
\GSM \ar[dr] &                                                    & \\
             & \U(F \oplus F^*) \ar[r]^-{\U(f)} \ar[d]_{\U(h)} & \U(\Ex \C^5) \ar[d]^1 \\
             & \U(\Ex \C^2 \otimes \Ex \C^3) \ar[r]^-{\U(g)}    & \U(\Ex \C^5) \\
}
\]

The work of Section~\ref{sec:su(5)} through Section~\ref{sec:route} 
showed that the vertical faces of the cube commute.
We can thus conclude from diagrammatic reasoning that the two maps from
$\GSM$ to $\U(\Ex \C^5)$ going around the top face are equal:
\[
\xymatrix{
\GSM \ar[r]^\phi \ar[d]_\theta         & \SU(5) \ar[d]^\psi & \\
\Spin(4) \times \Spin(6) \ar[r]^-\eta  & \Spin(10) \ar[dr]  & \\
                                       &                    & \U(\Ex \C^5) \\
}
\]
Since the Dirac spinor representation is faithful, the map $\Spin(10) \to
\U(\Ex \C^5)$ is injective. This means we can drop it from the above diagram,
and the remaining square commutes. But this is exactly the top face of the
cube.  So, the proof is done. \qed

\section{Conclusion} \label{sec:conclusion}

We have studied three different grand unified theories:
the $\SU(5)$, $\Spin(4) \times \Spin(6)$ and $\Spin(10)$
theories. The $\SU(5)$ and $\Spin(4) \times \Spin(6)$ theories were
based on different visions about how to extend the Standard
Model. However, we saw that both of these theories can be extended
to the $\Spin(10)$ theory, which therefore unites these visions.

The $\SU(5)$ theory is all about treating isospin and color on an
equal footing: it combines the two isospins of $\C^2$ with
the three colors of $\C^3$, and posits an $\SU(5)$ symmetry
acting on the resulting $\C^5$.  The particles and antiparticles in a single
generation of fermions are described by vectors in $\Ex \C^5$.  So, we
can describe each of these particles and antiparticles by a binary
code indicating the presence or absence of \textsl{up, down, red,
green} and \textsl{blue}.

In doing so, the $\SU(5)$ theory introduces unexpected relationships 
between matter and antimatter. The irreducible representations of $\SU(5)$
\[ \Ex^0 \C^5 \oplus \Ex^1 \C^5 \oplus \Ex^2 \C^5 \oplus 
   \Ex^3 \C^5 \oplus \Ex^4 \C^5 \oplus \Ex^5 \C^5    \]
unify some particles we normally consider to be `matter' with some
we normally consider `antimatter', as in
\[ \Ex^1 \C^5 \iso \angantilep \oplus \langle d_R \rangle. \]
In the Standard Model representation, we can think of the matter-antimatter
distinction as a $\Z_2$-grading, because the Standard Model representation $F
\oplus F^*$ splits into $F$ and $F^*$. By failing to respect this
grading, the $\SU(5)$ symmetry group fails to preserve the usual
distinction between matter and antimatter.

But the Standard Model has another $\Z_2$-grading that $\SU(5)$
\emph{does} respect.  This is the distinction between left- and
right-handedness.  Remember, the left-handed particles and
antiparticles live in the even grades:
\[ \Exev \C^5 \iso \langle \nubar_L \rangle \oplus \langle e^+_L \rangle \oplus \angquark \oplus \langle \ubar_L \rangle \oplus \anglep \oplus \langle \dbar_L \rangle \]
while the right-handed ones live in the odd grades:
\[ \Exodd \C^5 \iso \langle \nu_R \rangle \oplus \langle e^-_R \rangle \oplus \angantiquark \oplus \langle u_R \rangle \oplus \angantilep \oplus \langle d_R \rangle. \]
The action of $\SU(5)$ automatically
preserves this $\Z_2$-grading, because it comes
from the $\Z$-grading on $\Ex \C^5$, which $\SU(5)$ already respects.

This characteristic of the $\SU(5)$ theory lives on in its extension to
$\Spin(10)$.  There, the distinction between left and right is the only
distinction among particles and antiparticles that $\Spin(10)$ knows about,
because $\Exev \C^5$ and $\Exodd \C^5$ are irreducible.  This says the
$\Spin(10)$ theory unifies \emph{all} left-handed particles and antiparticles,
and \emph{all} right-handed particles and antiparticles.

In contrast, the $\Spin(4) \times \Spin(6)$ theory was all about adding a
fourth `color', $w$, to represent leptons, and restoring a kind of symmetry
between left and right by introducing a right-handed $\SU(2)$ that treats
right-handed particles like the left-handed $\SU(2)$ treats left-handed
particles.

Unlike the $\SU(5)$ theory, the $\Spin(4) \times \Spin(6)$ theory respects
\emph{both} $\Z_2$-gradings in the Standard Model: the matter-antimatter
grading, and the right-left grading.  
The reason is that $\Spin(4) \times \Spin(6)$ respects the
$\Z_2 \times \Z_2$-grading on $\Ex \C^2 \otimes \Ex \C^3$, and we have:
\[
\begin{array}{ccccc}
	F_L      & \;\iso\; & \Exodd \C^2 & \otimes & \Exodd \C^3 \\
	F_R      & \;\iso\; & \Exev \C^2  & \otimes & \Exodd \C^3  \\
        F_L^* & \;\iso\; & \Exodd \C^2 & \otimes & \Exev \C^3 \\
	F_R^* & \;\iso\; & \Exev \C^2  & \otimes & \Exev \C^3 
\end{array}
\]
Moreover, the matter-antimatter grading and the right-left grading
are \emph{all} that $\Spin(4) \times \Spin(6)$ respects, since
each of the four spaces listed is an irrep of this group.

When we extend $\Spin(4) \times \Spin(6)$ to the $\Spin(10)$ theory, 
we identify $\Ex \C^2 \otimes \Ex \C^3$ with $\Lambda \C^5$.  Then the
$\Z_2 \times \Z_2$-grading on $\Ex \C^2 \otimes \Ex \C^3$ gives 
the $\Z_2$-grading on $\Lambda \C^5$ using addition in $\Z_2$.
This sounds rather technical, but it is as simple as ``even + odd = odd'':
\[ \Exodd \C^5 \quad \iso \quad 
\left(\Exev \C^2 \otimes \Exodd \C^3\right) \; \oplus \; 
\left(\Exodd \C^2 \otimes \Exev \C^3 \right) \quad \iso \quad
F_R \oplus F_L^* \]
and ``odd + odd = even'', ``even + even = even'':
\[ \Exev \C^5 \quad \iso \quad \left(\Exodd \C^2 \otimes \Exodd \C^3\right) 
\;\oplus\; \left(\Exev \C^2 \otimes \Exev \C^3\right) 
\quad \iso \quad F_L \oplus F_R^*  .\]
Recall that $F_R \oplus F_L^*$ consists of all the fermions
and antifermions that are {\it right-handed}, while 
$F_L \oplus F_R^*$ consists of the {\it left-handed} ones.

Furthermore, the two routes to the $\Spin(10)$ theory that we have described,
one going through $\SU(5)$ and the other through $\Spin(4) \times \Spin(6)$,
are compatible.  In other words, this cube commutes:
\[
\xymatrix{
& \GSM \ar[rr]^\phi \ar[dl]_\theta \ar'[d] [ddd] & & \SU(5) \ar[dl]_\psi \ar[ddd] \\
\Spin(4) \times \Spin(6) \ar[rr]^\eta \ar[ddd] & & \Spin(10) \ar[ddd] & \\
& & & \\
& \U(F \oplus F^*) \ar'[r]^(0.8){U(f)}[rr] \ar[dl]_-{\U(h)} & & \U(\Ex \C^5) \ar[dl]_-1 \\
\U(\Ex \C^2 \otimes \Ex \C^3) \ar[rr]^-{\U(g)}	& & \U(\Ex \C^5) & \\
}
\] 

So, all four theories fit together in an elegant algebraic pattern.
What this means for physics---if
anything---remains unknown.  Yet we cannot resist feeling that it
means something, and we cannot resist venturing a guess:
\emph{the Standard Model is exactly the theory that
reconciles the visions built into the $\SU(5)$ and $\Spin(4) \times \Spin(6)$
theories.}  

What this might mean is not yet precise, but since all these
theories involve symmetries and representations, the `reconciliation' must
take place at both those levels---and we can see \emph{this} 
in a precise way.
First, at the level of symmetries, our Lie groups are related by the
commutative square of homomorphisms:
\[
\xymatrix{
\GSM \ar[r]^\phi \ar[d]_\theta        & \SU(5) \ar[d]^\psi \\
\Spin(4) \times \Spin(6) \ar[r]^-\eta & \Spin(10)
}
\]
Because this commutes, the image of $\GSM$ lies in the intersection of the
images of $\Spin(4) \times \Spin(6)$ and $\SU(5)$ inside $\Spin(10)$.  But
we claim it is \emph{precisely} that intersection!

To see this, first recall that the image of a group under 
a homomorphism is just the quotient group formed by
modding out the kernel of that homomorphism.  If we do this for each of our
homomorphisms above, we get a commutative square of inclusions:
\[ 
\xymatrix{
\GSM/\Z_6 \ar@{^{(}->}[r] \ar@{^{(}->}[d] & \SU(5) \ar@{^{(}->}[d] \\
\displaystyle\frac{\Spin(4) \times \Spin(6)}{\Z_2}  \ar@{^{(}->}[r] & \Spin(10) \\
}
\]
This implies that
\[ \GSM/\Z_6 \subseteq \SU(5) \cap \left( \frac{\Spin(4) \times \Spin(6)}{\Z_2} \right) \]
as subgroups of $\Spin(10)$.   To make good on our claim, we must
show these subgroups are equal:
\[ \GSM/\Z_6 = 
\SU(5) \cap \left( \frac{\Spin(4) \times \Spin(6)}{\Z_2} \right) .
\]
In other words, our commutative square of inclusions is a `pullback
square'.

As a step towards showing this, first consider what happens when we pass from 
the spin groups to the rotation groups. We can accomplish this by modding out 
by an additional $\Z_2$ above.  We get another commutative square of
inclusions:
\[
\xymatrix{
\GSM/\Z_6 \ar@{^{(}->}[r] \ar@{^{(}->}[d] & \SU(5) \ar@{^{(}->}[d] \\
\SO(4) \times \SO(6)  \ar@{^{(}->}[r] & \SO(10) \\
}
\]
Here the reader may wonder why we could quotient $(\Spin(4) \times
\Spin(6))/\Z_2$ and $\Spin(10)$ by $\Z_2$ without having to do the same for
their respective subgroups, $\GSM/\Z_6$ and $\SU(5)$. It is because $\Z_2$
intersects both of those subgroups trivially.  We can see this for $\SU(5)$ 
because we know the inclusion $\SU(5) \inclusion \Spin(10)$ 
is just the lift of the inclusion $\SU(5) \inclusion \SO(10)$
to universal covers, so it makes this diagram commute:
\[
\xymatrix{
\SU(5) \ar@{^{(}->}[r] \ar@{^{(}->}[dr] & \Spin(10) \ar[d]^p \\
                                        & \SO(10) \\
}
\]
But this means that $\SU(5)$ intersects $\Z_2 = \ker p$ in
only the identity.  The subgroup $\GSM/\Z_6$ therefore 
intersects $\Z_2$ trivially as well.

Now, let us show:
\begin{thm}\et
\label{thm:pullbackSO(10)}
$\GSM/\Z_6 = \SU(5) \cap ( \SO(4) \times \SO(6) ) \subseteq \SO(10).$ 
\end{thm}

\emph{Proof.} We can prove this in the same manner that we showed, in
Section~\ref{sec:su(5)}, that
\[ \GSM/\Z_6 \iso \S(\U(2) \times \U(3)) \subseteq \SU(5) \]
is precisely the subgroup of $\SU(5)$ that preserves the $2 + 3$ splitting of
$\C^5 \iso \C^2 \oplus \C^3$.

To begin with, the group $\SO(10)$ is the group of orientation-preserving
symmetries of the 10-dimensional real inner product space $\R^{10}$. But
$\R^{10}$ is suspiciously like $\C^5$, a 5-dimensional complex inner product
space. Indeed, if we forget the complex structure on $\C^5$, we get an
isomorphism $\C^5 \iso \R^{10}$, a real inner product space with symmetries
$\SO(10)$.  We can consider the subgroup of $\SO(10)$ that preserves the
original complex structure. This is $\U(5) \subseteq \SO(10)$. If we further
pick a volume form on $\C^5$, i.e. a nonzero element of $\Ex^5 \C^5$, and look
at the symmetries fixing that volume form, we get a copy of $\SU(5) \subseteq
\SO(10)$.

Then we can pick a $2+3$ splitting on $\C^5 \iso \C^2 \oplus \C^3$.  The
subgroup of $\SU(5)$ that also preserves this is
\[ \S(\U(2) \times \U(3)) \inclusion \SU(5) \inclusion \SO(10). \]
These inclusions form the top and right sides of our square: 
\[
\xymatrix{
\GSM/\Z_6 \ar@{^{(}->}[r] \ar@{^{(}->}[d] & \SU(5) \ar@{^{(}->}[d] \\
\SO(4) \times \SO(6)  \ar@{^{(}->}[r] & \SO(10) \\
}
\]

We can also reverse the order of these processes. Imposing a $2+3$
splitting on $\C^5$ yields a $4 + 6$ splitting on the underlying real vector
space, $\R^{10} \iso \R^4 \oplus \R^6$. The subgroup of $\SO(10)$ that
preserves this splitting is $\S(\O(4) \times \O(6))$: the
block diagonal matrices with $4 \times 4$ and $6 \times 6$ orthogonal blocks
and overall determinant 1.  The connected component of this subgroup
is $\SO(4) \times \SO(6)$. 

The direct summands in $\R^4 \oplus \R^6$ came from forgetting the complex
structure on $\C^2 \oplus \C^3$. The subgroup of $\S(\O(4) \times \O(6))$
preserving the original complex structure is $\U(2) \times \U(3)$, 
and the subgroup of this that also fixes a volume
form on $\C^5$ is $\S(\U(2) \times \U(3))$.
This group is connected, so it must lie entirely
in the connected component of the identity, and we get the inclusions:
\[ 
\S(\U(2) \times \U(3)) \inclusion \SO(4) \times \SO(6) \inclusion \SO(10). \]
These maps form the left and bottom sides of our square. 

It follows that $\GSM/\Z_6$ is precisely the subgroup of $\SO(10)$
that preserves a complex structure on $\R^{10}$, a chosen volume form on the
resulting complex vector space, and a $2 + 3$ splitting on this space. But this
$2 + 3$ splitting is the same as a \emph{compatible} $4 + 6$ splitting of
$\R^{10}$, one in which each summand is a complex vector subspace as well as a
real subspace. This means that
\[ \GSM/\Z_6 = \SU(5) \cap \S(\O(4) \times \O(6)) \subseteq \SO(10), \]
and since $\GSM/\Z_6$ is connected,
\[ \GSM/\Z_6 = \SU(5) \cap ( \SO(4) \times \SO(6) ) \subseteq \SO(10) \]
as desired.  \qed

From this, a little diagram chase proves our earlier claim:

\begin{thm}\et 
\label{thm:pullbackSpin(10)}
$\GSM/\Z_6 = \SU(5) \cap \left(\Spin(4) \times \Spin(6)\right)/\Z_2
\subseteq \Spin(10)$.
\end{thm}

\emph{Proof.}  By now we have built the following commutative diagram:
\[ 
\xymatrix{
\GSM/\Z_6 \ar@{^{(}->}[r]^-{\tilde \phi} \ar@{^{(}->}[d]_{\tilde \theta} 
& \SU(5) \ar@{^{(}->}[d]^\psi \\
(\Spin(4) \times \Spin(6))/{\Z_2} \ar@{^{(}->}[r]^-{\tilde \eta}
\ar[d]_q & \Spin(10) \ar[d]^p \\
\SO(4) \times \SO(6)  \ar@{^{(}->}[r]^-i & \SO(10) 
}
\]
where both the bottom vertical arrows are two-to-one, but 
the composite vertical maps $q \tilde{\theta}$ 
and $p \psi$ are one-to-one.  Our previous theorem says that the 
big square with $q \tilde{\theta}$ and $p \psi$ as vertical sides 
is a pullback.  Now we must show that the upper square is also a pullback.  
So, suppose we are given $g \in (\Spin(4) \times \Spin(6))/{\Z_2}$
and $g' \in \SU(5)$ with 
\[  \tilde{\eta}(g) = \psi(g'). \] 
We need to show there exists $x \in \GSM/\Z_6$ such that 
\[    \tilde{\theta} (x) = g, \qquad  \tilde{\phi}(x) = g'.  \]
Now, we know that 
\[ i q (g) =  p \tilde{\eta}(g) = p \psi(g') \]
so since the big square is a pullback,
there exists $x \in  \GSM/\Z_6$ with
\[    q \tilde{\theta} (x) = q(g), \qquad  \tilde{\phi}(x) = g'.  \]
The second equation is half of what we need to show.  So, we only
need to check that the first equation implies $\tilde{\theta} (x) = g$.  

The kernel of $q$ consists of two elements, which we will 
simply call $\pm 1$.  Since $q \tilde{\theta} (x) = q(g)$, we know
\[       \pm \tilde{\theta}(x) = g . \]
Since $\tilde{\eta}(g) = \psi(g')$, we thus have 
\[  \tilde{\eta} (\pm \tilde{\theta}(x)) = \psi(g') = \psi \tilde{\phi}(x) .\]
The one-to-one map $\tilde{\eta}$ sends the kernel of $q$ to the kernel of
$p$, which consists of two elements that we may again call $\pm 1$.  So,
$\pm \tilde{\eta} \tilde{\theta}(x) = \psi \tilde{\phi}(x)$. 
On the other hand, since the top square commutes we know
$\tilde{\eta} \tilde{\theta}(x) = \psi \tilde{\phi}(x)$. 
Thus the element $\pm 1$ must actually be $1$,
so $g = \tilde{\theta}(x)$ as desired.  \qed

In short, the Standard Model has precisely
the symmetries shared by both the $\SU(5)$ theory and the 
$\Spin(4) \times \Spin(6)$ theory.   Now let us see what this
means for the Standard Model representation. 

We can `break the symmetry' of the $\Spin(10)$ theory in two different ways.
In the first way, we start by picking the subgroup of $\Spin(10)$ that 
preserves the $\Z$-grading and volume form in $\Ex \C^5$.  This is $\SU(5)$.  
Then we pick the subgroup of $\SU(5)$ that respects 
the splitting of $\C^5$ into $\C^2 \oplus \C^3$.   This subgroup is
the Standard Model gauge group, modulo a discrete subgroup,
and its representation on $\Ex \C^5$ is the Standard Model representation.

We can draw this symmetry breaking process in the following diagram:
\[
\xymatrix{
\GSM \ar[r]^-\phi \ar[d] & \SU(5) \ar[r]^-\psi \ar[d] & \Spin(10) \ar[d] \\
\U(F \oplus F^*) \ar[r]^-{\U(f)} & \U(\Ex \C^5) \ar[r]^-1 & \U(\Ex \C^5) \\
\ar@{<~}[r]^{\mbox{ splitting}} & \ar@{<~}[r]^{\mbox{grading and}}_{\mbox{volume form}} &
}
\]
The $\SU(5)$ theory shows up as a `halfway house' here.

We can also break the symmetry of $\Spin(10)$ in a way that uses
the $\Spin(4) \times \Spin(6)$ theory as a halfway house.
We do essentially the same two steps as before, but in the reverse order! 
This time we start by picking the subgroup of $\Spin(10)$
that respects the splitting of $\R^{10}$ as $\R^4 \oplus \R^6$.
This subgroup is $\Spin(4) \times \Spin(6)$ modulo a discrete subgroup.
The two factors in this subgroup act separately on the factors of
$\Ex \C^5 \iso \Ex \C^2 \otimes \Ex \C^3$.  Then we pick the
subgroup of $\Spin(4) \times \Spin(6)$ that respects the $\Z$-grading 
and volume form on $\Ex \C^5$. 
This subgroup is the Standard Model gauge group, modulo a discrete subgroup,
and its representation on $\Ex \C^2 \otimes \Ex \C^3$ is the Standard Model
representation.

We can draw this alternate symmetry breaking process in the following
diagram:
\[
\xymatrix{
\GSM \ar[r]^-\theta \ar[d] & \Spin(4) \times \Spin(6) \ar[r]^-\eta \ar[d] & \Spin(10) \ar[d] \\  
\U(F \oplus F^*) \ar[r]^-{\U(h)} & \U(\Ex \C^2 \otimes \Ex \C^3) \ar[r]^-{\U(g)} & \U(\Ex \C^5) \\
\ar@{<~}[r]^{\mbox{grading and}}_{\mbox{volume form}} & \ar@{<~}[r]^{\mbox{ splitting}} &
}
\]
Will these tantalizing patterns help us understand physics 
beyond the Standard Model?   Only time will tell.

\subsubsection*{Acknowledgements}

We thank the denizens of the $n$-Category Caf\'e for catching mistakes
and suggesting stylistic improvements.  This research was supported by
a grant from the Foundational Questions Institute.


\begin{thebibliography}{99}

\bibitem{adams:lelg}
J.\ Frank Adams, {\em {Lectures on Exceptional Lie Groups}},  eds. Zafer\ Mahmoud
and Mamoru\ Mimura, University of Chicago Press, Chicago, 1996.

\bibitem{ABS}
Michael F.\ Atiyah, Raoul Bott and Arnold Shapiro, Clifford modules,
{\em Topology} {\bf 3} (1964), 3--38. 

\bibitem{BaezMuniain}
John\ C.\ Baez and Javier\ P.\ Muniain, {\em {Gauge Fields, Knots and
Gravity}}, World Scientific, Singapore, 1994.

\bibitem{BLM}
Stefano Bertolini, Luca Di Luzio and Michal Malinsky, 
Intermediate mass scales in the non-supersymmetric SO(10) grand 
unification: a reappraisal, available as
\href{http://arxiv.org/abs/hep-ph/0903.4049}
{\texttt{arXiv:0903.4049}}. 

\bibitem{Brown}
Lowell Brown, {\em {Quantum Field Theory}}, Cambridge U.\ Press, Cambridge,
1994.  

\bibitem{Chevalley} Claude Chevalley, {\em {The Algebraic Theory of
Spinors and Clifford Algebras}}, Springer, Berlin, 1996.

\bibitem{CassenCondon:nuclearforces}
Benedict\ Cassen and Edward\ U.\ Condon, On Nuclear Forces, 
{\em {Phys.\ Rev.\ }}{\bf 50} (1936), 846; reprinted in 
D.\ M.\ Brink, {\em {Nuclear Forces}}, Pergamon, Oxford, 1965, pp. 193--201.

\bibitem{CreaseMann:sc}
Robert\ P.\ Crease and Charles\ C.\ Mann, {\em {The Second Creation: Makers of
the Revolution in Twentieth-Century Physics}}, Rutgers University Press, New
Brunswick, New Jersey, 1996.

\bibitem{derdzinski}
Andrzej Derdzinski, {\em {Geometry of the Standard Model of Elementary
Particles}}, Springer, Berlin, 1992.

\bibitem{georgi:so(10)}
Howard Georgi, The state of the art---gauge theories, in {\em Particles and
Fields---1974}, ed.\ Carl\ E.\ Carlson, AIP Conference Proceedings 23, 1975,
pp.\  575--582.

\bibitem{georgi:lie}
Howard Georgi, {\em {Lie Algebras in Particle Physics: from 
Isospin to Unified Theories}}, Westview Press, Boulder, Colorado, 1999. 

\bibitem{GeorgiGlashow:su(5)}
Howard\ Georgi and Sheldon\ Glashow, Unity of all elementary-particle forces,
{\em Phys.\ Rev.\ Lett.\ }{\bf 32(8)} {Feb\ 1974}, 438--441. 

\bibitem{griffiths:intro}
David Griffiths, {\em {Introduction to Elementary Particles}}, Wiley, New
York 1987.

\bibitem{hall}
Brian Hall, {\em {Lie Groups, Lie Algebras, and Representations}}, 
Springer, Berlin, 2003. 

\bibitem{heisenberg:77} Werner Heisenberg, \"Uber den Bau der Atomkerne.\ I., {\em {Zeitschr.\ f.\ Phys.\
}}{\bf 77} (1932), 1--11; English translation in D.\ M.\ Brink, {\em {Nuclear
Forces}}, Pergamon, Oxford, 1965, pp. 144--154.

\bibitem{hoddeson} Laurie Brown, Max Dresden, Lillian Hoddeson and 
Michael Riordan, eds., {\em {The Rise of the Standard Model}}, 
Cambridge U.\ Press, Cambridge, 1997.
 
\bibitem{huang:qlgf}
Kerson\ Huang, {\em {Quarks, Leptons \& Gauge Fields}}, World Scientific,
Singapore, 1992.

\bibitem{isham}
Chris Isham, {\em {Modern Differential Geometry for Physicists}}, 
World Scientific, Singapore, 1999.

\bibitem{lee:pp}
T.\ D.\ Lee, {\em {Particle Physics and Introduction to Field Theory}}, 
Harwood, 1981.

\bibitem{lipkin}
Harry J.\ Lipkin, {\em {Lie Groups for Pedestrians}}, Dover, Mineola,
New York, 2002.

\bibitem{mohapatra:us}
R.\ N.\ Mohapatra, {\em {Unification and Supersymmetry: The Frontiers of
Quark-Lepton Physics}}, Springer, 1992.

\bibitem{naber:foundations}
Gregory L.\ Naber, {\em{Topology, Geometry and Gauge Fields: Foundations}}, 
Springer, Berlin, 1997. 

\bibitem{naber:interactions}
Gregory L.\ Naber, {\em{Topology, Geometry and Gauge Fields: Interactions}}, 
Springer, Berlin, 2000.

\bibitem{nakahara}
Mikio Nakahara, {\em {Geometry, Topology, and Physics}}, Academic Press,
1983.


\bibitem{pais:ib}
Abraham\ Pais, {\em {Inward Bound: Of Matter and Forces in the Physical World}},
Oxford University Press, 1988.

\bibitem{pati:decay}
Jogesh\ C.\ Pati, Proton decay: a must for theory, a challenge for experiment,
available as
\href{http://arxiv.org/abs/hep-ph/0005095}{\texttt{arXiv:hep-ph/0005095}}. 

\bibitem{pati:probing}
Jogesh\ C.\ Pati, Probing grand unification through neutrino oscillations,
leptogenesis, and proton decay, {\em {Int.\ J.\ Mod.\ Phys.\ A\ }}{\bf 18}
(2003), 4135--4156. Also available as
\href{http://arxiv.org/abs/hep-ph/0305221}{\texttt{arXiv:hep-ph/0305221}}.  

\bibitem{PatiSalam:model}
Jogesh\ C.\ Pati and Abdus\ Salam, Lepton number as the fourth ``color", {\em
{Phys.\ Rev.\ D\ }}{\bf 10} (1974), 275--289.

\bibitem{Peskin}
Michael\ E.\ Peskin, Beyond the Standard Model, available as 
\href{http://arxiv.org/abs/hep-ph/970549}{\texttt{arXiv:hep-ph/970549}}.

\bibitem{PeskinSchroeder:qft}
Michael\ E.\ Peskin and Dan\ V.\ Schroeder, {\em {An Introduction to Quantum
Field Theory}}, Westview Press, 1995.

\bibitem{ross:gut}
Graham G.\ Ross, {\em {Grand Unified Theories}}, Benjamin/Cummings, 1985.

\bibitem{Ryder}
Lewis H.\ Ryder, {\em {Quantum Field Theory}}, Cambridge U.\ Press, Cambridge,
1996.

\bibitem{segre:modern}
Emilio Segr{\`e}, {\em {From X-Rays to Quarks: Modern Physicists and Their
Discoveries}}, W.H.\ Freeman, San Francisco, 1980.

\bibitem{sternberg} Shlomo Sternberg, {\em {Group Theory and Physics}},
Cambridge U.\ Press, Cambridge, 1995.

\bibitem{srednicki:qft}
Mark Srednicki, {\em {Quantum Field Theory}}, Cambridge U. Press, 2007.
Also available at \href{http://www.physics.ucsb.edu/$\sim$mark/qft.html}
{\texttt{http://www.physics.ucsb.edu/$\sim$mark/qft.html}}.

\bibitem{sudbery}
Anthony Sudbery, {\em {Quantum Mechanics and the Particles of Nature: 
an Outline for Mathematicians}}, Cambridge U.\ Press, Cambridge, 1986.

\bibitem{ticciati:qft}
Robin Ticciati, {\em {Quantum Field Theory for Mathematicians}}, Cambridge U.\
Press, 1999.

\bibitem{tinkham}
Michael Tinkham, {\em {Group Theory and Quantum Mechanics}}, 
Dover, Mineola, New York, 2003.

\bibitem{witten:grandunification}
Edward Witten, Grand unification with and without supersymmetry, in {\em
{Introduction to supersymmetry in particle and nuclear physics}}, eds. O.
Castanos, A. Frank, L. Urrutia, Plenum Press, 1984, pp. 53--76.

\bibitem{zee:nutshell}
Anthony Zee, {\em {Quantum Field Theory in a Nutshell}}, Princeton U. Press,
Princeton, 2003.

\end{thebibliography}
\end{document}